\documentclass[12pt]{article}
\usepackage[T1]{fontenc}
\usepackage{lmodern}
\usepackage{graphics,lscape}
\usepackage{color}
\usepackage{multibib}
\newcites{OA}{Online Appendix References}
\usepackage{natbib}
\usepackage{multirow}
\usepackage{amsmath,amssymb,bm}
\usepackage{amsthm} 
\usepackage{graphicx}
\usepackage{enumitem}
\usepackage{float}
\usepackage{setspace}
\usepackage{booktabs}
\usepackage{epstopdf}
\usepackage{footnote}
\usepackage{geometry}
\usepackage{rotating}
\usepackage{titlesec}
\usepackage{lipsum} 
\usepackage{mathtools}
\usepackage{MnSymbol}
\usepackage[title]{appendix}
\usepackage[hyperfootnotes=false]{hyperref}
\usepackage{tikz}

\usepackage{adjustbox}

\hypersetup{
    colorlinks=true,
    citecolor=blue,
    linkcolor = red,
    filecolor = green,
    urlcolor = red,
}
\usepackage{fancyhdr}
\usepackage[section]{placeins}
\makesavenoteenv{tabular}
\usepackage[bottom]{footmisc}
\usepackage{fmtcount} 
\usepackage{caption}
\usepackage{srcltx} 
\usepackage{chngcntr}
\newcommand{\vectornorm}[1]{\left|\left|\right|\right|}
\makeatletter
\long\def\@makefigcaption#1#2{%
\vskip\abovecaptionskip
\sbox\@tempboxa{\textbf{#1 #2}}%
\global \@minipagefalse
\hb@xt@\hsize{\hfil\box\@tempboxa\hfil}%
\vskip\belowcaptionskip}
\long\def\@maketblcaption#1#2{%
\vskip\abovecaptionskip
\sbox\@tempboxa{\textbf{#1 #2}}%
\global \@minipagefalse
\hb@xt@\hsize{\hfil\box\@tempboxa\hfil}%
\vskip\belowcaptionskip}
\long\def\symbolfootnote[#1]#2{\begingroup%
\def\thefootnote{\fnsymbol{footnote}}\footnote[#1]{#2}\endgroup}

\newtheorem{lemma}{Lemma}

\newtheorem*{restrictionsimpulse responses}{Identification Based on Impulse Responses}

\newtheorem{Algorithm}{Algorithm}
 
\usepackage{microtype}
\usepackage[normalem]{ulem}
\usepackage{tabularx}
\newcolumntype{Y}{>{\centering\arraybackslash}X}

\usepackage{algorithm}
\usepackage{algpseudocode}
\usepackage{dsfont}
\DeclareMathOperator{\bfZ}{\mathbf{Z}}
\DeclareMathOperator{\bfZB}{\mathbf{Z}_{\bfB}}
\DeclareMathOperator{\bfZS}{\mathbf{Z}_{\bfS}}
\DeclareMathOperator{\bfZQ}{\mathbf{Z}_{\bfQ}}
\DeclareMathOperator{\bfZiB}{\mathbf{Z}^{i}_{\bfB}}
\DeclareMathOperator{\bfZimB}{\mathbf{Z}^{i-1}_{\bfB}}
\DeclareMathOperator{\bfZiS}{\mathbf{Z}^{i}_{\bfS}}
\DeclareMathOperator{\bfZimS}{\mathbf{Z}^{i-1}_{\bfS}}
\DeclareMathOperator{\bfZiQ}{\mathbf{Z}^{i}_{\bfQ}}
\DeclareMathOperator{\bfZimQ}{\mathbf{Z}^{i-1}_{\bfQ}}

\DeclareMathOperator{\wtA}{{\mathcal{A}}}
\DeclareMathOperator{\wtB}{{\mathcal{B}}}
\DeclareMathOperator{\wtC}{{\mathcal{C}}}
\DeclareMathOperator{\wtZ}{\mathcal{Z}}
\DeclareMathOperator{\bfQ}{\mathbf{Q}}
\DeclareMathOperator{\bfq}{\mathbf{q}}

\DeclareMathOperator{\bfV}{\mathbf{V}}
\DeclareMathOperator{\bfB}{\mathbf{B}}
\DeclareMathOperator{\bfA}{\mathbf{A}}
\DeclareMathOperator\bfOme{\mathbf{\Omega}}
\DeclareMathOperator{\bfL}{\mathbf{L}}
\DeclareMathOperator{\bfK}{\mathbf{K}}
\DeclareMathOperator{\bfy}{\mathbf{y}}
\DeclareMathOperator{\bfY}{\mathbf{Y}}
\DeclareMathOperator{\bfX}{\mathbf{X}}
\DeclareMathOperator{\bfx}{\mathbf{x}}
\DeclareMathOperator{\bfc}{\mathbf{c}}
\DeclareMathOperator{\bfd}{\mathbf{d}}
\DeclareMathOperator{\bfm}{\mathbf{m}}
\DeclareMathOperator{\bfb}{\mathbf{b}}
\DeclareMathOperator{\bfu}{\mathbf{u}}
\DeclareMathOperator{\bfw}{\mathbf{w}}
\DeclareMathOperator{\bfI}{\mathbf{I}}
\DeclareMathOperator{\bfe}{\boldsymbol{\varepsilon}}
\DeclareMathOperator{\bftheta}{\boldsymbol{\theta}}
\DeclareMathOperator{\bfbeta}{\boldsymbol{\beta}}
\DeclareMathOperator{\bfalpha}{\boldsymbol{\alpha}}
\DeclareMathOperator{\bfsigma}{\boldsymbol{\sigma}}
\DeclareMathOperator\bfS{\mathbf{\Sigma}}
\newcommand{\bfyT}{({\mathbf{y}}_t)_{t=1}^{T}}
\DeclareMathOperator\bfPsi{\mathbf{\Psi}}
\DeclareMathOperator\bfPhi{\mathbf{\Phi}}
\DeclareMathOperator{\bfmu}{\boldsymbol{\mu}}

\RequirePackage{subcaption} 
\usepackage{threeparttable,tablefootnote}
\allowdisplaybreaks

\newcommand{\be}{\begin{equation*}}
\newcommand{\ee}{\end{equation*}}

\usepackage{tabnotes}
\usepackage{multicol}
\newcounter{casenum}

\begin{document}
\hypersetup{pageanchor=false}
\begin{titlepage}

\title{\LARGE{\bf Large SVARs}\thanks{This paper supersedes ``A Gibbs Sampler for Efficient Bayesian Inference in Sign-Identified SVARs''. We thank Frank Schorfheide and Taeyoung Doh  for helpful comments. The views expressed in this paper are solely those of the authors and do not necessarily reflect the views of the Federal Reserve Bank of Atlanta, the Federal Reserve Bank of Philadelphia, or the Federal Reserve System. Any errors or omissions are the responsibility of the authors.}}

\author{
    \small{Jon{a}s E. Arias} \\ \small{Federal Reserve Bank of Philadelphia}  \and \small{Juan F. Rubio-Ram\'irez}\thanks{Corresponding author: Juan F. Rubio-Ram\'irez
      $<$\href{mailto:Juan_Rubio_Ramirez}{juan.rubio-ramirez@emory.edu}$>$, Economics Department, Emory University, Rich Memorial Building, Room 306,
Atlanta, Georgia 30322-2240.} \\\small{Emory University} \\ \small{Federal Reserve Bank of Atlanta} \and \small{Daniel Rudolf} \\
    \small{University of Passau}    \and \small{Minchul Shin} \\
\small{Federal Reserve Bank of
    Philadelphia}  }  
\date{\today}
\maketitle
\thispagestyle{empty}
\sloppy
\singlespacing
\begin{abstract}
We develop a new algorithm for inference in structural vector autoregressions (SVARs) identified with sign restrictions that can accommodate big data and modern identification schemes. The key innovation of our approach is to move beyond the traditional accept-reject framework commonly used in sign-identified SVARs. We show that an elliptical slice within Gibbs sampler can deliver dramatic gains in computational speed and render previously infeasible applications tractable. We also prove that the algorithm is well-defined, in the sense that its stationary distribution coincides with the posterior distribution of interest. To illustrate the approach in the context of sign-identified SVARs, we use a tractable example. We further assess the performance of our algorithm through two applications: a well-known small-SVAR model of the oil market featuring a tight identified set, and a large SVAR model with more than ten shocks and 100 sign restrictions.


\end{abstract}
\newpage
\end{titlepage}
\hypersetup{pageanchor=true}
\onehalfspacing
\newgeometry{tmargin=1.25in,bmargin=1.25in,lmargin=1.25in,rmargin=1.25in}
\onehalfspacing
\section{Introduction}\label{sec:theconcern}
The growing availability of large datasets has led to a renewed interest in the use of large-scale time-series models in economics. In univariate settings, \citet*{illusionofsparsity} show that densely parameterized models equipped with appropriate shrinkage priors typically outperform sparse alternatives in terms of predictive accuracy. In multivariate settings, \citet*{banbura2010} and \citet*{koop2013} have demonstrated that Bayesian shrinkage enables the estimation of large vector autoregressions (VARs) without compromising out-of-sample performance, and more recently, \citet*{crump2025large} show how this approach can be used for monetary policy analysis. These insights have direct implications for structural vector autoregressions (SVARs)---one of the major workhorses for studying the propagation of structural shocks in macroeconomics. If prediction tasks are better handled by large models, then inference on structural shocks using SVARs should similarly benefit from broader information sets, in line with early arguments in favor of using large SVARs to understand the macroeconomic effects of monetary policy \citep*[e.g.,][]{leeper1996,bernankefavar}.

Within the SVAR paradigm, sign restrictions have become a particularly popular method for identifying the parameters of interest, typically impulse responses. The conventional Bayesian approach to implementing sign restrictions—pioneered by \citet*{faust1998robustness}, \citet*{canova2002monetary}, and \citet*{uhlig2005effects} and extended by \citet*{RRWZ2010}—relies on sampling from the reduced-form posterior and a uniform prior over the set of orthogonal matrices, combined with an accept-reject approach to impose the restrictions. The method is straightforward to implement and produces independent draws. However, it becomes increasingly infeasible in large systems due to the vanishing probability of sampling admissible orthogonal matrices—especially as the number of sign restrictions needed to identify multiple structural shocks increases, tightening the identified set.\footnote{We use the term admissible to refer to orthogonal matrices that satisfy the sign restrictions given the reduced-form parameters.} Recent work by \citet*{chan2025largestructuralvarsmultiple} introduces algorithmic refinements by exploiting symmetry and permutation invariance in the space of orthogonal matrices, but even these improved methods also face limitations under tight identification. Notably, such computational burdens are not unique to high-dimensional models. These challenges have coincided with advances in identification strategies that also lead to tighter identified sets---even in low-dimensional SVARs---such as ranking restrictions and elasticity bounds \citep*{kilian2012agnostic,AmirAhmadiDrautzburg}, as well as narrative sign restrictions \citep*{antolinrubionarrative,sydneymaserena}. These approaches go beyond traditional sign restrictions while preserving their intuitive appeal and further reducing the volume of the admissible space of orthogonal matrices. Together, these trends---the adoption of larger information sets and the use of tighter identification strategies---and the limitations of accept-reject sampling methods underscore the need for alternative algorithms for Bayesian inference in SVARs identified with sign restrictions.

In this paper, we break with the state-of-the-art accept-reject tradition and show that embedding the elliptical slice sampling method of \citet*{Murray2010} within a Gibbs sampler---yielding an elliptical slice within Gibbs sampler---delivers substantial gains in computational speed, rendering previously infeasible applications tractable. Like in the conventional approach, using the uniform prior over the set of orthogonal matrices, our goal is to draw from the posterior distribution of the orthogonal reduced-form parameters conditional on sign restrictions. However, by eliminating the accept-reject step and directly conditioning on the sign restrictions within the Gibbs sampler, our algorithm overcomes the bottlenecks that arise under tight identification---thus enabling dynamic structural analysis with big data and rich identification schemes. We also show that the algorithm is well-defined by establishing that the stationary distribution of the elliptical slice within Gibbs sampler is the posterior distribution of interest. To illustrate the advantages of our approach, we consider a very simple example similar to the one in \citet*{granziera2018inference} and demonstrate that the efficiency of the {state-of-the-art} accept-reject algorithm hinges critically on the size of the identified set. As the identified set becomes tighter, the accept-reject algorithm slows down dramatically. In contrast, our Gibbs sampler efficiently shrinks the support of the candidate impulse responses toward the identified set, maintaining speed even under stringent restrictions.

For clarity and comparability with the literature, when describing our proposed Gibbs sampler algorithm, we adopt the conjugate normal-inverse-Wishart prior for the reduced-form parameters as our baseline. While this prior is popular due to its analytical convenience, it precludes cross-variable shrinkage. To address this limitation, we extend our algorithm to accommodate alternative priors, including the independent normal-inverse-Wishart and the asymmetric prior of \citet*{Chanassymetric}, both of which support cross-variable shrinkage. We evaluate the performance of our approach using two applications. In the first, we replicate \citet*{kilianmurphy2014}, a model of the world oil market in which the standard accept-reject algorithm fails. To address this infeasibility, \citet*{kilianmurphy2014} adopt an approach similar to that of \citet*{chan2025largestructuralvarsmultiple}, exploiting permutations and sign alternations. Our algorithm handles this application multiple times faster than the accept-reject approach, though the computational times of both approaches are within a range most practitioners would find acceptable. However, once we tighten the identified set by adding a restriction on the price elasticity of oil demand---motivated by \citet*{CALDARA20191}---the difference in performance becomes substantial: The {state-of-the-art} accept-reject algorithm moves from requiring about 20 minutes to produce 1,000 draws to nearly eight hours, whereas the computational time of our Gibbs sampler remains roughly constant, increasing only from about 2 minutes to 5 minutes for the same number of effective draws. In the second application, we revisit the structural analysis in \citet*{chan2025largestructuralvarsmultiple}, who use \citeauthor{crump2025large}’s \citeyearpar{crump2025large} large SVAR model of the U.S. economy to identify eight structural shocks. While the latter uses the Minnesota prior, the former relies on the asymmetric prior of \citet*{Chanassymetric}. To simplify the comparison, we revert to the Minnesota prior when applying both the accept-reject algorithm and the Gibbs sampler. We show that as the number of shocks under analysis increases, the efficiency of the algorithm in \citet*{chan2025largestructuralvarsmultiple} declines markedly, eventually becoming impractical. With ten shocks, it would take several days to obtain 1,000 draws. In contrast, the computational time of our Gibbs sampler is largely insensitive to the number of identified structural shocks. Even with ten shocks, it would take only a few minutes to obtain 1,000 effective draws. We also show that these striking differences are robust to using the asymmetric prior instead of the Minnesota prior.

We also highlight the contemporaneous contribution of \citet*{readzhu2025}, who propose an algorithm based on slice sampling, but it is limited to the conditionally uniform prior described in \citet*{uhlig2017shocks} and \citet*{AmirAhmadiDrautzburg}. While this prior can deliver substantial speed gains without the need for a Gibbs sampler, it does not satisfy the requirements set out in \citet*{arias2025uniform}. As we explain in Section~\ref{sec:pitfall}, this has important consequences: it implicitly alters the prior over the impulse responses in a way that depends on the identification scheme. This entanglement of inference and identification makes it difficult to know whether differences in posterior inference reflect genuine differences in identification or are merely artifacts of unintended changes in the prior distribution. By contrast, the uniform prior over orthogonal matrices satisfies the requirements in \citet*{arias2025uniform} and, therefore, guarantees that inference remains invariant to the set of imposed restrictions and allows researchers to cleanly separate the role of prior beliefs from the role of identification assumptions. 

The remainder of the paper is organized as follows. Sections \ref{sec:model} through \ref{sec:PriorsPosteriors} introduce the SVAR model, the sign restrictions, and the baseline conjugate uniform-normal-inverse-Wishart prior. Section~\ref{sec:simple} describes the problem in a simple environment. Section \ref{sec:newalgorithm} presents our elliptical slice within Gibbs sampler and outlines its theoretical properties. Section \ref{sec:applications} applies the algorithm to two empirical settings: a small SVAR model of the world oil market and a large SVAR model of the U.S. economy. Section~\ref{sec:pitfall} shows the shortcomings of the conditionally uniform prior. Section \ref{sec:conclusion} concludes. The Appendix adapts the algorithm to two popular priors in SVAR analysis, the independent normal-inverse-Wishart prior and the asymmetric prior proposed by \citet*{Chanassymetric}, and provides a robustness analysis.

\section{The Model}\label{sec:model}
Consider the SVAR with the general form,
\begin{equation}\label{eqn:systemmatrix}
\bfy_t^{\prime}\bfA_0 = \bfx_t^{\prime}\bfA_+ + \bfe_t^{\prime}, \quad 1 \leq t \leq T,
\end{equation}
where $\bfA_+^{\prime}=\left[\bfA_{1}^{\prime} \;\cdots\;\bfA_{p}^{\prime} \;\;{\bfc}^{\prime}\right]$ and
$\bfx_t^{\prime}=\left[\bfy_{t-1}^{\prime} \;\cdots\; \bfy_{t-p}^{\prime} \;\;1 \right]$ for $1 \leq t \leq T$, and where $\bfy_t$ is an $n \times 1$ vector of endogenous variables, $\bfe_t$ is an $n \times 1$ vector of exogenous structural shocks, $\bfA_{\ell}$ is an $n \times n$ matrix of parameters for $0 \leq \ell \leq p$ with $\bfA_0$ invertible, $\bfc$ is a $1 \times n$ vector of parameters, $p$ is the lag length, and $T$ is the sample size. Hence, the dimension of $\bfA_+$ is $m \times n$, where $m = np + 1$. The vector $\bfe_t$, conditional on past information and the initial conditions $\bfy_0, \ldots, \bfy_{1-p}$, is Gaussian with mean zero and covariance matrix $\bfI_n$, the $n \times n$ identity matrix.

The reduced-form representation implied by Equation~\eqref{eqn:systemmatrix} is
\begin{equation}\label{eqn:reducedform}
\bfy_t^{\prime}= \bfx_t^{\prime}\bfB + \bfu_t^{\prime}, \text{ for } 1 \leq t \leq T,
\end{equation}
where $\bfB =\bfA_+\bfA_0^{-1}$, $\bfu_t^{\prime}=\bfe_t^{\prime}\bfA_0^{-1}$, and $\mathbb{E}\left[\bfu_t\bfu_t^{\prime}\right]=\bfS=\left(\bfA_0\bfA_0^{\prime}\right)^{-1}$. The matrices $\bfB$ and $\bfS$ are the reduced-form parameters, while $\bfA_0$ and $\bfA_+$ are the structural parameters. While $\bfB$ is an $m \times n$ matrix, $\bfS$ belongs to the set $\mathcal{S}(n)$, which is the set of $n \times n$ positive definite matrices. It will be useful to partition $\bfB$ as follows: $\bfB=\left[\bfB_{1}^{\prime} \;\cdots\; \bfB_{p}^{\prime} \;\;{\bf{d}}^{\prime}\right]^{\prime}$ where $\bfB_{\ell}$ is an $n \times n$ matrix of parameters for $1 \le \ell \le p$, and ${\bf{d}}$ is a $1 \times n$ vector of parameters.

It is well known that for linear Gaussian models of the type studied in this paper, $(\bfA_0, \bfA_+)$ and $(\tilde\bfA_0, \tilde\bfA_+)$ are observationally equivalent if and only if they have the same reduced-form representation. This implies that the structural parameters $(\bfA_0, \bfA_+)$ and $(\tilde\bfA_0, \tilde\bfA_+)$ are observationally equivalent if and only if $\bfA_0 = \tilde\bfA_0 \bfQ$ and $\bfA_+ = \tilde\bfA_+ \bfQ$ for some $\bfQ \in \mathcal{O}(n)$, where $\mathcal{O}(n)$ is the set of all $n \times n$ orthogonal matrices. To solve the identification problem, one often imposes sign restrictions on either the structural parameters or some function of the structural parameters, such as the impulse responses. To simplify the notation, we summarize the sign restrictions by $\mathbf{S}_S(\bfA_0, \bfA_+) > \bf0$, where $\bf0$ is a column vector of the appropriate dimension, and let $\left[\mathbf{S}_S(\bfA_0, \bfA_+) > \bf0\right]$ be an indicator function that equals one if the sign restrictions are satisfied and zero otherwise. We will consider continuous functions $\mathbf{S}_S$.

\section{The Orthogonal Reduced-Form Parameterization}\label{sec:orthogonalreducedform}
Equation~\eqref{eqn:systemmatrix} represents the SVAR in terms of the structural parameterization, which is characterized by $(\bfA_0,\bfA_+)$. The SVAR can alternatively be written in what we call the orthogonal reduced-form parameterization; see \citet*{arrw2018}. This parameterization is characterized by the reduced-form parameters $(\bfB,\bfS)$ together with an orthogonal matrix $\bfQ$, and is given by the following equation:
\begin{equation}\label{eqn:systemmatrixorthogonalreducedform}
\bfy_t^{\prime} = \bfx_t^{\prime}\bfB + \bfe_t^{\prime}\bfQ^{\prime}h(\bfS), \text{ for } 1 \le t \le T,
\end{equation}
where the $n \times n$ matrix $h(\bfS)$ is any decomposition of the covariance matrix $\bfS$ satisfying $h(\bfS)^{\prime}h(\bfS) = \bfS$. We take $h$ to be the Cholesky decomposition, though any differentiable decomposition would suffice.  

Given Equations~\eqref{eqn:systemmatrix} and \eqref{eqn:systemmatrixorthogonalreducedform}, we can define a mapping between $(\bfB, \bfS, \bfQ)$ and $(\bfA_0, \bfA_+)$ by
\begin{equation*}
f(\bfB, \bfS, \bfQ) = \big( \underbrace{h(\bfS)^{-1} \bfQ}_{\bfA_0}, \underbrace{\bfB h(\bfS)^{-1} \bfQ}_{\bfA_+} \big).
\end{equation*}
This mapping makes clear how the structural parameters depend on the reduced-form parameters and orthogonal matrices. Given the reduced-form parameters, each value of $\bfQ \in \mathcal{O}(n)$ can be viewed as a particular choice among observationally equivalent structural parameters. Thus, we can always write the sign restrictions in terms of the orthogonal reduced-form parameterization. Hence, let $\left[\mathbf{S}_R(\bfB, \bfS, \bfQ) > {\bf0}\right]$ be an indicator function in terms of the orthogonal reduced-form parameterization that equals one if the sign restrictions are satisfied and zero otherwise, where $\mathbf{S}_R(\bfB, \bfS, \bfQ)=\mathbf{S}_S(f(\bfB, \bfS, \bfQ))$. Because both $\mathbf{S}_S$ and $f$ are continuous, $\mathbf{S}_R$ is also continuous.  

We can also define the impulse responses. Let $\bfu_t = \bfL_0 \bfe_t$ for $1 \le t \le T$, where $\bfL_0$ is an $n \times n$ invertible matrix that represents the impulse responses at horizon zero. Given $\bfL_0$ and $\bfB$, it is possible to obtain the impulse responses beyond horizon zero recursively as
\begin{equation}\label{eqn:RtoL}
\bfL_{\ell} = \sum_{k=1}^{\min \{\ell, p\}} \bfB_k^{\prime} \bfL_{\ell - k}, \quad \ell > 0.
\end{equation}
We combine the impulse responses from horizons one through $p$ and the constant term ${\bfc}$ into a single matrix, $\bfL_+ = \left[ \bfL_1^{\prime} \;\cdots\; \bfL_p^{\prime} \;\; \bfc^{\prime} \right]^{\prime}$, where the maximum horizon of the impulse response in ${\bf L}_+$ matches the lag length in Equation~\eqref{eqn:systemmatrix}. The impulse response parameterization is characterized by $(\bfL_0,\bfL_+)$. Given the function $f$ and Equation~\eqref{eqn:RtoL}, we can also define a mapping from $(\bfB, \bfS, \bfQ)$ to $(\bfL_0, \bfL_+)$ by
\begin{equation}\label{eqn:RQtoL}
\phi(\bfB, \bfS, \bfQ) = \big( \underbrace{h(\bfS)^{\prime} \bfQ}_{\bfL_0}, 
\underbrace{\left[ \bfL_1(\bfB, \bfS, \bfQ)^{\prime} \; \cdots \; \bfL_p(\bfB, \bfS, \bfQ)^{\prime} \;\; \bfQ^{\prime} (h(\bfS)^{-1})^{\prime} \bfd^{\prime} \right]^{\prime}}_{\bfL_+} \big),
\end{equation}
where ${\bf L}_\ell(\bfB, \bfS, \bfQ)$ for $1 \le \ell \le p$ is implicitly defined by Equation~\eqref{eqn:RtoL}, and $\bfd = \bfc\bfL_0^{\prime}$. The functions $f$ and $\phi$ are invertible, and $f$, $\phi$, and their inverses are differentiable.

\section{Conjugate Priors and Posteriors}\label{sec:PriorsPosteriors}
For the reduced-form representation in Equation~\eqref{eqn:reducedform}, the normal-inverse-Wishart family of distributions is conjugate. A conjugate normal-inverse-Wishart distribution over the reduced-form parameters is characterized by four parameters: a scalar $\nu\geq n$, an $n\times n$ symmetric and positive definite matrix $\bfPhi$, an $m\times n$ matrix $\bfPsi$, and an $m\times m$ symmetric and positive definite matrix $\bfOme$. We denote this distribution by $NIW(\nu,\bfPhi,\bfPsi,\bfOme)$ and its density by $NIW_{(\nu,\bfPhi,\bfPsi,\bfOme)}(\bfB,\bfS)$. Furthermore, 
\be
\begin{gathered}
NIW_{(\nu,\bfPhi,\bfPsi,\bfOme)}(\bfB,\bfS)= \\
\adjustbox{max width=\textwidth}{$\displaystyle c_{\mathrm{NIW}}(\nu,\bfPhi,\bfOme)\underbrace{|\det(\bfS)|^{-\frac{\nu+n+1}{2}} e^{-\frac{1}{2}\operatorname{tr}(\bfPhi\bfS^{-1})}}_{\text{inverse-Wishart}}\times\underbrace{|\det(\bfS)|^{-\frac{m}{2}} e^{-\frac{1}{2}\operatorname{vec}(\bfB - \bfPsi)^{\prime}(\bfS\otimes\bfOme)^{-1}\operatorname{vec}(\bfB - {\bf \Psi})}}_{\text{conditionally normal}}.$}
\end{gathered}
\ee
where
\begin{equation*}
c_{\mathrm{NIW}}(\nu,\bfPhi,\bfOme)=\frac{|\det(\bfPhi)|^{\nu/2}}{2^{\nu n/2}\Gamma_{n}(\nu/2)(2\pi)^{mn/2}|\det(\bfOme)|^{n/2}}.
\end{equation*}
If the prior {distribution} {over the reduced-form parameters} is $NIW(\bar{\nu},\bar\bfPhi,\bar\bfPsi,\bar\bfOme)$, then the posterior {distribution} over the reduced-form parameters is $NIW(\tilde{\nu},\tilde\bfPhi,\tilde\bfPsi,\tilde\bfOme)$, where
\begin{gather}
\tilde{\nu}=T+\bar{\nu}, \nonumber \\
\tilde\bfOme=({\bf X}^\prime{\bf X}+\bar\bfOme^{-1})^{-1}, \nonumber \\
\tilde\bfPsi=\tilde\bfOme({\bf X}^\prime{\bf Y} + \bar\bfOme^{-1}\bar\bfPsi), \nonumber \\ 
\tilde\bfPhi={\bf Y}^\prime{\bf Y} + \bar\bfPhi+\bar\bfPsi^\prime\bar\bfOme^{-1}\bar\bfPsi - \tilde\bfPsi^\prime\tilde\bfOme^{-1}\tilde\bfPsi, \nonumber
\end{gather}
for $\bfY=[\bfy_1\;\;\cdots\;\;\bfy_T]^\prime$ and $\bfX=[\bfx_1\;\;\cdots\;\;\bfx_T]^\prime$. 

The conjugate normal-inverse-Wishart prior is widely used in Bayesian VARs due to its computational convenience and desirable properties \citep*[see][]{uhlig1994macroeconomists,faust1998robustness,uhlig2005effects,sims1998bayesian,RRWZ2010,kilian2012agnostic,kilianmurphy2014}. When combined with the conventional accept-reject approach, it produces independent draws from the posterior, which makes it especially attractive. However, it also imposes a Kronecker structure on the prior distribution of $\bfB$, thereby constraining its covariance matrix, and it rules out {cross-variable} shrinkage. Consequently, researchers {often} consider: (i) the independent normal-inverse-Wishart prior, which avoids the Kronecker covariance structure and allows for greater flexibility; and (ii) the asymmetric priors proposed by \citet*{Chanassymetric}, {which are becoming popular because they accommodate} cross-variable shrinkage. {Although} we present the methodology using the conjugate normal-inverse-Wishart prior, due to its prevalence in the literature, in Online Appendix \ref{app:alt_priors} we also adapt the algorithm to these alternative priors.

Given the results in \citet*{arias2025uniform}, we will combine the conjugate prior with the following uniform density over the set of orthogonal matrices:
\be
\pi(\bfQ \mid \bfB, \bfS) =
\begin{cases}
\kappa & \text{if } \bfQ \in \mathcal{O}(n), \\
0 & \text{otherwise},
\end{cases}
\ee
where $\int_{\mathcal{O}(n)} \kappa d\bfQ = 1$. This choice can be motivated by the fact that it assigns equal prior weight to both observationally equivalent vectors of impulse responses and observationally equivalent structural parameters \citep*[see][]{arias2025uniform}. We call this combination the conjugate uniform-normal-inverse-Wishart distribution over the orthogonal reduced-form parameterization; denote it by $UNIW(\nu,\bfPhi,\bfPsi,\bfOme)$, and denote its density over the orthogonal reduced-form parameterization by $UNIW_{(\nu,\bfPhi,\bfPsi,\bfOme)}(\bfB,\bfS,\bfQ)$. It is the case that 
\begin{equation}\label{eqn:priorcorrect}
UNIW_{(\nu,\bfPhi,\bfPsi,\bfOme)}(\bfB,\bfS,\bfQ) =
\begin{cases}
\kappa \, NIW_{(\nu,\bfPhi,\bfPsi,\bfOme)}(\bfB, \bfS) & \text{if } \bfQ \in \mathcal{O}(n), \\
0 & \text{otherwise}.
\end{cases}
\end{equation}

\subsection*{Inference Based on Sign Restrictions}\label{sec:inference}
Our objective will be to draw from the posterior of the orthogonal reduced-form parameters conditional on the sign restrictions,
\begin{equation}\label{eqn:posteriorSigns}
p(\bfB,\bfS,\bfQ \mid \bfyT, \mathbf{S}_R(\bfB,\bfS,\bfQ) > {\bf0}) =
\frac{\left[\mathbf{S}_R(\bfB,\bfS,\bfQ) > {\bf0}\right] UNIW_{(\tilde{\nu}, \tilde\bfPhi, \tilde\bfPsi, \tilde\bfOme)}(\bfB,\bfS,\bfQ)}{\Pr\left(\mathbf{S}_R(\bfB,\bfS,\bfQ) > {\bf0} \mid \bfyT\right)}, 
\end{equation}
where
\begin{align*}
    &\Pr\left(\mathbf{S}_R(\bfB,\bfS,\bfQ) > {\bf0} \mid \bfyT\right)=\\
    &\smashoperator[r]{\int_{\mathbb{R}^{m\times n}\times\mathcal{S}(n)\times\mathcal{O}(n)}} \left[\mathbf{S}_R(\bfB,\bfS,\bfQ) > {\bf0}\right]UNIW_{(\tilde{\nu}, \tilde\bfPhi, \tilde\bfPsi, \tilde\bfOme)}(\bfB,\bfS,\bfQ) d\bfB d\bfS d\bfQ,
\end{align*}
and then use $f$ and $\phi$ to transform the draws to the desired impulse responses. The traditional approach to obtain draws from Equation~\eqref{eqn:posteriorSigns} uses the following accept-reject algorithm:
\begin{Algorithm}\label{alg:basicdraw}
This algorithm independently draws from the conditional posterior in Equation~\eqref{eqn:posteriorSigns}.
\begin{enumerate}
\item \label{alg:BSigmabasicdraw} Draw $(\bfB,\bfS)$ independently from the $NIW(\tilde{\nu},\tilde\bfPhi,\tilde\bfPsi,\tilde\bfOme)$ distribution.
\item \label{alg:uniformQdraw} Draw $\bfQ$ independently from the uniform distribution over $\mathcal{O}(n)$.
\item Keep $(\bfB,\bfS,\bfQ)$ if $\left[\mathbf{S}_R(\bfB,\bfS,\bfQ) > {\bf0}\right]=1$. 
\item Return to Step \ref{alg:BSigmabasicdraw} until the required number of draws has been obtained.
\end{enumerate}
\end{Algorithm}

As mentioned above, while this algorithm has been widely adopted, it is well known that there are cases in which the identified set is narrow, limiting the efficiency of the algorithm \citep*[{see, e.g.,}][]{kilianmurphy2014,Baumeister_Hamilton_2024,chan2025largestructuralvarsmultiple,readzhu2025}. In the next section, we use a simple example to show its shortcomings. We will also show how a carefully designed elliptical slice within Gibbs sampler is not subject to this limitation and delivers dramatic speed gains. Importantly, \citet*{chan2025largestructuralvarsmultiple} present a numerically efficient version of Algorithm~\ref{alg:basicdraw}. Therefore, when comparing our algorithm to the traditional accept-reject approach, we will use this efficient version as the benchmark.

\section{The Problem with Accept-Reject Sampling}\label{sec:simple}
For the purposes of demonstrating the limitations of the accept-reject approach, it suffices to work with a simple example similar to the one explored by \citet*{granziera2018inference}. Thus, consider the following SVAR, with $n = 2$, $p = 0$, and ${\bfc}={\bf 0}$ written under the orthogonal reduced-form parameterization:
\be
\bfy_t^{\prime} = \left(y_{t,1}, y_{t,2}\right) = {\bm{\varepsilon}}_t^{\prime} (\bfS_{tr} \bfQ)^{\prime},
\ee
where we let $\bfS_{tr} = h(\bfS)^{\prime}$. Initially, we assume $\bfS_{tr}$ is known, but we will later relax this assumption. Let $\sigma_{tr, ij}$ denote the $i$-th row and $j$-th column entry of $\bfS_{tr}$. For simplicity, we set $\sigma_{tr, 11} = \sigma_{tr, 22} = 1$ and $\sigma_{tr, 21} = -0.9$. Note that the contemporaneous impact matrix ${\bf L}_{0}$ is defined as ${\bf L}_{0} = \bfS_{tr} \bfQ$. Henceforth, we focus on the impulse responses to the first shock—it is straightforward to extend our analysis to the second shock.

\begin{figure}[t!]
  \centering
  \begin{subfigure}[t]{0.46\linewidth}
    \centering
    \caption{Identified set}
    \label{fig:simple_iset}
    \includegraphics[width=\linewidth]{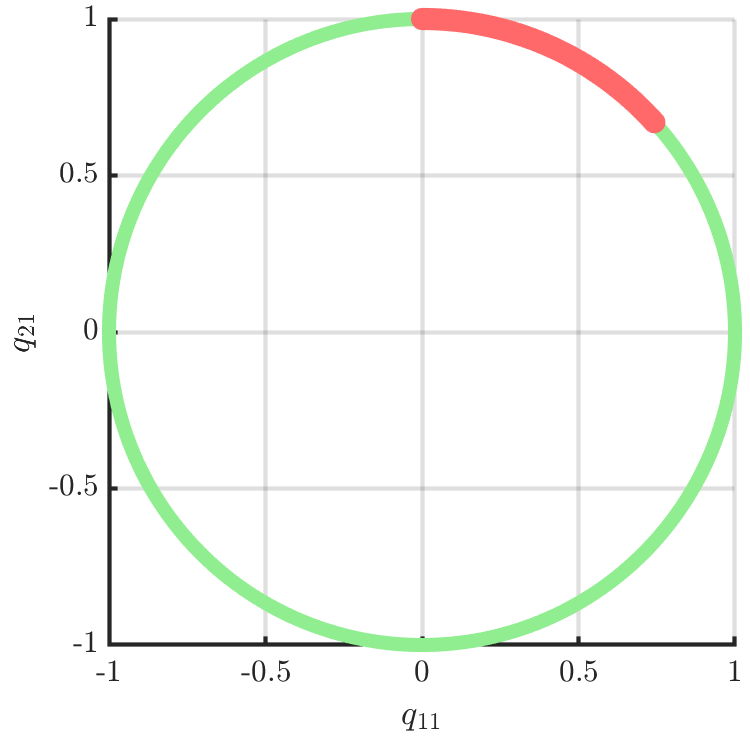}
  \end{subfigure}
  \hfill
  \begin{subfigure}[t]{0.46\linewidth}
    \centering
    \caption{Number of $\bfQ_1$ draws vs.\ arc length}
    \label{fig:simple_loop}
    \includegraphics[width=\linewidth]{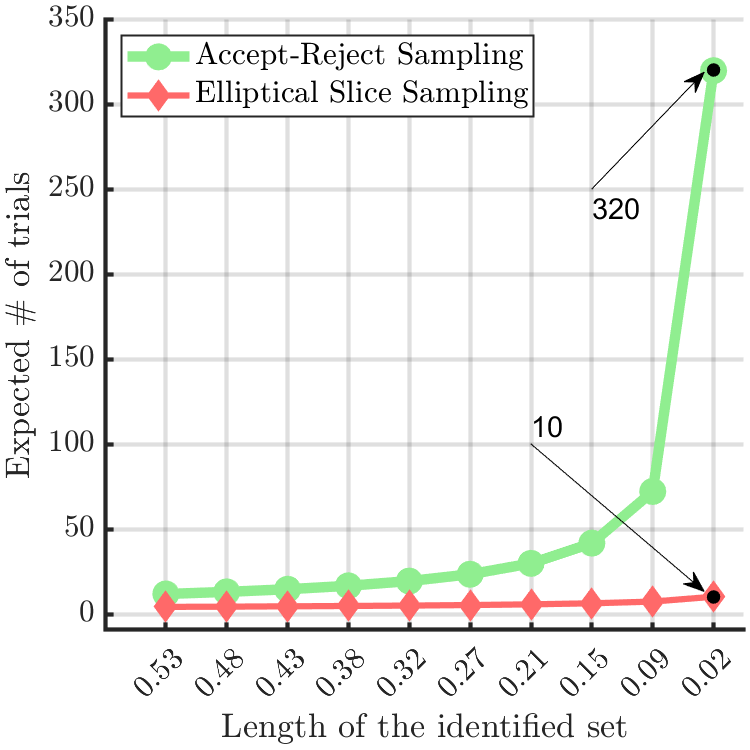}
  \end{subfigure}
  \caption{(a)~Identified set (red) and domain of $\left(q_{11}, q_{21}\right)^{\prime}$ (green).  
  (b)~Expected number of draws required to meet the sign restrictions as a function of the identified set size (arc length).}
  \label{fig:simple}
\end{figure}

Given the above, it is easy to see that the impact of the first shock on $y_{t,1}$ and $y_{t,2}$ can be written as $\ell_{11} = q_{11}$ and $\ell_{21} = -0.9 q_{11} + q_{21}$, where $\ell_{i1}$ and $q_{i1}$ are the $(i,1)$ entries of ${\bf L}_{0}$ and $\bfQ$, respectively. We now impose sign restrictions requiring that $\ell_{11}$ and $\ell_{21}$ are nonnegative. These sign restrictions imply $q_{11} \geq 0$ and $q_{21} \geq 0.9 q_{11}$. Figure~\ref{fig:simple_iset} illustrates this setup graphically. The green circle represents the domain of $\bfQ_1 = \left(q_{11}, q_{21}\right)^{\prime}$, while the red arc highlights the identified set that satisfies the imposed sign restrictions.

When using the popular accept-reject sampling approach described in Algorithm~\ref{alg:basicdraw}, obtaining a draw from the posterior distribution of impulse responses satisfying the sign restrictions involves drawing a $2 \times 1$ vector $\bfx_1$ from a $N({\bf 0}, \bfI_{2})$ distribution and converting it into a unit vector $\bfq_1$ via the normalization $\bfq_1 = \bfx_1 / ||\bfx_1||$. The draw is accepted only if $\bfq_1$ satisfies the sign restrictions. Unrestricted draws $\left(q_{11}, q_{21}\right)^{\prime}$ lie uniformly on the entire unit circle (depicted in green), whereas the accepted draws are uniformly distributed only over the subset of the unit circle that meets the sign restrictions (the red arc).

The efficiency of the posterior simulator based on this type of accept-reject algorithm depends heavily on the size of the identified set. As the identified set becomes tighter, we naturally expect to discard a larger number of draws. Indeed, the expected number of draws required to satisfy the sign restrictions is inversely proportional to the probability of meeting those restrictions. Figure~\ref{fig:simple_loop} illustrates this relationship analytically: the green line plots the expected number of draws needed to satisfy the sign restrictions as a function of the size of the identified set (i.e., the length of the red arc). More specifically, we generate smaller identified sets by gradually moving the left endpoint of the red arc toward its right endpoint. As shown in the figure, the expected number of draws required increases hyperbolically as the identified set shrinks. In realistic scenarios, as illustrated later in our empirical applications, the number of draws required can become quite large, rendering the algorithm inefficient.

In this paper, we propose an elliptical slice within Gibbs sampler, which draws from the identified set more efficiently. This method can be viewed as an {advanced}
Metropolis-Hastings algorithm that transitions from the previous draw $\bfx_1^{(0)}$ to a new draw using the following elliptical slice sampling proposal:
\be
\bfx_1^{(\star)} = \boldsymbol{\nu} \sin(\theta) + \bfx_1^{(0)} \cos(\theta)\text{, where } \theta \in [0, 2\pi],
\ee
where $\boldsymbol{\nu}$ is a $2 \times 1$ vector drawn from $N({\bf 0}, \bfI_{2})$. The scalar parameter $\theta$ controls the step size of the proposed move. For instance, when $\theta$ is close to $0$, the proposal is closer to the previous draw $\bfx_1^{(0)}$, whereas when $\theta$ approaches $\pi/2$, the proposal is closer to the newly drawn random vector $\boldsymbol{\nu}$. Unlike a conventional Metropolis-Hastings algorithm, the elliptical slice sampling {automatically} searches for a suitable step size to guarantee acceptance of the proposed draw at every iteration. Intuitively, given that the previous draw lies within the identified set, the elliptical slice sampling ensures that the new proposal $\bfx_1^{(\star)}$ also remains within the identified set by uniformly drawing $\theta$ from a candidate set that shrinks exponentially. Under appropriate regularity conditions, this procedure ensures the validity and convergence of the algorithm as long as the random variable of interest, in this case $\bfq_1$, can be written as a transformation of a normally distributed random variable \citep*[see][for details]{Murray2010,natarovskii2021,hasenpflug2025reversibility}. 

The fact that the candidate set for $\theta$ shrinks exponentially is an appealing feature, as it significantly reduces the number of candidate draws of $\bfq_1$ needed to satisfy the restrictions. This efficiency gain becomes particularly important as the dimension of the model increases, since generating new draws of $\bfq_1$ is computationally costly. Figure~\ref{fig:simple_loop} (red line) displays the average number of trials required by the elliptical slice sampling to generate an accepted draw of $\bfq_1$ within the identified set as a function of the length of the identified set. The number of required trials for the conventional accept-reject sampler grows hyperbolically, whereas that for elliptical slice sampling increases at a much slower rate.\footnote{To make a fair comparison, one should account for the autocorrelation introduced by {the elliptical slice within Gibbs sampler}, since the accept-reject algorithm generates independent draws. As demonstrated in Section \ref{sec:applications}, we compute the effective sample size and find that in this example the number of draws required to obtain one effective draw ranges from 1.04 to 1.35. Therefore, this adjustment does not alter the main conclusion illustrated in the figure.}

In the following section, we extend this simple example into a more realistic and useful setting by: (1) identifying multiple shocks simultaneously rather than just a single shock; (2) allowing sign restrictions to take a general form; and (3) developing an elliptical slice within Gibbs sampler that targets the posterior of the orthogonal reduced-form parameters conditional on the sign restrictions.

\section{Inference}\label{sec:newalgorithm}
In this section, we develop an elliptical slice within Gibbs sampler to obtain draws that approximate the posterior of the orthogonal reduced-form parameters conditional on the sign restrictions defined in Equation \eqref{eqn:posteriorSigns}. As we will demonstrate in Section \ref{sec:applications}, this algorithm can achieve order-of-magnitude gains when compared with the state-of-the-art accept-reject algorithm. For ease of exposition, we organize the section into four parts. Section \ref{sec:changeofvariable} establishes a mapping between the posterior distribution of interest and its implied posterior over a transformed random vector of matrices (defined below) that forms the basis of our algorithm. Section \ref{sec:samplerdesign} describes the design of our sampler. Section \ref{sec:algorithmmainGibbs} describes the algorithm. Finally, Section \ref{sec:welldefinedness} provides an argument for the well-definedness of our elliptical slice within Gibbs sampler.

\subsection{Change of Variables}\label{sec:changeofvariable}
Let $\bfZ=(\bfZB,\bfZS,\bfZQ)\in \wtZ$, with $\wtZ = \mathbb{R}^{m\times n} \times \mathbb{R}^{n\times \tilde{\nu}} \times \mathbb{R}^{n \times n}$, and let $p_{\bfZ}(\bfZ\mid \bfyT, \mathbf{S}_R(T(\bfZ)) > {\bf0})$ denote the following conditional Lebesgue density:
\begin{equation}\label{eqn:transf_post_dens}
 \frac{\left[\mathbf{S}_R(T(\bfZ)) > {\bf0}\right]
 N_{(\tilde\bfPsi, \tilde\bfOme, \zeta(\bfZS))}(\bfZB) N_{(\mathbf{0}_{n \times \tilde{\nu}}, \tilde{\bfPhi}^{-1}, \bfI_{\tilde{\nu}})}(\bfZS) N_{(\mathbf{0}_{n \times n}, \bfI_n, \bfI_n)}(\bfZQ)}{\Pr\left(\mathbf{S}_R(T(\bfZ)) > {\bf0} \mid \bfyT\right)},
\end{equation}
where 
\begin{equation*}
\begin{gathered}
\Pr\left(\mathbf{S}_R(T(\bfZ)) > {\bf0} \mid \bfyT\right)=\\
\adjustbox{max width=\textwidth}{$\displaystyle \smashoperator[r]{\int_{\mathbb{R}^{m\times n}\times\mathbb{R}^{n\times \tilde{\nu}}\times\mathbb{R}^{n\times n}}} \left[\mathbf{S}_R(T(\bfZ)) > {\bf0}\right] N_{(\tilde\bfPsi, \tilde\bfOme, \zeta(\bfZS))}(\bfZB) N_{(\mathbf{0}_{n \times \tilde{\nu}}, \tilde{\bfPhi}^{-1}, \bfI_{\tilde{\nu}})}(\bfZS) N_{(\mathbf{0}_{n \times n}, \bfI_n, \bfI_n)}(\bfZQ)\, d\bfZB\, d\bfZS\, d\bfZQ,$}
\end{gathered}
\end{equation*}
\noindent and $T(\bfZ)= (\bfZB,\zeta(\bfZS),\gamma(\bfZQ))$ is a transformed random vector in which $\gamma(\bfZQ)$ denotes the orthogonal matrix obtained from the QR decomposition of $\bfZQ$ normalized so that the diagonal of the associated upper triangular matrix is positive, and $\zeta(\bfZS) = (\bfZS \bfZS^{\prime} )^{-1}$ denotes a positive definite matrix.

Considering the transformed random matrices $\zeta(\bfZS)$ and $\gamma(\bfZQ)$, the measurable sets $\wtA\subseteq \mathbb{R}^{m\times n}$, $\wtB\subseteq  \mathcal{S}(n)$, and $\wtC\subseteq \mathcal{O}(n)$, and defining $(\bfB,\bfS,\bfQ)= T(\bfZ)$,
we have 
\begin{eqnarray}
\label{eqn:distBSQ}
&& \Pr\left(T(\bfZ)\in \wtA\times \wtB\times \wtC \mid  \bfyT, \mathbf{S}_R(T(\bfZ)) > {\bf0}\right)= \\
&& \adjustbox{max width=\textwidth}{$\displaystyle \smashoperator[r]{\int_{\wtC\times\wtB\times\wtA}} \frac{\left[\mathbf{S}_R(T(\bfZ)) > {\bf0}\right]N(\tilde\bfPsi, \tilde\bfOme, \bfS)(d \bfB)\, IW_{(\tilde{\nu}, \tilde\bfPhi)}(d\bfS)\kappa\, d\bfQ}{\Pr\left(\mathbf{S}_R(T(\bfZ)) > {\bf0} \mid \bfyT\right)},$}\nonumber
\end{eqnarray}
where
\begin{equation*}
IW_{(\tilde{\nu}, \tilde\bfPhi)}(d\bfS) = (\zeta_{\#} N(\mathbf{0}_{n \times \tilde{\nu}}, \tilde{\bfPhi}^{-1}, \bfI_{\tilde{\nu}}))(d\bfS)
\quad \text{and} \quad
\kappa\, d\bfQ = (\gamma_{\#} N(\mathbf{0}_{n \times n}, \bfI_n, \bfI_n))(d\bfQ),
\end{equation*}
denote the push-forward measures associated with the mappings $\zeta \colon \mathbb{R}^{n\times \tilde{\nu}} \to \mathcal{S}(n)$ and $\gamma\colon \mathbb{R}^{n\times n} \to\mathcal{O}(n)$.\footnote{The push-forward characterization is introduced only for notational consistency. The explicit forms of these measures follow immediately from Theorem 2 in \cite{arrw2018}: $\zeta$ yields an inverse-Wishart distribution and $\gamma$ yields the Haar measure on $\mathcal{O}(n)$.} As a consequence, the density of the random vector $T(\bfZ)$ associated with Equation \eqref{eqn:distBSQ} coincides with the density in Equation \eqref{eqn:posteriorSigns}, that is, the posterior distribution of interest. This implies that we can draw from Equation~\eqref{eqn:transf_post_dens} and transform the draws using the mapping $T$ to obtain approximate draws from Equation \eqref{eqn:posteriorSigns}.

\subsection{Design}\label{sec:samplerdesign}
The sampling strategy has at its core an elliptical slice within Gibbs sampler that generates approximate draws from the target density $p_{\bfZ}$. More specifically, the sampler produces draws of a Markov chain $(\bfZ^{i})_{i\in\mathbb{N}}$ whose {single random variable distribution} $Law(\bfZ^{i})$ is close to the target law $Law(\bfZ)$ induced by $p_{\bfZ}$. 
Our ultimate object of interest, however, is the posterior distribution of $(\bfB,\bfS,\bfQ)$ with density $p$ in Equation \eqref{eqn:posteriorSigns}. Hence, to accomplish our goal we obtain draws in this parameterization of interest via the transformation $T(\bfZ^{i})$. For now, we {justify this approach by assessing} the approximation error using the total variation distance between the target $Law(\bfB,\bfS,\bfQ)$ and the induced $Law(T(\bfZ^{i}))$. Since $(\bfB,\bfS,\bfQ)=T(\bfZ)$,  $Law(\bfB,\bfS,\bfQ)$ is the push-forward of $Law(\bfZ)$ under $T$, and similarly for $Law(T(\bfZ^{i}))$ and $Law(\bfZ^{i})$. A standard contraction property of total variation under measurable maps implies
\begin{equation*}
\bigl\| Law(\bfB,\bfS,\bfQ) - Law(T(\bfZ^{i})) \bigr\|_{tv}
\;\le\;
\bigl\| Law(\bfZ) - Law(\bfZ^{i}) \bigr\|_{tv}.
\end{equation*}
Therefore, whenever the right-hand side is small, the transformed draws $T(\bfZ^{i})$ approximate the target posterior $Law(\bfB,\bfS,\bfQ)$ in total variation distance.

\subsection{Elliptical Slice within Gibbs Sampler}\label{sec:algorithmmainGibbs}
This section describes the aforementioned elliptical slice within Gibbs sampler. We {begin by describing} the three building blocks of the approach: the conditional posterior $\bfZQ$ given $(\bfZB,\bfZS)$, the conditional posterior for $\bfZS$ given $(\bfZB,\bfZQ)$, and the conditional posterior for $\bfZB$ given $(\bfZS,\bfZQ)$. Then we combine these posteriors into a single algorithm.

\paragraph{Conditional Posterior for $\bfZQ$.}
We first derive the posterior for $\bfZQ$ conditional on $\bfZB, \bfZS$, and the sign restrictions. Using Bayes’ rule and independence of $(\bfZB, \bfZS)$ and $\bfZQ$, Equation~\eqref{eqn:transf_post_dens} implies:
\begin{equation*}
p_{\bfZ}(\bfZQ \mid \bfZB, \bfZS, \bfyT, \mathbf{S}_R(T(\bfZ)) > {\bf0}) \propto \left[\mathbf{S}_R(T(\bfZ)) > {\bf0}\right]N_{(\mathbf{0}_{n \times n}, \bfI_n, \bfI_n)}(\bfZQ).
\end{equation*}
The right-hand side determines a well-defined conditional Lebesgue density. To see this, define the set
\begin{equation*}
\mathcal{P} = \{\bfZ \in \wtZ : \left[\mathbf{S}_R(T(\bfZ)) > {\bf0}\right] = 1\},
\end{equation*}
and notice that by the continuity of $\zeta$, $\gamma$ and $\mathbf{S}_R$ the set $\mathcal{P}$ is open. Consequently, for any fixed $(\bfZB, \bfZS)$, the set 
\begin{equation*}
\mathcal{P}(\bfZB, \bfZS) = \{\bfZQ \in \mathbb{R}^{n\times n} : \bfZ \in \mathcal{P}\}
\end{equation*}
is open in $\mathbb{R}^{n \times n}$.

\paragraph{Conditional Posterior for $\bfZS$.}
We next derive the posterior of $\bfZS$ conditional on $(\bfZB, \bfZQ)$ and the sign restrictions:
\begin{equation*}
\adjustbox{max width=\textwidth}{$\displaystyle p_{\bfZ}(\bfZS \mid \bfZB, \bfZQ, \bfyT,\mathbf{S}_R(T(\bfZ)) > {\bf0}) \propto \left[\mathbf{S}_R(T(\bfZ)) > {\bf0}\right] N_{(\tilde\bfPsi, \tilde\bfOme, \zeta(\bfZS))}(\bfZB) N_{(\mathbf{0}_{n \times \tilde{\nu}}, \tilde{\bfPhi}^{-1}, \bfI_{\tilde{\nu}})}(\bfZS).$}
\end{equation*}
As before, the proportionality argument (well-definedness of the denominator) follows from the fact that, for any fixed 
$(\bfZB, \bfZQ)$, the set
\begin{equation*}
\mathcal{P}(\bfZB, \bfZQ)
= \{\bfZS \in \mathbb{R}^{n\times \tilde{\nu}} : \bfZ \in \mathcal{P}\}
\end{equation*}
is open in $\mathbb{R}^{n\times \tilde{\nu}}$.

\paragraph{Conditional Posterior for $\bfZB$.} 
Finally, the posterior of $\bfZB$ conditional on $(\bfZS, \bfZQ)$ and the sign restrictions is:
\begin{equation*}
 p_{\bfZ}(\bfZB \mid \bfZS, \bfZQ, \bfyT, \mathbf{S}_R(T(\bfZ)) > {\bf0}) 
  \propto \left[\mathbf{S}_R(T(\bfZ)) > {\bf0}\right]
  N_{(\tilde\bfPsi, \tilde\bfOme, \zeta(\bfZS))}(\bfZB)  
\end{equation*}
Finally, the proportionality argument in this case follows from the fact that, for any fixed 
$(\bfZS, \bfZQ)$, the set
\begin{equation*}
\mathcal{P}(\bfZS, \bfZQ)
= \{\bfZB \in \mathbb{R}^{m \times n} : \bfZ \in \mathcal{P}\}
\end{equation*}
is open in $\mathbb{R}^{m \times n}$.

\paragraph{Algorithm.} Crucially, each of the conditional densities defined above shares the feature of being proportional to the product of a multivariate Gaussian and an arbitrary function that we can evaluate. This is exploited in our elliptical slice sampling-within-Gibbs approach. We derive a Markov chain $(\bfZ^i)_{i\in\mathbb{N}}=(\bfZiB,\bfZiS,\bfZiQ)_{i\in\mathbb{N}}$ and the transformed chain $(\bfB^i,\bfS^i,\bfQ^i)_{i\in\mathbb{N}}=(T(\bfZ^i))_{i\in\mathbb{N}}$ for targeting the distribution of interest determined by Equation~\eqref{eqn:posteriorSigns}.

\begin{Algorithm}\label{alg:algorithmnovel}
This algorithm draws samples $(\bfZ^i)_{0\leq i \leq I}$ from an elliptical slice within Gibbs sampler and it returns transformed samples $(\bfB^i,\bfS^i,\bfQ^i)_{1\leq i \leq I}$ for targeting the distribution of interest determined by Equation~\eqref{eqn:posteriorSigns}.
\begin{enumerate}
\item  Set $I > 1$, initialize $i = 1$, and assign initial values to $\bfZ^{0} 
\in \mathcal{P}$.
\item \label{algo:stepQ}
Use elliptical slice sampling to draw $\bfZiQ$ from $P_{\bfZimB, \bfZimS}(\bfZimQ,d\bfZiQ)$, which approximately samples from
\begin{equation*}
    p_{\bfZ}(\bfZQ\mid \bfZimB, \bfZimS, \bfyT, \mathbf{S}_R(\bfZimB, \zeta(\bfZimS), \gamma(\bfZQ)) > {\bf0})
\end{equation*}
and set $\bfQ^i = \gamma(\bfZiQ)$.\footnote{$P_{\bfZimB,\bfZimS}(\bfZimQ,d\bfZiQ)$ denotes the transition kernel (as a probability measure in $d\bfZiQ$) induced by elliptical slice sampling for the conditional target specified in Step~\ref{algo:stepQ}. We use analogous notation in the two subsequent steps of this algorithm.}
\item \label{algo:stepS}
Use elliptical slice sampling to draw $\bfZiS$ from $P_{\bfZimB, \bfZiQ}(\bfZimS,d\bfZiS)$, which approximately samples from
\begin{equation*}
    p_{\bfZ}(\bfZS \mid \bfZimB, \bfZiQ, \bfyT, \mathbf{S}_R(\bfZimB, \zeta(\bfZS), \gamma(\bfZiQ)) > {\bf0}) 
\end{equation*}
and set $\bfS^i = \zeta(\bfZiS)$.
\item \label{algo:stepB} Use elliptical slice sampling to draw $\bfZiB$ from $P_{\bfZiS, \bfZiQ}(\bfZimB,d\bfZiB)$, which approximately samples from
\begin{equation*}
    p_{\bfZ}(\bfZB \mid \bfZiS, \bfZiQ, \bfyT, \mathbf{S}_R(\bfZB, \zeta(\bfZiS), \gamma(\bfZiQ)) > {\bf0}) 
\end{equation*}
and set $\bfB^{i} = \bfZiB$.
\item If $i < I$, increment $i$ and return to Step~2.
\end{enumerate}
\end{Algorithm}

As mentioned above, we have used the conjugate normal-inverse-Wishart prior over the reduced-form parameters to describe the algorithm. 
Appendix~\ref{app:alt_priors} shows that the approach can be easily adapted to two alternative priors: the independent normal-inverse-Wishart prior and the asymmetric conjugate priors of \citet{Chanassymetric}. 

As in other MCMC algorithms, our algorithm must start from parameter values that satisfy the sign restrictions. In particular, the elliptical slice sampling in Step~2 must be initialized at values $\bfZ^{0} \in \mathcal{P}$. Depending on the sign restrictions imposed, finding such parameter values to initialize the algorithm can be computationally costly. Therefore, we develop a novel and robust method that combines the i.i.d.\ sampler with our proposed Gibbs sampler. Appendix~\ref{app:initialization} describes the procedure in detail.

\subsection{Well-Definedness}\label{sec:welldefinedness}
In this section, we justify the well-definedness of Algorithm \ref{alg:algorithmnovel} by establishing stationarity of the elliptical slice sampling-within-Gibbs transition mechanism with respect to the desired distribution $p_{\bfZ}$ given in Equation~\eqref{eqn:transf_post_dens}. To motivate the argument, suppose that $\bfZ^{i-1}$ follows the desired posterior distribution. We then want to construct the transition to $\bfZ^{i}$ so that this distribution is preserved exactly. If so, $\bfZ^{i}$ also follows the posterior of interest.
Lemma \ref{lemma:stationarity} formalizes the intuition provided above.

\begin{lemma}\label{lemma:stationarity}
Let the initial values $\bfZ^{i-1}$ follow from the desired posterior distribution given in Equation \eqref{eqn:transf_post_dens}, then, after a single transition from Algorithm~\ref{alg:algorithmnovel}, the result $\bfZ^{i}$ also follows from our desired posterior distribution.
\end{lemma}
\begin{proof}
See Appendix \ref{app:lemma1andlemma2}.
\end{proof}

\section{Applications}\label{sec:applications}
We now illustrate the performance of our algorithm using two empirical applications. The first is a small-scale SVAR of the global oil market, based on the model in \citet*{kilianmurphy2014}, which identifies flow supply, flow demand, and speculative demand shocks using a combination of sign and elasticity bounds. The tight identifying assumptions in this model render traditional accept-reject methods computationally intensive, whereas our algorithm improves efficiency while replicating the main results. The second application revisits the large-scale SVAR model of the U.S. economy developed by \citet*{crump2025large} and analyzed structurally by \citet*{chan2025largestructuralvarsmultiple}, which includes 35 macroeconomic and financial variables and identifies up to eight structural shocks. We show that our algorithm remains computationally stable as the number of restrictions increases, in contrast to the exponential rise in computation time exhibited by the accept-reject method. Both applications highlight the scalability of our approach in distinct empirical settings. For each application, we first demonstrate that our approach replicates the main results reported in the original papers and then analyze the computational timing to show that our method can be more efficient than the traditional accept-reject algorithm.

\subsection{Small SVAR of the World Oil Market}\label{sec:smallSVAR}
In our first application, we replicate the results of \citet{kilianmurphy2014}, who extend the framework of \citet{kilian2012agnostic} by incorporating oil inventories to identify speculative demand shocks. The identification strategy in \citet{kilianmurphy2014} relies on tight sign and elasticity bound restrictions, which result in a small identified set and may render standard accept-reject algorithms slow. Notably, \citet{kilianmurphy2014} adopt an approach similar to that of \citet*{chan2025largestructuralvarsmultiple}, relying on permutations and sign alternations. Therefore, when using the same set of sign and elasticity bound restrictions, the computation times we report for the accept-reject algorithm are comparable to those in the original study and similar to those obtained with our {Gibbs-based} approach. However, when an additional elasticity bound is introduced, the accept-reject approach becomes computationally unfeasible, while the Gibbs sampler remains as fast, highlighting the speed advantages of our algorithm in such settings.

\subsubsection*{Model Specification and Impulse Responses}
We begin by describing the model specification in \citet*{kilianmurphy2014}. They model the global market for crude oil using a four-variable SVAR featuring the percent change in global crude oil production, a measure of global real activity, the real price of crude oil, and the change in global above-ground crude oil inventories. The SVAR is specified at a monthly frequency, with an estimation sample covering 1973M2--2009M8. The model includes 24 lags, a constant, and seasonal dummies to remove seasonal variation. \citet*{kilianmurphy2014} adopt a weak conjugate normal-inverse-Wishart prior distribution \citep*[see, e.g.,][]{uhlig2005effects} for the reduced-form parameters.

Turning to identification, the goal of \citet*{kilianmurphy2014} is to identify three structural shocks using a combination of sign restrictions on impact impulse responses, sign restrictions at horizons 1 through 12, and elasticity bounds. Table~\ref{table:sign_restrictions_kilianmurphy} summarizes the identifying assumptions. The structural shocks are labeled flow supply shock, flow demand shock, and speculative demand shock.

\begin{table}[!htbp]
\centering
\scalebox{1}{
\begin{tabular}{lccc}
\toprule
\multicolumn{4}{c}{\underline{Sign Restrictions on Impact Impulse Responses}} \\ 
Variable/Shock & Flow supply & Flow demand & Speculative demand  \\
\hline
Oil production & $-1$ & $+1$ & $+1$ \\
Real activity & $-1$ & $+1$ & $-1$ \\
Real price of oil & $+1$ & $+1$ & $+1$ \\
Inventories &  * & * & $+1$ \\ \hline
\multicolumn{4}{c}{\underline{Elasticity Bounds}} \\ 
 & Flow supply  & Flow demand  & Speculative demand  \\ \hline
 Price Elasticity of Oil Supply & * & (0, 0.025) & (0, 0.025) \\  \hline
\multicolumn{4}{c}{\underline{Sign Restrictions on Impulse Responses at Horizons 1 through 12}} \\ 
 & Flow supply & Flow demand & Speculative demand  \\ \hline
Real activity & $-1$ & * & * \\
Real price of oil & $+1$ & * & * \\ \hline
\end{tabular}}
\caption{Sign and Elasticity Bound Restrictions}
\label{table:sign_restrictions_kilianmurphy}
\vspace{0.5em}
\begin{minipage}{\textwidth}
\small\textit{Note:} All shocks raise the real oil price. An entry equal to ±1 indicates positive or negative sign restrictions, and * indicates no restriction.
\end{minipage}
\end{table}

\begin{figure}[h!]
    \centering
    \hspace*{-2cm}
\includegraphics[scale=0.5]{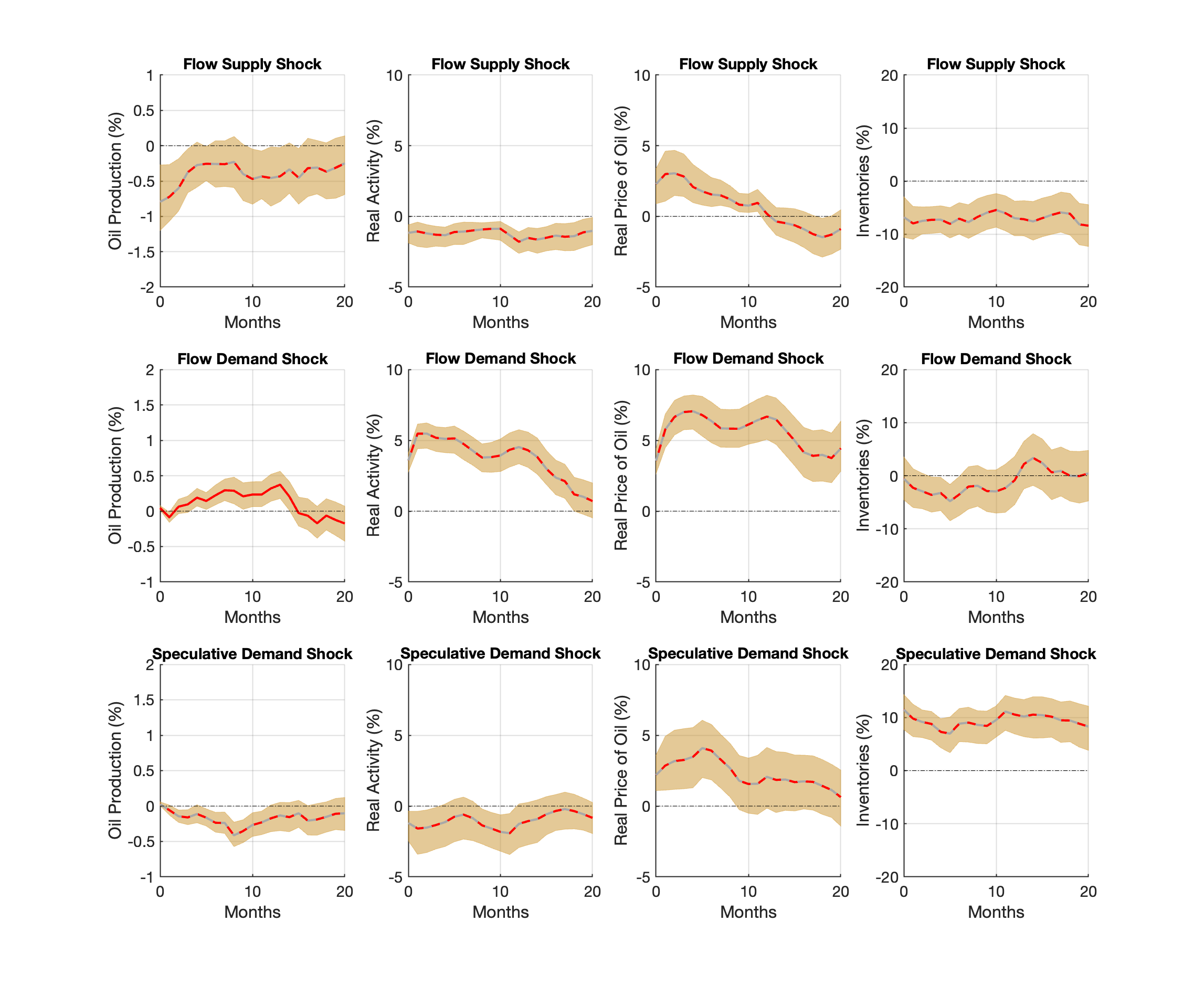}
    \caption{Impulse Responses}
    \label{fig:kilianmurphy}
    \vspace{0.5em}
\begin{minipage}{\textwidth}
\small\textit{Note:} The solid red lines and the orange-shaded areas depict the point-wise posterior median and 68 percent probability bands implied by the Gibbs sampler algorithm; the dashed gray lines and the gray shaded areas depict the point-wise 68 percent posterior probability bands implied by the accept-reject algorithm.
\end{minipage}
\end{figure}

We adopt the exact same specification and use Algorithm~\ref{alg:algorithmnovel}. We obtain one million draws, saving one every 10; hence, the figures are produced using one hundred thousand draws. Figure~\ref{fig:kilianmurphy} presents the impulse responses to the three shocks implied by the Gibbs sampler and the conventional accept-reject approach in \cite{RRWZ2010} with the efficiency gains proposed by \cite{chan2025largestructuralvarsmultiple}. {As can be seen,} the results are nearly identical and they are in line with those in \citet*{kilianmurphy2014}. In particular, a negative flow supply shock causes a persistent decline in global economic activity and oil inventories, and a persistent increase in the real price of oil. The response of oil production is persistently negative. A positive flow demand shock is associated with a persistent increase in global economic activity, a persistent increase in the real price of oil, and a positive response of oil production. Oil production increases sluggishly, given the imposed elasticity bounds, and peaks at about one year after the shock before declining to pre-shock levels. Finally, a positive speculative demand shock causes a persistent increase in the real price of oil and a large increase in inventories. Global real activity and oil production decline persistently in response to this shock, although the effects are modest.

\subsubsection*{Timing}
We next compare the computational time of our Gibbs sampler to that of the accept-reject algorithm. Table~\ref{tab:oiltimes} reports the time (in hours) per 1,000 effective draws---defined as total computation time divided by the effective sample size and scaled by 1,000---using Algorithm~\ref{alg:algorithmnovel} and the accept-reject method.\footnote{Since the accept-reject approach produces independent draws, the number of effective draws equals the total number of draws in that case.} As highlighted above, we implement the efficient variant of the accept-reject algorithm proposed by \citeauthor{chan2025largestructuralvarsmultiple}. We approximate the effective sample size using the multivariate effective sample size metric of \citet*{vats2019multivariate}, which is well suited for SVAR analysis where inference often targets high-dimensional objects such as vectors of impulse responses. In particular, we estimate effective sample size under the impulse response parameterization using the multivariate batch means approach described in that work, with a batch size of $N^{1/4}$, where $N$ denotes the number of stored draws.\footnote{We consider the first three columns of the impulse response parameters, which correspond to the shocks of interest in this application. Results are robust to using impulse responses for all shocks.} Because the accept-reject algorithm produces independent draws, effective draws and sample size are the same.

The column ``Benchmark Model'' in Table~\ref{tab:oiltimes} compares the time (in hours) required to obtain 1,000 effective draws under the specification of \citet*{kilianmurphy2014}. As shown in the table, the Gibbs sampler requires less than 2 minutes to produce 1,000 effective draws. In contrast, the accept-reject algorithm takes approximately 20 minutes to achieve the same number of draws.\footnote{All computations were performed in MATLAB on an Intel Xeon Platinum 8488C processor with 16 active cores running at $2.4$ GHz on an x86\_64 architecture.} To further illustrate the gains of our proposed Gibbs sampler, we consider a scenario in which a researcher imposes an additional restriction on the price elasticity of oil demand in response to a flow supply shock. As emphasized by \citet*{CALDARA20191}, such a restriction is empirically important in SVAR models of the oil market. Following their work, we constrain the price elasticity of crude oil demand to lie within a narrow interval around the point estimate of $-0.08$ reported by \citet*{CALDARA20191}; specifically, we impose the restriction that the elasticity must lie in the interval $(-0.09,-0.07)$. The column ``Benchmark Model + Additional Restriction'' in Table~\ref{tab:oiltimes} reports the results under this added constraint. The time under the Gibbs sampler is under 6 minutes, while the performance of the accept-reject algorithm deteriorates sharply, requiring nearly 8 hours to obtain 1,000 draws.

\begin{table}[!htbp]
\centering
    \begin{tabular}{lcc} \toprule
     Specification    & Benchmark Model & Benchmark Model +  \\
                      &                 & Additional Restriction \\
     \hline
    Gibbs Sampler     & $0.03$  & $0.10$ \\
     Accept-Reject    & $0.33$  & $7.92$ \\
     \bottomrule
    \end{tabular}
    \caption{Time (Hours) Per 1,000 Effective Draws}
    \label{tab:oiltimes}
\end{table}

These results indicate that, for this model, the accept-reject algorithm is already near its computational limit under the benchmark specification, and that introducing even a single additional restriction dramatically reduces its efficiency. In contrast, the Gibbs sampler maintains its performance even as additional identifying restrictions are introduced.

\subsection{Large SVAR of the U.S. Economy}\label{sec:largeSVAR}
In our second application, we replicate and extend the analysis of \citet*{chan2025largestructuralvarsmultiple}, who build on the large-scale SVAR framework of \citet*{crump2025large} to study the structural dynamics of the U.S. economy. Their model incorporates 35 macroeconomic and financial variables commonly monitored by the Federal Reserve and identifies eight structural shocks using an extensive set of sign and ranking restrictions. \citet*{chan2025largestructuralvarsmultiple} employ an accept-reject algorithm, which becomes computationally intensive as the number of identifying restrictions increases.\footnote{We thank Christian Matthes for sharing their replication files and data with us.}  As we show below, our algorithm is more efficient. 

To demonstrate this, we extend the baseline model by identifying two additional shocks---an oil price shock and a consumer sentiment shock---bringing the total number of sign restrictions from 105 to 129. This provides a stringent test of our algorithm's performance relative to the accept-reject method. Importantly, \citet*{chan2025largestructuralvarsmultiple} use the asymmetric priors defined in \citet*{Chanassymetric} for the reduced-form parameters, instead of the Minnesota prior used in \citet*{crump2025large}. To simplify the comparison, we revert to the Minnesota prior when applying both the accept-reject algorithm and the Gibbs sampler. Online Appendix~\ref{app:largeSVARasymmetric} compares both approaches under the asymmetric prior.
 
\subsubsection*{Model Specification and Impulse Responses}
The SVAR used in this section is specified at a quarterly frequency, includes a constant and five lags, and uses an estimation sample that spans from 1977Q4 to 2019Q4.\footnote{The starting date corresponds to the first date with data availability for all variables after taking into account the initial observations required by the specification.} As mentioned, we assume a Minnesota prior for the reduced-form parameters and set the hyperparameters following \citet*{giannone2015prior}. Turning to the identification, Table~\ref{tab:sign_restrictions} summarizes the variables and the sign and ranking restrictions imposed on the contemporaneous impulse responses.
\begin{table}[htbp!]
\centering
\small
\setlength{\tabcolsep}{3pt}
\scalebox{0.9}{
\begin{tabular}{lcccccccccc}
\hline
\multicolumn{1}{l}{Sign restrictions} & Dem & Inv & Fin & Mon & Gov & Tec & Lab & Wag &  Oil & Con\\
\hline
GDP                                  & +1 & +1 & +1 & -1 & +1 & +1 & +1 & +1 & +1 & +1 \\
PCE     & *  & *  & *  & *  & *  & +1 & *  & *  & +1 & +1\\
Residential investment              & *  & *  & *  & *  & *  & *  & *  & *  & *  & +1\\
Nonresidential investment            & *  & +1 & *  & *  & *  & +1 & *  & *  & +1 & +1\\
Exports                              & *  & *  & *  & *  & *  & *  & *  & *  & * & *\\
Imports                              & *  & *  & *  & *  & *  & *  & *  & *  & * & *\\
Government spending                  & *  & *  & *  & *  & +1 & *  & *  & *  & * & *\\
Fed. budget surplus/deficit       & *  & *  & *  & *  & -1 & *  & *  & *  & * & *\\
Fed. tax receipts                 & *  & *  & *  & *  & +1 & *  & *  & *  & * & *\\
GDP deflator                         & +1 & +1 & +1 & -1 & +1 & -1 & -1 & -1 & -1& +1\\
PCE index                            & +1 & +1 & +1 & -1 & +1 & -1 & -1 & -1 & -1& +1\\
PCE index less F\&E       & +1 & +1 & +1 & -1 & +1 & -1 & -1 & -1 & -1& +1\\
CPI index                            & +1 & +1 & +1 & -1 & +1 & -1 & -1 & -1 & -1& +1\\
CPI index less F\&E         & +1 & +1 & +1 & -1 & +1 & -1 & -1 & -1 & -1& +1\\
Real hourly wage                          & *  & *  & *  & *  & *  & +1 & -1 & -1 & +1& *\\
Labor productivity                   & *  & *  & *  & *  & *  & +1 & *  & *  & +1& *\\
Utilization-adjusted TFP             & *  & *  & *  & *  & *  & +1 & *  & *  & +1& *\\
Employment                           & *  & *  & *  & -1 & *  & *  & * & *  & *& *\\
Unemployment rate                    & -1 & -1 & -1 & +1 & -1 & -1 & +1 & -1 & +1& +1\\
Industrial production index          & +1 & +1 & +1 & -1 & *  & *  & *  & *  & *& *\\
Capacity utilization                 & +1 & +1 & +1 & -1 & *  & *  & *  & *  & *& *\\
Housing starts                       & *  & *  & *  & *  & *  & *  & *  & *  & *& *\\
Disposable income                    & *  & *  & *  & *  & *  & *  & *  & *  & *& *\\
Consumer sentiment                   & *  & *  & *  & *  & *  & *  & *  & *  & *& *\\
Fed funds rate                       & +1 & +1 & +1 & +1 & +1 & *  & *  & *  & *& *\\
3-month T-bill rate                   & +1 & +1 & +1 & +1 & +1 & *  & *  & *  & *&* \\
2-year T-note rate                    & *  & *  & *  & +1 & *  & *  & *  & *  & *&* \\
5-year T-note rate                    & *  & *  & *  & +1 & *  & *  & *  & *  & *&* \\
10-year T-note rate                   & *  & *  & *  & +1 & *  & *  & *  & *  & *&* \\
Prime rate                           & +1 & +1 & +1 & +1 & +1 & *  & *  & *  & *& *\\
Aaa corporate bond yield             & *  & *  & *  & +1 & *  & *  & *  & *  & *& *\\
Baa corporate bond yield             & *  & *  & *  & +1 & *  & *  & *  & *  & *& *\\
Trade-weighted US index              & *  & *  & *  & *  & *  & *  & *  & *  & *& *\\
S\&P 500                             & *  & -1 & +1 & -1 & *  & *  & *  & *  & *& +1\\
Spot oil price                       & *  & *  & *  & *  & *  & *  & *  & *  & -1& *\\
\hline
Ranking restrictions & & & & & & & & \\
\hline
Nonresidential investment/GDP        & -1 & +1 & +1 & *  & *  & *  & *  & *  & *&* \\
Government spending/GDP              & -1 & -1 & -1 & *  & +1 & *  & *  & *  & *& *\\
\hline
$\mathbb{N}^0$ of restrictions & 14& 15 & 15 & 19 & 14 &12 & 8 & 8 & 13 & 11 \\
Cum. $\mathbb{N}^0$ of restrictions & 14& 29 & 44 & 63 & 77 & 89 & 97 & 105 & 118 & 129 \\
\hline
\end{tabular}}
\caption{Restrictions on the contemporaneous impulse responses}
\vspace{0.5em}
\begin{minipage}{\textwidth}
\small\textit{Note:} The mnemonics for the shocks are as follows. Dem: demand, Inv: Investment, Fin: Financial, Mon: Monetary Policy, Gov: Government Spending, Tec: Technology, Lab: Labor Supply, Wag: Wage Bargaining, Oil: Oil Price, Con: Consumer Sentiment.
\end{minipage}
\label{tab:sign_restrictions}
\end{table}
A table entry equal to $+1$ indicates a positive sign restriction, a $-1$ indicates a negative one, and $*$ indicates that the corresponding contemporaneous impulse response has been left unrestricted.
\citet*{chan2025largestructuralvarsmultiple} consider only the first eight shocks (demand, investment, financial, monetary, government spending, technology, labor supply, and wage bargaining). In total, 105 sign restrictions are imposed in their baseline specification. We have added two additional shocks (labeled oil price and consumer sentiment) to assess the performance of our algorithm. With the inclusion of these two shocks, the total number of sign restrictions increases to 129. When using the Gibbs sampler, we obtain one million draws and retain one every 10.

\begin{figure}[h!]
  \centering
  \begin{subfigure}[t]{0.48\linewidth}
    \centering
    \caption{Demand shock}
    \includegraphics[width=\linewidth]{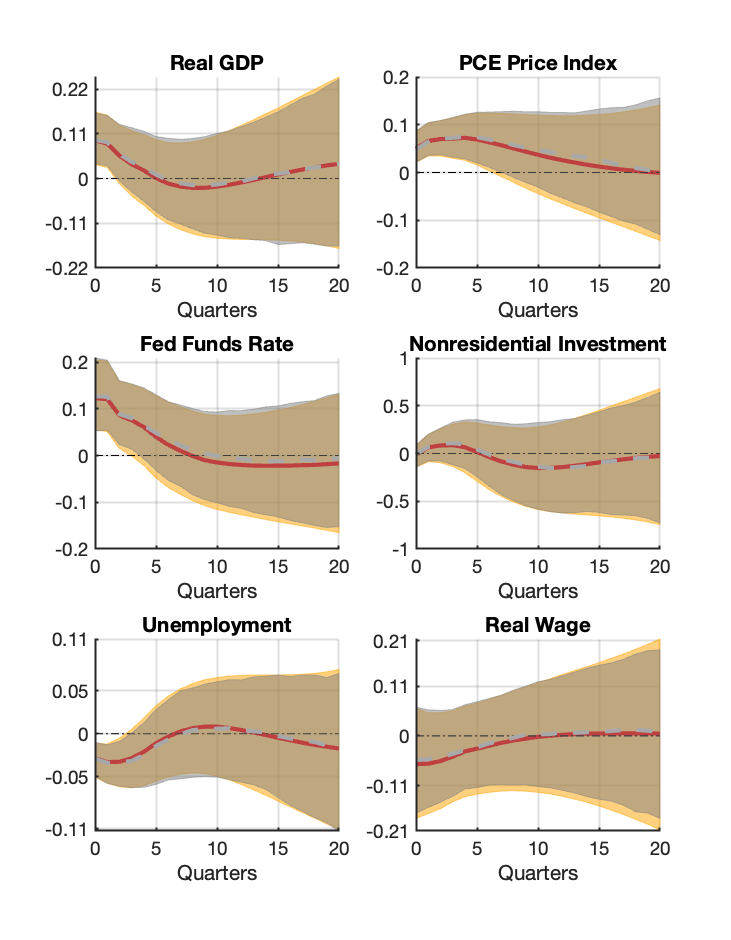}
    \label{fig:demand}
  \end{subfigure}
  \hfill
  \begin{subfigure}[t]{0.48\linewidth}
    \centering
    \caption{Investment shock}
    \includegraphics[width=\linewidth]{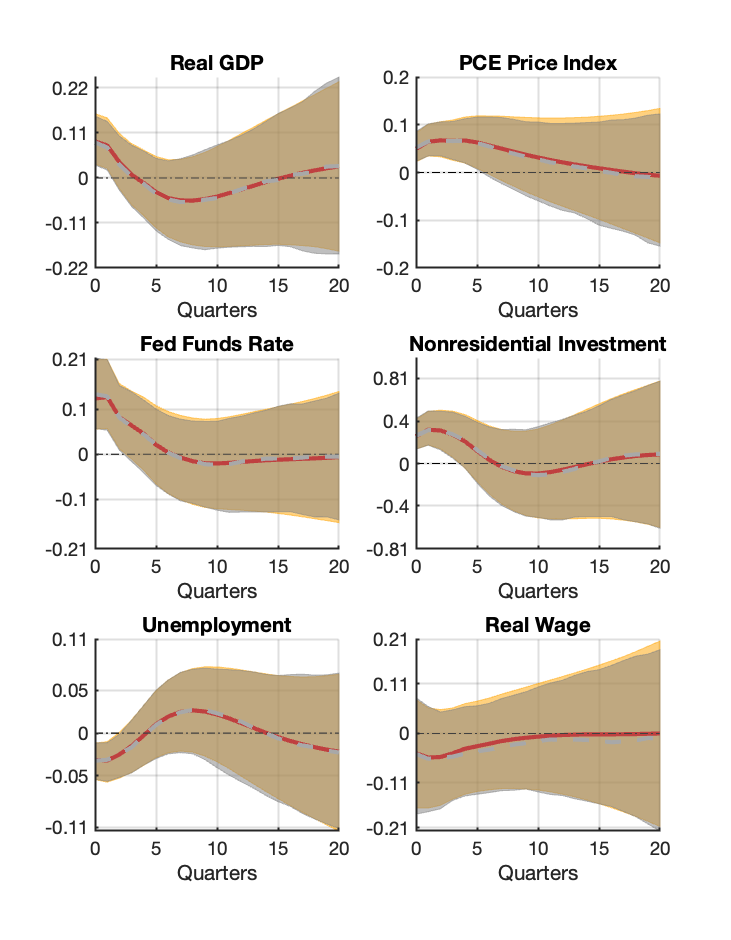}
    \label{fig:investment}
  \end{subfigure}
  \vspace{-28pt} 
  \caption{Impulse Responses}
  \label{fig:demand_side_shocks}
  \begin{minipage}{\textwidth}
\small\textit{Note:} The solid red lines and the orange-shaded areas depict the point-wise posterior median and 68 percent probability bands implied by the Gibbs sampler algorithm; the dashed gray lines and the gray shaded areas depict the point-wise 68 percent posterior probability bands implied by the accept-reject algorithm.
\end{minipage}
\end{figure}

\begin{figure}[h!]
  \centering
  \begin{subfigure}[t]{0.48\linewidth}
    \centering
    \caption{Financial shock}
    \includegraphics[width=\linewidth]{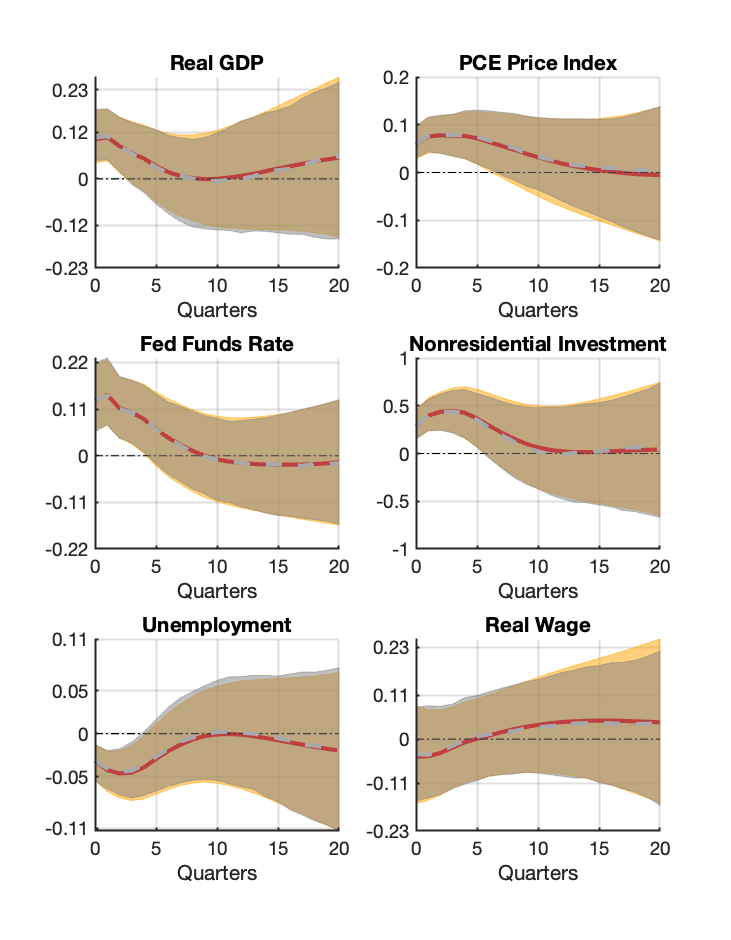}
    \label{fig:financial}
  \end{subfigure}
  \hfill
  \begin{subfigure}[t]{0.48\linewidth}
    \centering
    \caption{Monetary policy shock}
    \includegraphics[width=\linewidth]{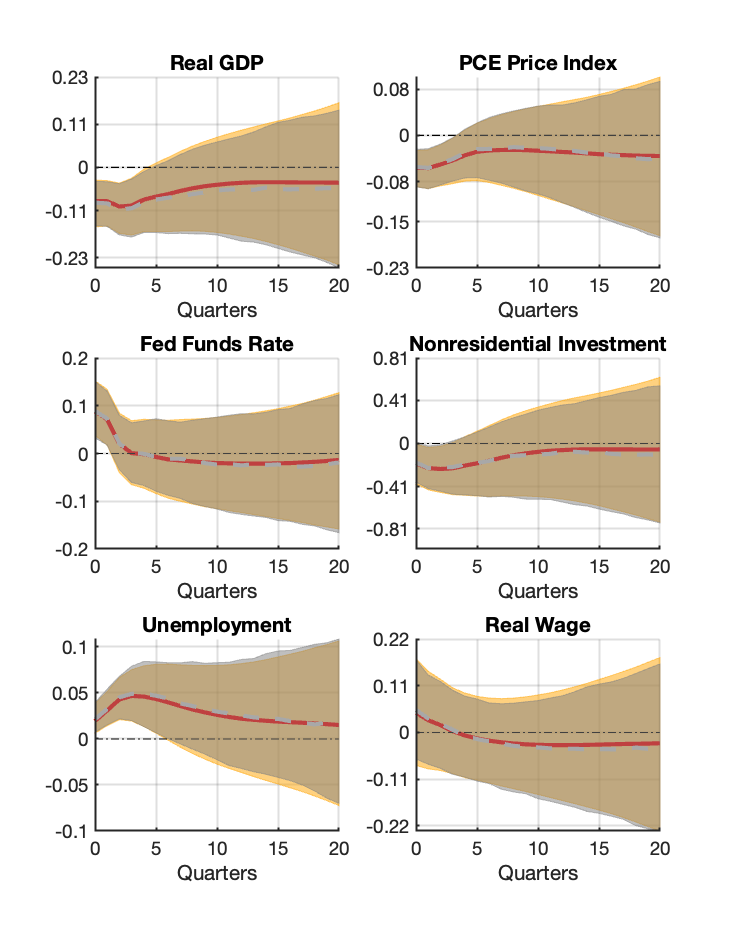}
    \label{fig:monetary}
  \end{subfigure}  
  \vspace{-28pt}
  \caption{Impulse Responses}
  \label{fig:financial_monetary_shocks}
    \begin{minipage}{\textwidth}
\small\textit{Note:} The solid red lines and the orange-shaded areas depict the point-wise posterior median and 68 percent probability bands implied by the Gibbs sampler algorithm; the dashed gray lines and the gray shaded areas depict the point-wise 68 percent posterior probability bands implied by the accept-reject algorithm.
\end{minipage}
\end{figure}

\begin{figure}[h!]
  \centering
  \begin{subfigure}[t]{0.48\linewidth}
    \centering
    \caption{Government spending shock}
    \includegraphics[width=\linewidth]{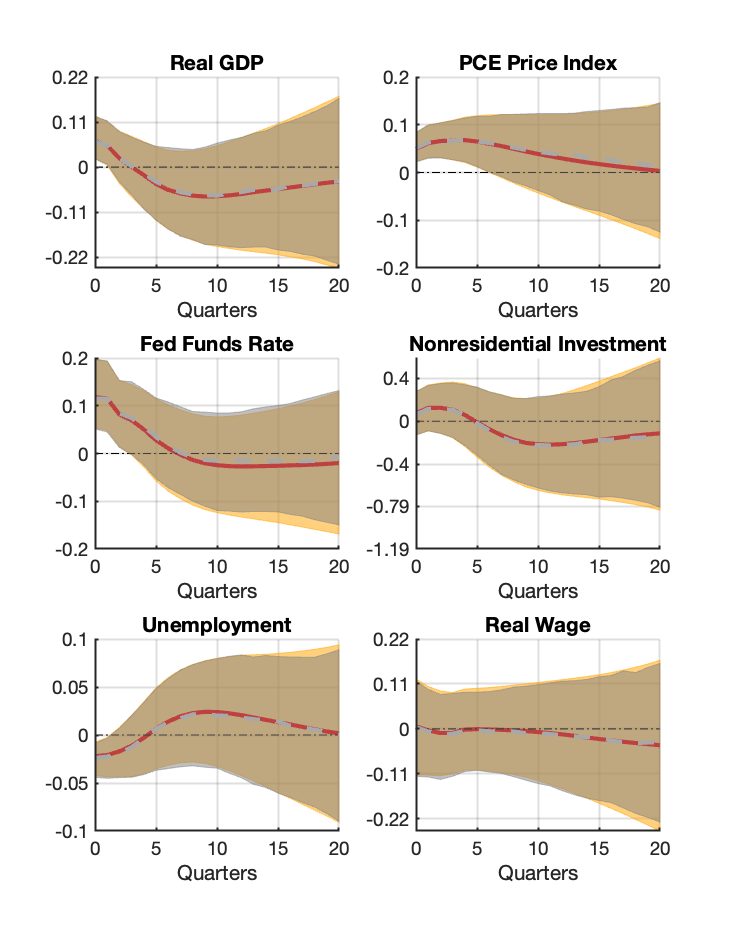}
    \label{fig:Gspending}
  \end{subfigure}
  \hfill
  \begin{subfigure}[t]{0.48\linewidth}
    \centering
    \caption{Technology shock}
    \includegraphics[width=\linewidth]{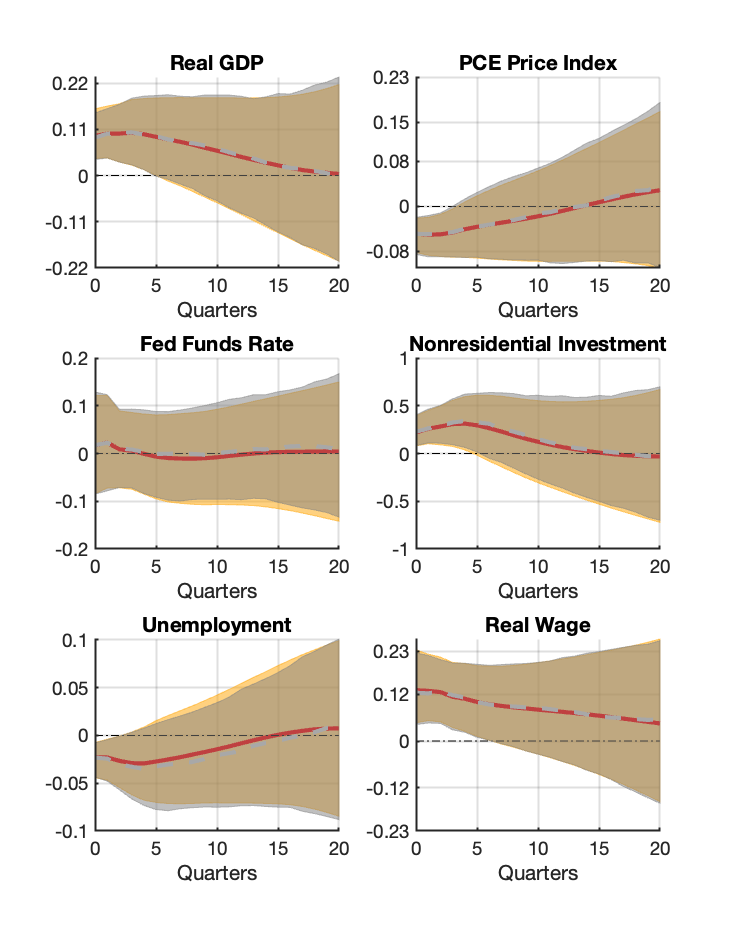}
    \label{fig:technology}
  \end{subfigure}  \vspace{-28pt}
  \caption{Impulse Responses}
  \label{fig:government_technology_shocks}
    \begin{minipage}{\textwidth}
\small\textit{Note:} The solid red lines and the orange-shaded areas depict the point-wise posterior median and 68 percent probability bands implied by the Gibbs sampler algorithm; the dashed gray lines and the gray shaded areas depict the point-wise 68 percent posterior probability bands implied by the accept-reject algorithm.
\end{minipage}
\end{figure}

\begin{figure}[h!]
  \centering
  \begin{subfigure}[t]{0.48\linewidth}
    \centering
    \caption{Labor supply shock}
    \includegraphics[width=\linewidth]{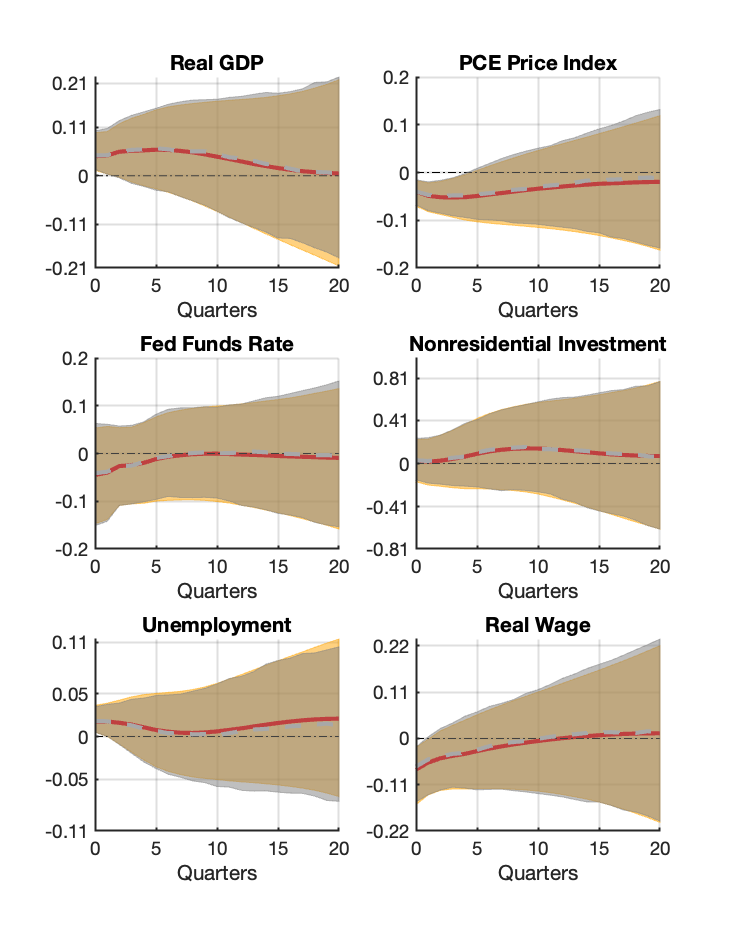}
    \label{fig:laborsupply}
  \end{subfigure}
  \hfill
  \begin{subfigure}[t]{0.48\linewidth}
    \centering
    \caption{Wage bargaining shock}
    \includegraphics[width=\linewidth]{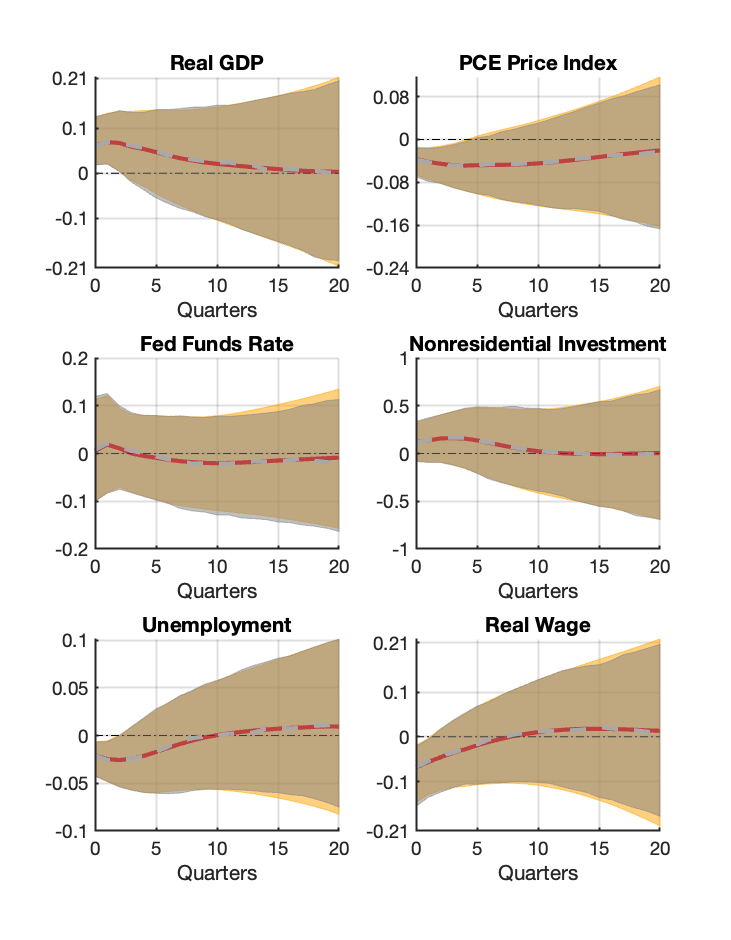}
    \label{fig:wagebargaining}
  \end{subfigure}  \vspace{-28pt}
  \caption{Impulse Responses}
  \label{fig:labor_market_shocks}
    \begin{minipage}{\textwidth}
\small\textit{Note:} The solid red lines and the orange-shaded areas depict the point-wise posterior median and 68 percent probability bands implied by the Gibbs sampler algorithm; the dashed gray lines and the gray shaded areas depict the point-wise 68 percent posterior probability bands implied by the accept-reject algorithm.
\end{minipage}
\end{figure}

Figures \ref{fig:demand_side_shocks} through \ref{fig:labor_market_shocks}
report the impulse responses for each of the eight structural shocks identified by \citet*{chan2025largestructuralvarsmultiple} obtained using the Gibbs sampler described in  Algorithm \ref{alg:algorithmnovel}. The solid red lines and the orange-shaded areas depict the point-wise posterior median and 68 percent probability bands, respectively. For comparison, we also report the point-wise posterior median (gray dashed lines) and point-wise 68\% probability bands (gray-shaded area) obtained with \citeauthor{chan2025largestructuralvarsmultiple}'s \citeyearpar{chan2025largestructuralvarsmultiple} accept-reject algorithm.\footnote{When using their algorithm we use 1,000 draws to report the posterior impulse responses.} The results are virtually identical as expected given that both algorithms are sampling from the same posterior distribution. 
Let us begin by describing the selected impulse responses to a unit standard deviation expansionary demand shock, shown in Figure~\ref{fig:demand}. The signs of the impact responses of real GDP, the PCE price index, the federal funds rate, and the unemployment rate are restricted. The remaining horizons, as well as the responses of non-residential investment and the real wage, are unrestricted. As can be seen, the demand shock causes a transient increase in output and prices, and a decrease in the unemployment rate. The federal funds rate increases in response to the shock. The restrictive stance of monetary policy eventually lowers economic activity, as seen, for example, in the decline of non-residential investment. The real wage decreases in the short run in response to the shock, as nominal wage increases are not sufficient to offset higher prices---possibly due to sluggish nominal wage adjustment.

The investment shock, shown in Figure~\ref{fig:investment}, appears similar to the demand shock in terms of economic consequences for real GDP, the federal funds rate, the price level, and the unemployment rate. However, the impulse response of non-residential investment is substantially different. In particular, the investment shock causes a short-run boom in non-residential investment. This finding partly reflects the ranking restriction requiring that the impact response of non-residential investment be larger than the impact response of real GDP. As with the demand shock, the investment shock causes a persistently negative response of the real wage.

Turning to the financial shock, shown in Figure~\ref{fig:financial}, it is worth highlighting that this shock is identified using the same sign restrictions as the investment shock, except for the impact response of the S\&P 500, which is assumed to be positive instead of negative. Overall, the impulse responses are similar, except that the decline in non-residential investment after five quarters is slightly less pronounced under the financial shock, consistent with the positive response of asset prices.

The impulse responses to a unit standard deviation contractionary monetary policy shock are depicted in Figure~\ref{fig:monetary}. This shock causes the federal funds rate to remain above zero for more than two years, reflecting inertia in the conduct of monetary policy. Real GDP and prices decline persistently, and the unemployment rate jumps upon impact before slowly returning to baseline. Non-residential investment drops on impact and recovers after about one year, in line with a less restrictive monetary policy stance. The real wage increases, driven by a decrease in the price level. A notable aspect of these responses is that they suggest monetary policy can operate with shorter lags than traditionally assumed under the ``long and variable lags'' view.

The government spending shock is shown in Figure~\ref{fig:Gspending}. An expansionary one unit standard deviation government spending shock leads to an increase in real GDP for about two quarters and to a long-lasting increase in the price level. To conclude, we discuss the impulse responses to the supply-related structural shocks, that is, the technology, labor supply, and wage bargaining shocks. A unit standard deviation positive technology shock leads to a protracted increase in real GDP and non-residential investment (see Figure~\ref{fig:technology}). The higher level of output is accompanied by a sustained decline in the unemployment rate and a sustained increase in the real wage. The federal funds rate rises marginally, indicating that monetary policy remains roughly neutral in response to technology shocks.

The responses to a unit standard deviation positive labor supply shock are shown in Figure~\ref{fig:laborsupply}. This shock induces a hump-shaped response of real GDP and leads to persistently lower prices. The responses to a unit standard deviation negative wage bargaining shock are shown in Figure~\ref{fig:wagebargaining}. The identifying assumptions for this shock are identical to those of an expansionary labor supply shock, except that the unemployment rate is assumed to decrease upon impact. When a negative wage bargaining shock occurs, workers experience a decline in their nominal wage alongside a decrease in the unemployment rate. The real wage remains unaffected on impact, as the lower wages are offset by the assumed decrease in the price level. Subsequently, the price level remains below zero, inducing an increase in the real wage.

\subsubsection*{Timing}
\begin{figure}[h!]
  \centering
  \begin{subfigure}[t]{0.325\linewidth}
    \centering
    \caption{Gibbs Sampler}
    \label{fig:timing_ess_large_SVAR}
    \includegraphics[width=\linewidth]{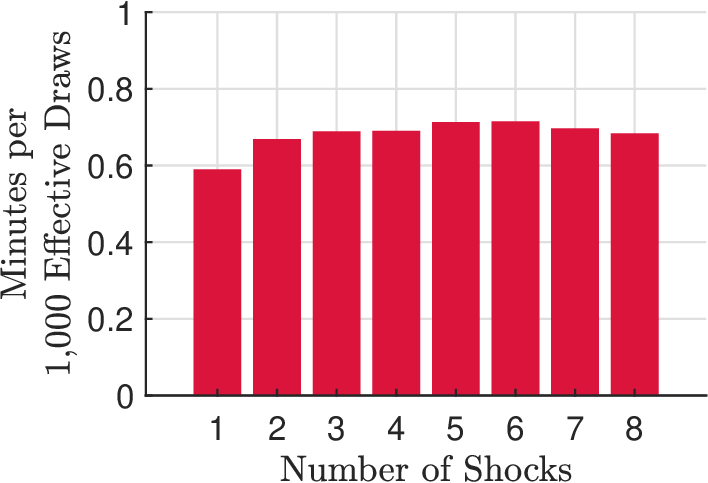}
  \end{subfigure}
  \hfill
  \begin{subfigure}[t]{0.325\linewidth}
    \centering
    \caption{Accept-Reject}
    \label{fig:timing_Talgo_large_SVAR}
    \includegraphics[width=\linewidth]{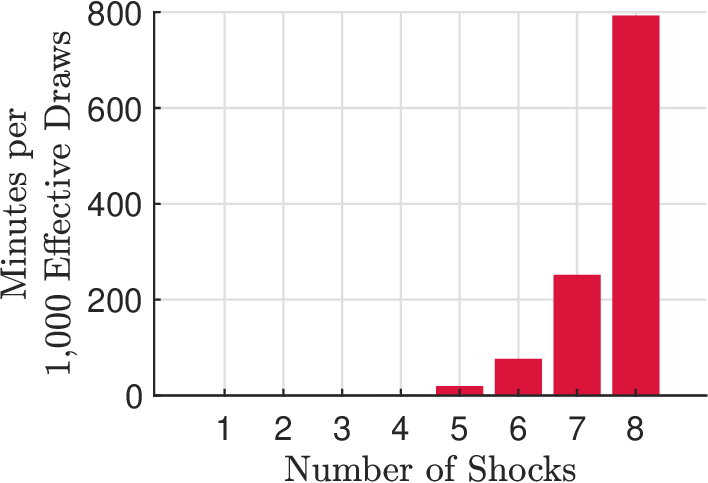}
  \end{subfigure}
  \hfill
  \begin{subfigure}[t]{0.325\linewidth}
    \centering
    \caption{Comparison}
    \label{fig:timing_BOTH_large_SVAR}
    \includegraphics[width=\linewidth]{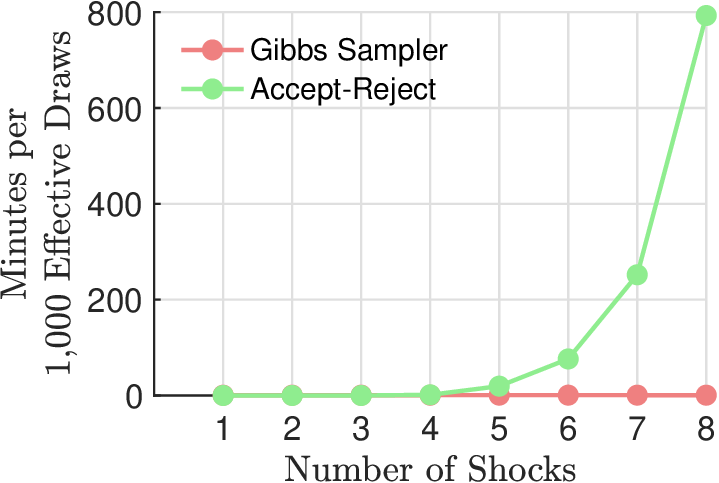}
  \end{subfigure}
  \caption{Time Per 1,000 Effective Draws}
  \label{fig:timing_three_panel}
\end{figure}

We begin by comparing the efficiency of the Gibbs sampler algorithm relative to the accept-reject algorithm when replicating the identification scheme in \citet*{chan2025largestructuralvarsmultiple}. Figure~\ref{fig:timing_ess_large_SVAR} reports the time (in minutes) per 1,000 effective draws using Algorithm~\ref{alg:algorithmnovel} as a function of the number of identified shocks.\footnote{When computing the  multivariate effective sample size, we only consider the columns of the impulse response parameters corresponding to the shocks of interest in this application.}

To assess the computational time as a function of the size of the identified set, we proceed incrementally: we first obtain draws by identifying only the demand shock, then add the investment shock, the financial shock, and so on, until all eight shocks in Table~\ref{tab:sign_restrictions} are included. As shown, the time per 1,000 effective draws remains computationally feasible even as the number of sign restrictions increases.

\begin{figure}[h!]
  \centering
  \begin{subfigure}[t]{0.435\linewidth}
    \centering
    \caption{Gibbs Sampler Shocks (1 to 9)}
    \label{fig:timing_ess_large_SVAR_1to9}
    \includegraphics[width=\linewidth]{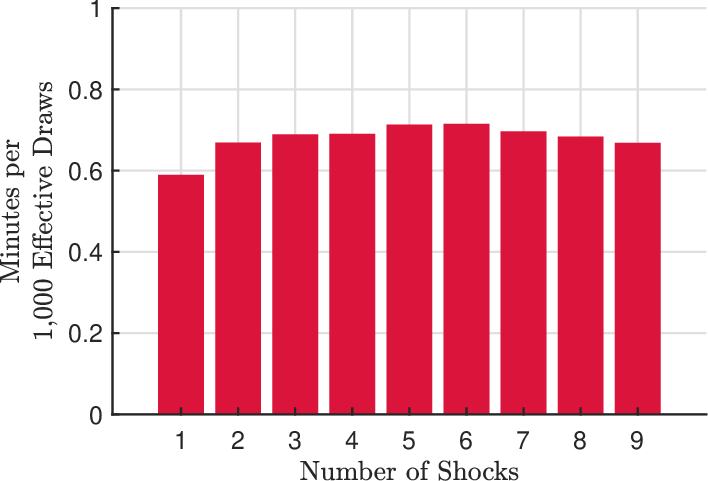}
  \end{subfigure}
  \hfill
  \begin{subfigure}[t]{0.435\linewidth}
    \centering
    \caption{Gibbs Sampler Shocks (1 to 10)}
    \label{fig:timing_ess_large_SVAR_1to10}
    \includegraphics[width=\linewidth]{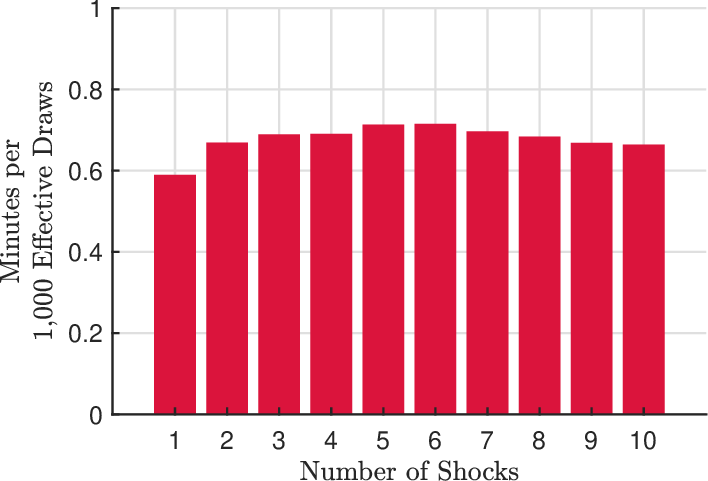}
  \end{subfigure}
  
  \vspace{1em} 
  
  \begin{subfigure}[t]{0.435\linewidth}
    \centering
    \caption{Accept-Reject Shocks (1 to 9)}
    \label{fig:timing_Talgo_large_SVAR_1to9}
    \includegraphics[width=\linewidth]{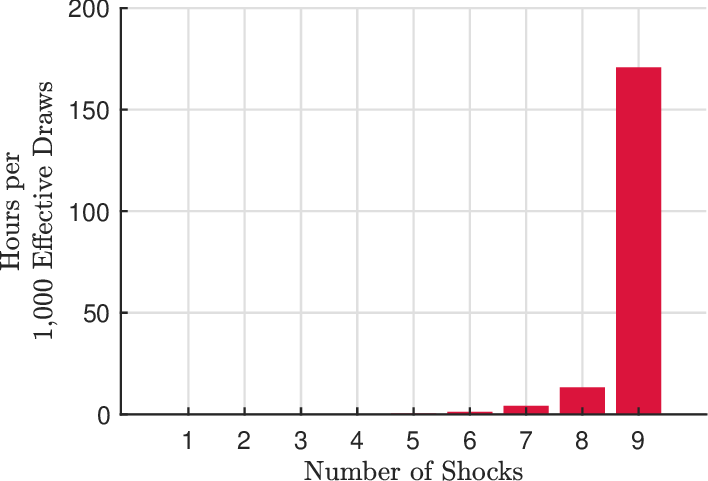}
  \end{subfigure}
  \hfill
  \begin{subfigure}[t]{0.435\linewidth}
    \centering
    \caption{Accept-Reject Shocks (1 to 10)}
    \label{fig:timing_Talgo_large_SVAR_1to10}
    \includegraphics[width=\linewidth]{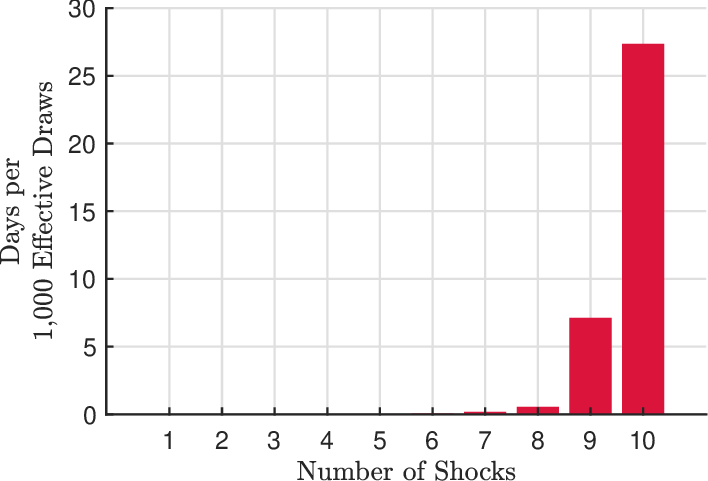}
  \end{subfigure}
  
  \caption{Gibbs Sampler vs. Accept-Reject}
  \label{fig:timing_comparison_all}
  \vspace{0.5em}
    \begin{minipage}{\textwidth}
\small\textit{Note:} The time of the accept-reject algorithm for shocks 9 and 10 {is extrapolated from 10 draws}.
\end{minipage}
\end{figure}

Figure~\ref{fig:timing_Talgo_large_SVAR} replicates the same figure but using the efficient accept-reject version of Algorithm~\ref{alg:basicdraw} proposed by \citet*{chan2025largestructuralvarsmultiple}. In this case, the computation time increases dramatically, as shown in the figure. Although the runtime will vary depending on the hardware architecture and the number of variables, the main conclusion from comparing Figures~\ref{fig:timing_ess_large_SVAR}–\ref{fig:timing_Talgo_large_SVAR} remains unchanged: the performance of the accept-reject algorithm can deteriorate sharply as the identified set narrows. Figure~\ref{fig:timing_BOTH_large_SVAR} combines the timings to facilitate visual comparison.

To further emphasize this point, we now consider additional shocks to illustrate that the accept-reject approach can eventually become impractical. Specifically, we extend the number of shocks identified in \citet*{chan2025largestructuralvarsmultiple} by adding the oil price shock and the consumer sentiment shock described in Table~\ref{tab:sign_restrictions}. Figures~\ref{fig:timing_ess_large_SVAR_1to9}–\ref{fig:timing_Talgo_large_SVAR_1to10} replicate the exercise shown in Figures~\ref{fig:timing_ess_large_SVAR}–\ref{fig:timing_Talgo_large_SVAR} for the cases of nine and ten shocks.\footnote{The runtime of the accept-reject algorithm for the nine- and ten-shock cases is extrapolated based on ten draws.}

\begin{figure}[h!]
  \centering
  \begin{subfigure}[t]{0.405\linewidth}
    \centering
    \caption{Comparison Shocks (1 to 9)}
    \label{fig:timing_BOTH_large_SVAR_1to9}
    \includegraphics[width=\linewidth]{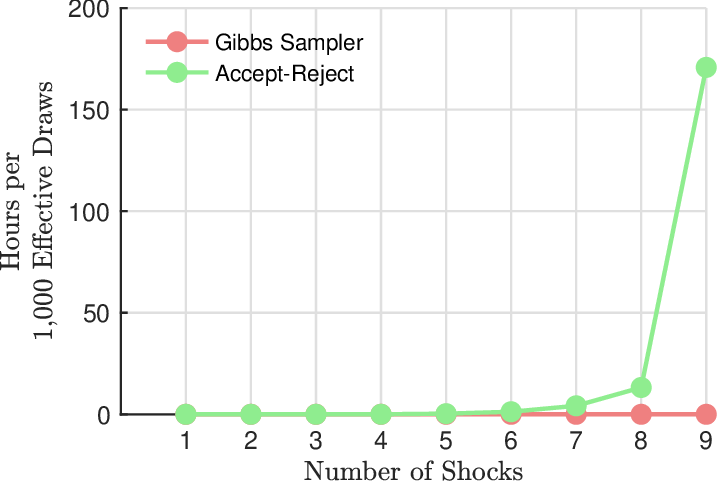}
  \end{subfigure}
  \hfill
  \begin{subfigure}[t]{0.405\linewidth}
    \centering
    \caption{Comparison Shocks (1 to 10)}
    \label{fig:timing_BOTH_large_SVAR_1to10}
    \includegraphics[width=\linewidth]{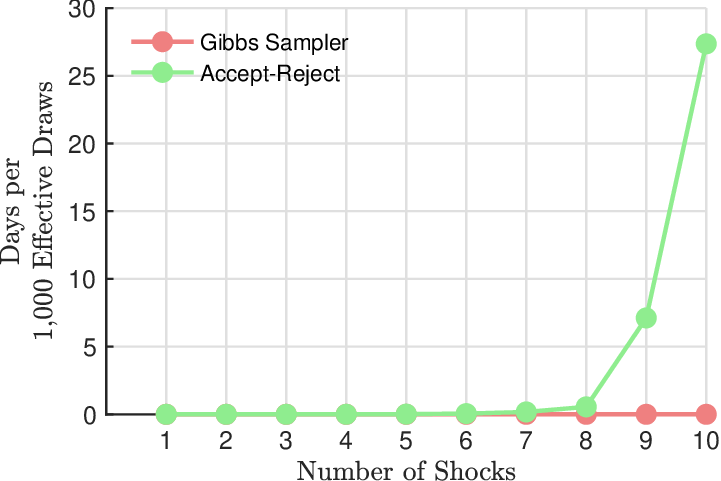}
  \end{subfigure}
  \caption{Gibbs Sampler vs. Accept-Reject}
  \label{fig:timing_both_1to9_1to10}
  \begin{minipage}{\textwidth}
\end{minipage}
\end{figure}

As the reader can see, computation time does not increase exponentially when we use the elliptical slice sampling-within-Gibbs approach. In contrast, when we consider nine shocks under the accept-reject approach, the times are now measured in hours, and when we consider ten shocks, the times are measured in days---Figures~\ref{fig:timing_Talgo_large_SVAR_1to9} and~\ref{fig:timing_Talgo_large_SVAR_1to10} provide the detailed timings. To facilitate the comparison, Figure~\ref{fig:timing_both_1to9_1to10} overlays both sets of timings to make clear that our algorithm can handle settings (in terms of the number of variables and shocks) that the traditional accept-reject approach cannot.

\section{Pitfalls of the Conditionally Uniform Prior }\label{sec:pitfall}
The main computational cost of our approach stems from running the Gibbs sampler, particularly in large models, since it produces {autocorrelated} draws. As a result, one might be tempted to bypass this cost and instead address the bottleneck issues inherent in the accept-reject approach by employing the conditionally uniform prior approach described in \citet*{uhlig2017shocks}, \citet*{AmirAhmadiDrautzburg}, and \citet*{readzhu2025}, among others. The main appeal of this simpler approach is that, like the accept-reject algorithm, it generally yields independent draws. While such a simplification is indeed attractive due to its lower computational burden, it is essential for the researcher to be aware of two critical drawbacks, which we will explain in this section. Before turning to these pitfalls, we first describe the conditionally uniform prior and outline the main algorithmic steps commonly used in its implementation.

Let $\mathbb{Q}_n(\bfB,\bfS)=\{\bfQ \in \mathcal{O}(n): \mathbf{S}_R(\bfB,\bfS,\bfQ)> {\bf0}\}$ and define $\kappa(\bfB,\bfS)$ implicitly by $\int_{\mathbb{Q}_n(\bfB,\bfS)} \kappa(\bfB,\bfS) d\bfQ = 1$. Unlike the case of the uniform prior, this constant of integration depends on $(\bfB,\bfS)$. The prior underlying the conditionally uniform approach is then given by:
\begin{equation}\label{eqn:priorincorrect}
CUNIW_{(\nu, \bfPhi, \bfPsi, \bfOme)}(\bfB, \bfS, \bfQ) =
\begin{cases}
\kappa(\bfB, \bfS) \, NIW_{(\nu, \bfPhi, \bfPsi, \bfOme)}(\bfB, \bfS) & \text{if } \bfQ \in \mathbb{Q}_n(\bfB, \bfS), \\
0 & \text{otherwise}.
\end{cases}
\end{equation}
Crucially, $\kappa(\bfB,\bfS)$ depends on $(\bfB,\bfS)$, while $\kappa$ in Equation~\eqref{eqn:priorcorrect} does not. This is due to the fact that the conditionally uniform approach combines the conjugate normal-inverse-Wishart prior over the reduced-form parameters with $\pi(\bfQ \mid \bfB, \bfS)$ of the form:
\be
\pi(\bfQ \mid \bfB, \bfS) =
\begin{cases}
\kappa(\bfB,\bfS) & \text{if } \bfQ \in \mathbb{Q}_n(\bfB,\bfS), \\
0 & \text{otherwise}.
\end{cases}
\ee
This conditional uniform prior has the property that it overweights reduced-form parameters with smaller identified sets, measured by $\kappa^{-1}(\bfB, \bfS)$ \citep*[see][]{uhlig2017shocks}. Unlike the prior in Equation~\eqref{eqn:priorcorrect}, the prior in Equation~\eqref{eqn:priorincorrect} cannot be justified using the results of \citet{arias2025uniform}, who aim to construct priors that, among other things, separate inference from identification, as it is undesirable for the prior to change when the restrictions are modified, since this makes it impossible to determine whether differences in results stem from changes in the prior or from changes in the identification restrictions.

Under the conditional uniform prior, the objective is to draw from the following posterior of the orthogonal reduced-form parameters conditional on the sign restrictions:
\begin{equation}\label{eqn:posteriorSignsCU}
\begin{gathered}
p(\bfB,\bfS,\bfQ \mid \bfyT, \mathbf{S}_R(\bfB,\bfS,\bfQ) > {\bf0}) = \\
\adjustbox{max width=\textwidth}{$\displaystyle \frac{\left[\mathbf{S}_R(\bfB,\bfS,\bfQ) > {\bf0}\right]}{\Pr\left(\mathbf{S}_R(\bfB,\bfS,\bfQ) > {\bf0} \mid \bfyT\right)} CUNIW_{(\tilde{\nu}, \tilde\bfPhi, \tilde\bfPsi, \tilde\bfOme)}(\bfB,\bfS,\bfQ),$}
\end{gathered}
\end{equation}
where
\begin{align*}
    &\Pr\left(\mathbf{S}_R(\bfB,\bfS,\bfQ) > {\bf0} \mid \bfyT\right)=\\
    &\smashoperator[r]{\int_{\mathbb{R}^{m\times n}\times\mathcal{S}(n)\times\mathcal{O}(n)}} \left[\mathbf{S}_R(\bfB,\bfS,\bfQ) > {\bf0}\right]CUNIW_{(\tilde{\nu}, \tilde\bfPhi, \tilde\bfPsi, \tilde\bfOme)}(\bfB,\bfS,\bfQ) d\bfB d\bfS d\bfQ,
\end{align*}
and then use $f$ and $\phi$ to transform the draws to the desired impulse responses. It is straightforward to adapt the traditional approach described in Algorithm~\ref{alg:basicdraw} to obtain draws from Equation~\eqref{eqn:posteriorSignsCU} as follows:
\begin{Algorithm}\label{alg:CU}
This algorithm independently draws from the conditional posterior in Equation~\eqref{eqn:posteriorSignsCU}.
\begin{enumerate}
\item Draw $(\bfB,\bfS)$ independently from the $NIW(\tilde{\nu},\tilde\bfPhi,\tilde\bfPsi,\tilde\bfOme)$ distribution.
\item Draw $\bfQ$ independently from the uniform distribution over $\mathcal{O}(n)$ until the sign restrictions are satisfied.
\item Repeat Steps 1 and 2 until the desired number of draws is obtained.
\end{enumerate}
\end{Algorithm}
While Algorithm~\ref{alg:CU} does not sample from the posterior distribution defined in Equation~\eqref{eqn:posteriorSigns}, it draws from the posterior distribution defined in Equation~\eqref{eqn:posteriorSignsCU} and it can be justified under a prior different from that in Equation~\eqref{eqn:priorcorrect}. 

\subsection{Possibility of Non-Termination}
The first issue with Algorithm~\ref{alg:CU} is that Step 2 may {never} succeed for some draws of $(\bfB,\bfS)$. Recall $\mathbb{Q}_n(\bfB,\bfS)=\{\bfQ\in\mathcal{O}(n):\mathbf{S}_R(\bfB,\bfS,\bfQ)> {\bf0}\}$. If, for a given $(\bfB,\bfS)$, $\mathbb{Q}_n(\bfB,\bfS)=\varnothing$,
then drawing $\bfQ$ uniformly over $\mathcal{O}(n)$ until $\mathbf{S}_R(\bfB,\bfS,\bfQ)> {\bf0}$ results in non-termination. 
This situation is more pervasive than one might expect. For example, requiring the impulse response of a given variable to a particular shock to maintain the same sign for more than one period rules out sign-flipping impulse-response patterns (i.e., responses that change sign every period). Consequently, any $\bfB$ that implies such every-period oscillations in the impulse responses (e.g., via complex eigenvalues of the companion-form VAR coefficient matrix) will be associated with an empty identified set.

In contrast,  our algorithm conditions on the open set $\mathcal{P}$, defined in Section \ref{sec:newalgorithm}. As a consequence, by construction it cannot produce draws outside {that set}.

\subsection{Intertwined Inference and Identification}
The second issue with the conditionally uniform prior approach is that inference and identification become intertwined. Consider an SVAR with $n = 3$ and $m = 0$, that is, without lags or constant terms. Suppose that Researcher A aims to identify three structural shocks using identification scheme A, as defined in Table~\ref{table:pitfalls}, while Researcher B employs identification scheme B, also defined in the same table. Clearly, any set of impulse responses satisfying scheme B will also satisfy scheme A.

\begin{table*}[!tbp]
\begin{center}
\begin{threeparttable}
\begin{tabular}{lccc}
\toprule
\multicolumn{4}{c}{\underline{Identification A}}\\
             & Shock 1 & Shock 2 & Shock 3  \\
\hline
Variable 1 & $+1$ & $+1$ & $+1$ \\
Variable 2 & $+1$ & $-1$ & $+1$ \\
Variable 3 & $+1$ &      & $-1$  \\\hline
       &  &  & \\
\multicolumn{4}{c}{\underline{Identification B}} \\
             & Shock 1 & Shock 2 & Shock 3  \\
\hline
Variable 1 & $+1$ & $+1$ & $+1$ \\
Variable 2 & $+1$ & $-1$ & $+1$ \\
Variable 3 & $+1$ & $-1$ & $-1$  \\\hline
\end{tabular}
\caption{Sign restrictions: $+1$ and $-1$ indicate positive and negative sign restrictions, respectively; blanks indicate no restriction.}
\label{table:pitfalls}
\end{threeparttable}
\end{center}
\end{table*}

Define $\mathbb{Q}_n^j(\bfS)=\{\bfQ \in \mathcal{O}(n): \mathbf{S}^j_R(\bfS,\bfQ)> {\bf0}\}$ as {the set of orthogonal matrices} consistent with the identification restrictions of researcher $j \in \{A,B\}$, with associated $\kappa^j(\bfS)$ implicitly defined by $\int_{\mathbb{Q}_n^j(\bfS)} \kappa^j(\bfS) d\bfQ = 1$. Even if both researchers specify the same reduced-form prior, changing the identification scheme alters the implied prior over the impulse responses. To illustrate, set $\nu=100$ and $\bfPhi=\bfI_n$, and consider ten values of $\{\bfS^i\}_{i=1}^{10}$ such that when evaluated at the prior distribution $IW_{(\nu,\bfPhi)}$ they all have identical prior density. Let $\{\bfL_0^i\}_{i=1}^{10}$ denote ten corresponding impact impulse response matrices {that happen to satisfy identification B}. Because the volume element from $(\bfS,\bfQ)$ to $\bfL_0$ only depends on the determinant of $\bfS$ and we have also restricted $\{\bfS^i\}_{i=1}^{10}$ to have the same determinant, all $\{\bfL_0^i\}_{i=1}^{10}$ are equally favored by the prior before any sign restrictions are imposed. After the sign restrictions are introduced, we have:
\be
    \frac{\pi^j(\bfL_0^i)}{\pi^j(\bfL_0^{i^{\prime}})}=\frac{\kappa^j(\bfS^i)}{\kappa^j(\bfS^{i^{\prime}})}\text{ for } i,i^{\prime}=1,\ldots,10, \text{ and } j \in \{A,B\},
\ee
where $\pi^j(\bfL_0)$ is the prior density over impulse responses under researcher $j \in \{A,B\}$. This expression implies that any variation in this ratio is solely attributable to the additional negative sign restriction on shock 2 to variable 3 associated with identification scheme B. As shown in Table~\ref{table:implied_prior}, the difference in the identification schemes leads to differences in the implied priors. On the one hand, Researcher A’s prior favors $\bfL^5_0$ $1.45$ times as much as $\bfL^{1}_{0}$, while Researcher B’s prior favors $\bfL^1_0$ about twice as much as $\bfL^4_0$. On the other hand, Researcher B’s prior favors $\bfL^3_0$ twice as much as $\bfL^{1}_{0}$, while Researcher A’s prior favors $\bfL^1_0$ $1.13$ times as much as $\bfL^3_0$. These findings demonstrate that, under the conditionally uniform approach, changing identification schemes alters the implied prior over parameters of interest (such as impulse responses), thereby entangling estimation and identification.

\begin{table*}[!tbp]
\begin{center}
\begin{threeparttable}
\scalebox{1}{
\begin{tabular}{cccccccccccc}
\hline
\hline
Draw         &  1  &  2  &  3  &  4  &  5  &  6  &  7  &  8  &  9  &  10  \\ \hline
$\pi^A(\bfL_{0}^{i})/\pi^A(\bfL_{0}^{1})$  &      1.00      &    1.29       &   0.89     &     0.62      &    1.45      &    1.52    &      0.46     &     0.07      &    1.24    &      0.41\\
$\pi^B(\bfL_{0}^{i})/\pi^B(\bfL_{0}^{1})$ &       1.00     &     1.60      &    1.88    &      0.25     &     0.58     &     0.83   &       0.26    &      0.03     &     1.00   &       0.31\\ \bottomrule    
\end{tabular}}
\caption{Ratio of priors across draws of $\bfL^{i}_0$.}
\label{table:implied_prior}
\end{threeparttable}
\end{center}
\end{table*}

The reason for this unfortunate result is that $\kappa(\bfS)$ varies across identification schemes in a way that disproportionately favors $\bfL_{0}$ values associated with $\bfS$ that induce smaller identified sets. Since different sign restrictions affect the size of the identified set differently for each $\bfS$, different identification restrictions will imply different priors over impulse response functions. This problem does not arise under the uniform prior described in Equation~\eqref{eqn:priorcorrect}, since in that case $\kappa$ does not depend on the reduced-form parameters. More broadly, this highlights the cost of not adopting a uniform prior over the orthogonal matrices as described in \citet*{arias2025uniform}. As shown in that paper, specifying a uniform prior over the set of orthogonal matrices {ensures that inference is disentangled from identification}.

\section{Conclusion}\label{sec:conclusion}
This paper proposes an elliptical slice within Gibbs sampler for inference based on structural vector autoregressions identified with sign restrictions. We show that the algorithm effectively overcomes the computational bottlenecks associated with conventional accept-reject methods, especially as the number of identifying restrictions increases or as the identified set becomes tight. Our empirical applications illustrate how the proposed algorithm can extend existing analyses in the literature, including SVARs with a {large number} of macroeconomic and financial variables. Overall, the paper provides contributions to the implementation of sign-restricted SVARs, offering tools that are applicable across a wide range of empirical models.
\bibliographystyle{chicago}	
\bibliography{myrefs}	

\newpage
\clearpage

\addappheadtotoc
\renewcommand{\thesection}{\Roman{section}}
\renewcommand{\theHsection}{app.\Roman{section}}
\renewcommand{\theequation}{A.\arabic{equation}}  
\renewcommand{\theHequation}{app.\arabic{equation}}
\setcounter{section}{0}
\setcounter{equation}{0}

\noindent {\huge \bf{Appendix}}

\section{Proof of Stationarity}\label{app:lemma1andlemma2}
Let us begin by noting that given 
$\bfZ^{i-1}$, we obtain $\bfZ^{i}$ by crucially drawing from the elliptical slice sampling in Steps~\eqref{algo:stepQ}-\eqref{algo:stepB} of Algorithm \ref{alg:algorithmnovel}. Recall that the transition kernels associated with each of the elliptical slice samplers are $P_{\bfZimB,\bfZimS}( \bfZimQ,d\bfZiQ)$, $P_{\bfZimB, \bfZiQ}(\bfZimS,d\bfZiS)$, and $P_{\bfZiS, \bfZiQ}(\bfZimB,d\bfZiB)$, respectively. 
Our proof of stationarity builds on reversibility and on three block-conditional representations of Equation \eqref{eqn:transf_post_dens}. We describe these properties in Sections \ref{sec:reversibility}-\ref{sec:usefulrelations}, and then we use them to establish stationarity in Section \ref{sec:stationarity}. For notational convenience in this proof, sometimes we use measure notation and other times we use density notation.

\subsection{Reversibility} \label{sec:reversibility}
Importantly, these {kernels} satisfy by construction a reversibility property, cf. \cite[Theorem~3.1]{hasenpflug2025reversibility}\footnote{In \cite{hasenpflug2025reversibility} Theorem~3.1 is formulated for densities that are strictly positive. The issue of applying the result to densities that may attain the value zero can be resolved by modifying the objects and arguments of \citet{hasenpflug2025reversibility} as in \cite[Theorem~8]{Durmus2026geodesic}.}. For the latter result to be applicable 
we exploit the facts, provided in Section~\ref{sec:algorithmmainGibbs}, that the sets $\mathcal{P}(\bfZ_{\bfB},\bfZ_{\bfS})$, $\mathcal{P}(\bfZ_{\bfB},\bfZ_{\bfQ})$ 
and $\mathcal{P}(\bfZ_{\bfS},\bfZ_{\bfQ})$ are open and therefore yield that the conditional posterior densities 
for $\bfZ_{\bfQ}$, $\bfZ_{\bfS}$ and $\bfZ_{\bfB}$ are lower semi-continuous. This is essential to guarantee that the shrinkage procedure used within elliptical slice sampling terminates almost surely, cf. \cite[Lemma~2.7 and Corollary~2.8]{hasenpflug2025reversibility}.
The aforementioned reversibility property implies the following crucial stationarity features with respect to the appearing conditionals:
\begin{align}
\label{al: stat_Q}
   & \smashoperator[r]{\int_{\mathbb{R}^{n\times n}}} P_{\bfZimB,\bfZimS}( \bfZimQ,d \bfZiQ)\; p_{\bfZ}(d \bfZimQ \mid H^{\bfQ}_{\bfZimB, \bfZimS,\bfZimQ}) 
    =  p_{\bfZ}(d\bfZiQ \mid H^{\bfQ}_{\bfZimB, \bfZimS,\bfZiQ}), \\
   \label{al: stat_Sigma}
   & \smashoperator[r]{\int_{\mathbb{R}^{n\times \tilde{\nu}}}}P_{\bfZimB, \bfZiQ}(\bfZimS, d\bfZiS)\,
    p_{\bfZ}(d \bfZimS \mid H^{\bfS}_{\bfZimB, \bfZimS,\bfZiQ}) 
    = p_{\bfZ}(d \bfZiS \mid H^{\bfS}_{\bfZimB, \bfZiS,\bfZiQ}),\\
\label{al: stat_B}
   & \smashoperator[r]{\int_{\mathbb{R}^{m\times n}}}P_{\bfZiS, \bfZiQ}(\bfZimB,d \bfZiB)\,
p_{\bfZ}(d\bfZimB \mid H^{\bfB}_{\bfZimB, \bfZiS,\bfZiQ})\, 
   = p_{\bfZ}(d \bfZiB \mid H^{\bfB}_{\bfZiB, \bfZiS,\bfZiQ}),
\end{align}
where we shorten the notation using a placeholder $\Delta\in \{
\bfB,\bfS,\bfQ\}$ to write $H^{\Delta}_{\bfZB, \bfZS,\bfZQ}$ for the ``conditioning expression''
\begin{equation*}
\{\bfZB, \bfZS,\bfZQ\}\setminus \{\bfZ_\Delta\}, H_{\bfZB, \bfZS,\bfZQ},
\end{equation*}
$H_{\bfZB, \bfZS,\bfZQ}$ for $\bfyT, \mathbf{S}_R(\bfZB, \zeta(\bfZS), \gamma(\bfZQ)) > {\bf0}$, and 
\begin{equation*}
    p_{\bfZ}(d \bfZ_{\Delta} \mid H_{\bfZB, \bfZS,\bfZQ}) = p_{\bfZ}(\bfZ_{\Delta} \mid H_{\bfZB, \bfZS,\bfZQ}) d \bfZ_{\Delta}.
\end{equation*}

\subsection{Block-Conditional Representations}\label{sec:usefulrelations}
In addition to Equations \eqref{al: stat_Q}-\eqref{al: stat_B}, we also need the following relations regarding the different conditional densities.
The next displayed expression rewrites the same joint density in three equivalent block-conditional forms, one for each Gibbs step.
\begin{align}
\label{al: Q_repres}
p_{\bfZ}(\bfZ \mid H_{\bfZB, \bfZS,\bfZQ}) = & \\
\notag
&\adjustbox{max width=\textwidth}{$\displaystyle \frac{C_{\bfQ}(\bfZB, \bfZS)}{C_{\bfZ}} \, p_{\bfZ}(\bfZQ \mid H^{\bfQ}_{\bfZB, \bfZS,\bfZQ}) N_{(\tilde\bfPsi, \tilde\bfOme, \zeta(\bfZS))}(\bfZB) N_{(\mathbf{0}_{n \times \tilde{\nu}}, \tilde{\bfPhi}^{-1}, \bfI_{\tilde{\nu}})}(\bfZS)$}\\
\label{al: Sigma_repres}
= & \adjustbox{max width=\textwidth}{$\displaystyle \frac{C_{\bfS}(\bfZB, \bfZQ)}{C_{\bfZ}}\, p_{\bfZ}(\bfZS\mid H^{\bfS}_{\bfZB, \bfZS,\bfZQ}) N_{(\mathbf{0}_{n \times n}, \bfI_n, \bfI_n)}(\bfZQ)$}\\
\label{al: B_repres}
= & \adjustbox{max width=\textwidth}{$\displaystyle \frac{C_{\bfB}(\bfZS, \bfZQ)}{C_{\bfZ}}\,p_{\bfZ}(\bfZB\mid H^{\bfB}_{\bfZB, \bfZS,\bfZQ}) N_{(\mathbf{0}_{n \times \tilde{\nu}}, \tilde{\bfPhi}^{-1}, \bfI_{\tilde{\nu}})}(\bfZS) N_{(\mathbf{0}_{n \times n}, \bfI_n, \bfI_n)}(\bfZQ)$}
\end{align}
where 
 \begin{align*}
    C_{\bfZ} & = \\
&\adjustbox{max width=\textwidth}{$\displaystyle \smashoperator[r]{\int_{\mathbb{R}^{n\times n}\times\mathbb{R}^{n\times \tilde{\nu}}\times\mathbb{R}^{m\times n}}} \left[\mathbf{S}_R(\bfZB, \zeta(\bfZS), \gamma(\bfZQ)) > {\bf0}\right] N(\tilde\bfPsi, \tilde\bfOme, \zeta(\bfZS))(d\bfZB) N(\mathbf{0}_{n \times \tilde{\nu}}, \tilde{\bfPhi}^{-1}, \bfI_{\tilde{\nu}})(d \bfZS) N(\mathbf{0}_{n \times n}, \bfI_n, \bfI_n)(d \bfZQ),$}\\
    C_{\bfQ}(\bfZB, \bfZS) & = \\
&\adjustbox{max width=\textwidth}{$\displaystyle \smashoperator[r]{\int_{\mathbb{R}^{n\times n}}} \left[\mathbf{S}_R(\bfZB, \zeta(\bfZS), \gamma(\bfZQ)) > {\bf0}\right] N_{(\mathbf{0}_{n \times n}, \bfI_n, \bfI_n)}(d \bfZQ)$},\\
    C_{\bfS}(\bfZB, \bfZQ) & = \\
&\adjustbox{max width=\textwidth}{$\displaystyle \smashoperator[r]{\int_{\mathbb{R}^{n\times \tilde{\nu}}}} \left[\mathbf{S}_R(\bfZB, \zeta(\bfZS), \gamma(\bfZQ)) > {\bf0}\right] N_{(\tilde\bfPsi, \tilde\bfOme, \zeta(\bfZS))}(\bfZB) N_{(\mathbf{0}_{n \times \tilde{\nu}}, \tilde{\bfPhi}^{-1}, \bfI_{\tilde{\nu}})}(d\bfZS)$} \text{, and}\\ 
    C_{\bfB}(\bfZS, \bfZQ) & = \\
&\adjustbox{max width=\textwidth}{$\displaystyle \smashoperator[r]{\int_{\mathbb{R}^{m\times n}}} \left[\mathbf{S}_R(\bfZB, \zeta(\bfZS), \gamma(\bfZQ)) > {\bf0}\right] N_{(\tilde\bfPsi, \tilde\bfOme, \zeta(\bfZS))}(d\bfZB)$}.
\end{align*} 

\subsection{Stationarity} \label{sec:stationarity}
Given that $\bfZ^{i-1}$ is distributed according to $p_{\bfZ}$, then the distribution of $\bfZ^{i}$ is determined by the probability of $(\bfZiB, \bfZiS,\bfZiQ) \in \mathcal{A} \times \mathcal{B} \times \mathcal{C}$, for arbitrary measurable sets $\mathcal{A}\subseteq \mathbb{R}^{m \times n}$, $\mathcal{B} \subseteq \mathbb{R}^{n \times \tilde{\nu}}$, $\mathcal{C} \subseteq \mathbb{R}^{n \times n}$, given by
\begin{equation*}
\adjustbox{max width=\textwidth}{$\displaystyle \smashoperator[r]{\int_{\mathbb{R}^{n\times n}\times\mathbb{R}^{n\times \tilde{\nu}}\times\mathbb{R}^{m\times n}}} \Pr\left((\bfZiB, \bfZiS,\bfZiQ)\in \mathcal{A} \times \mathcal{B} \times \mathcal{C} \mid \bfZimB, \bfZimS,\bfZimQ\right) p_{\bfZ}(\bfZ^{i-1} \mid H_{\bfZimB, \bfZimS,\bfZimQ}) d\bfZimB\, d\bfZimS\, d\bfZimQ$}
\end{equation*}
with
\begin{equation*}
\begin{aligned}
&\Pr\left((\bfZiB, \bfZiS,\bfZiQ)\in \mathcal{A} \times \mathcal{B} \times \mathcal{C} \mid  \bfZimB, \bfZimS,\bfZimQ\right)=\\
&\adjustbox{max width=\textwidth}{$\displaystyle \smashoperator[r]{\int_{\mathbb{R}^{n\times n}\times\mathbb{R}^{n\times \tilde{\nu}}\times\mathbb{R}^{m\times n}}} \mathbf{1}_{\mathcal{A}}(\bfZiB) P_{\bfZiS,\bfZiQ}(\bfZimB, d \bfZiB) \mathbf{1}_{\mathcal{B}}(\bfZiS) P_{\bfZimB,\bfZiQ}(\bfZimS, d \bfZiS) \mathbf{1}_{\mathcal{C}}(\bfZiQ) P_{\bfZimB,\bfZimS}(\bfZimQ, d \bfZiQ).$}
\end{aligned}
\end{equation*}
Combining the two equations above, we obtain:
\begin{equation*}
\begin{gathered}
\adjustbox{max width=\textwidth}{$\displaystyle \smashoperator[r]{\int_{(\mathbb{R}^{n\times n}\times\mathbb{R}^{n\times \tilde{\nu}}\times\mathbb{R}^{m\times n})\times(\mathbb{R}^{n\times n}\times\mathbb{R}^{n\times \tilde{\nu}}\times\mathbb{R}^{m\times n})}} \mathbf{1}_{\mathcal{A}}(\bfZiB) P_{\bfZiS,\bfZiQ}(\bfZimB, d \bfZiB) \mathbf{1}_{\mathcal{B}}(\bfZiS) P_{\bfZimB,\bfZiQ}(\bfZimS, d \bfZiS) \mathbf{1}_{\mathcal{C}}(\bfZiQ) P_{\bfZimB,\bfZimS}(\bfZimQ, d \bfZiQ)$}\\
\adjustbox{max width=\textwidth}{$\displaystyle p_{\bfZ}(\bfZ^{i-1}  \mid H_{\bfZimB, \bfZimS,\bfZimQ})\, d \bfZimB\, d \bfZimS\, d \bfZimQ.$}
\end{gathered}
\end{equation*}
Subsequently, we will exploit the equations described in Sections \ref{sec:reversibility}-\ref{sec:usefulrelations} along with multiple changes of the order of integration to demonstrate stationarity. Let us begin by substituting Equation~\eqref{al: Q_repres} for $p_{\bfZ}(\bfZ^{i-1}  \mid H_{\bfZimB, \bfZimS,\bfZimQ})$ in order to obtain:
\begin{equation*}
\begin{gathered}
\adjustbox{max width=\textwidth}{$\displaystyle \smashoperator[r]{\int_{(\mathbb{R}^{n\times \tilde{\nu}}\times\mathbb{R}^{m\times n})\times(\mathbb{R}^{n\times n}\times\mathbb{R}^{n\times \tilde{\nu}}\times\mathbb{R}^{m\times n})}} \mathbf{1}_{\mathcal{A}}(\bfZiB) P_{\bfZiS,\bfZiQ}(\bfZimB, d \bfZiB) \mathbf{1}_{\mathcal{B}}(\bfZiS) P_{\bfZimB,\bfZiQ}(\bfZimS, d \bfZiS) \mathbf{1}_{\mathcal{C}}(\bfZiQ)$}\\
\adjustbox{max width=\textwidth}{$\displaystyle \frac{C_{\bfQ}(\bfZimB, \bfZimS)}{C_{\bfZ}}\smashoperator[r]{\int_{\mathbb{R}^{n\times n}}} P_{\bfZimB,\bfZimS}(\bfZimQ, d \bfZiQ)\, p_{\bfZ}(d \bfZimQ \mid H^{\bfQ}_{\bfZimB, \bfZimS,\bfZimQ})\, N(\tilde\bfPsi, \tilde\bfOme, \zeta(\bfZimS))(d \bfZimB)\, N(\mathbf{0}_{n \times \tilde{\nu}}, \tilde{\bfPhi}^{-1}, \bfI_{\tilde{\nu}})(d \bfZimS).$}
\end{gathered}
\end{equation*}
Then, by Equation~\eqref{al: stat_Q} we obtain:
\begin{equation*}
\begin{gathered}
\adjustbox{max width=\textwidth}{$\displaystyle \smashoperator[r]{\int_{(\mathbb{R}^{n\times \tilde{\nu}}\times\mathbb{R}^{m\times n})\times(\mathbb{R}^{n\times n}\times\mathbb{R}^{n\times \tilde{\nu}}\times\mathbb{R}^{m\times n})}} \mathbf{1}_{\mathcal{A}}(\bfZiB) P_{\bfZiS,\bfZiQ}(\bfZimB, d \bfZiB) \mathbf{1}_{\mathcal{B}}(\bfZiS) P_{\bfZimB,\bfZiQ}(\bfZimS, d \bfZiS) \mathbf{1}_{\mathcal{C}}(\bfZiQ)$}\\
\adjustbox{max width=\textwidth}{$\displaystyle \frac{C_{\bfQ}(\bfZimB, \bfZimS)}{C_{\bfZ}} p_{\bfZ}(d \bfZiQ \mid H^{\bfQ}_{\bfZimB, \bfZimS,\bfZiQ}) N(\tilde\bfPsi, \tilde\bfOme, \zeta(\bfZimS))(d \bfZimB) N(\mathbf{0}_{n \times \tilde{\nu}}, \tilde{\bfPhi}^{-1}, \bfI_{\tilde{\nu}})(d \bfZimS).$}
\end{gathered}
\end{equation*}
From Equation~\eqref{al: Sigma_repres}, it follows that:
\begin{equation*}
\begin{gathered}
\adjustbox{max width=\textwidth}{$\displaystyle \smashoperator[r]{\int_{(\mathbb{R}^{m\times n}\times\mathbb{R}^{n\times n})\times(\mathbb{R}^{n\times \tilde{\nu}}\times\mathbb{R}^{m\times n})}} \mathbf{1}_{\mathcal{A}}(\bfZiB) P_{\bfZiS,\bfZiQ}(\bfZimB, d \bfZiB) \mathbf{1}_{\mathcal{B}}(\bfZiS) \mathbf{1}_{\mathcal{C}}(\bfZiQ)$}\\
\adjustbox{max width=\textwidth}{$\displaystyle \frac{C_{\bfS}(\bfZimB, \bfZiQ)}{C_{\bfZ}} \smashoperator[r]{\int_{\mathbb{R}^{n\times \tilde{\nu}}}} P_{\bfZimB,\bfZiQ}(\bfZimS, d \bfZiS)p_{\bfZ}(d \bfZimS \mid H^{\bfS}_{\bfZimB, \bfZimS,\bfZiQ})\, d \bfZimB\, N_{(\mathbf{0}_{n \times n}, \bfI_n, \bfI_n)}(d\bfZiQ).$}
\end{gathered}
\end{equation*}
By Equation~\eqref{al: stat_Sigma}, we can substitute out $\int_{\mathbb{R}^{n \times \tilde{\nu}}}
  P_{\bfZimB,\bfZiQ}
  (\bfZimS, d \bfZiS)P_{\bfZiS,\bfZiQ}
  (\bfZimB, d \bfZiB)$:
\begin{equation*}
\begin{gathered}
\adjustbox{max width=\textwidth}{$\displaystyle \smashoperator[r]{\int_{(\mathbb{R}^{m\times n}\times\mathbb{R}^{n\times n})\times(\mathbb{R}^{n\times \tilde{\nu}}\times\mathbb{R}^{m\times n})}} \mathbf{1}_{\mathcal{A}}(\bfZiB) P_{\bfZiS,\bfZiQ}(\bfZimB, d \bfZiB) \mathbf{1}_{\mathcal{B}}(\bfZiS) \mathbf{1}_{\mathcal{C}}(\bfZiQ)$}\\
\adjustbox{max width=\textwidth}{$\displaystyle \frac{C_{\bfS}(\bfZimB, \bfZiQ)}{C_{\bfZ}} p_{\bfZ}(d \bfZiS \mid H^{\bfS}_{\bfZimB, \bfZiS,\bfZiQ})\, d \bfZimB\, N_{(\mathbf{0}_{n \times n}, \bfI_n, \bfI_n)}(d\bfZiQ).$}
\end{gathered}
\end{equation*}
Using Equation~\eqref{al: B_repres}, we get:
\begin{equation*}
\begin{gathered}
\adjustbox{max width=\textwidth}{$\displaystyle \smashoperator[r]{\int_{\mathbb{R}^{n\times n}\times\mathbb{R}^{n\times \tilde{\nu}}\times\mathbb{R}^{m\times n}}} \mathbf{1}_{\mathcal{A}}(\bfZiB) \mathbf{1}_{\mathcal{B}}(\bfZiS) \mathbf{1}_{\mathcal{C}}(\bfZiQ)$}\\
\adjustbox{max width=\textwidth}{$\displaystyle \frac{C_{\bfB}(\bfZiS, \bfZiQ)}{C_{\bfZ}} \smashoperator[r]{\int_{\mathbb{R}^{m\times n}}} P_{\bfZiS,\bfZiQ}(\bfZimB, d \bfZiB)p_{\bfZ}(d \bfZimB \mid H^{\bfB}_{\bfZimB, \bfZiS,\bfZiQ}) N_{(\mathbf{0}_{n \times \tilde{\nu}}, \tilde{\bfPhi}^{-1}, \bfI_{\tilde{\nu}})}(d\bfZiS) N_{(\mathbf{0}_{n \times n}, \bfI_n, \bfI_n)}(d\bfZiQ).$}
\end{gathered}
\end{equation*}
Applying Equation~\eqref{al: stat_B}, it follows that:
\begin{equation*}
\begin{gathered}
\adjustbox{max width=\textwidth}{$\displaystyle \smashoperator[r]{\int_{\mathbb{R}^{n\times n}\times\mathbb{R}^{n\times \tilde{\nu}}\times\mathbb{R}^{m\times n}}} \mathbf{1}_{\mathcal{A}}(\bfZiB) \mathbf{1}_{\mathcal{B}}(\bfZiS) \mathbf{1}_{\mathcal{C}}(\bfZiQ)$}\\
\adjustbox{max width=\textwidth}{$\displaystyle \frac{C_{\bfB}(\bfZiS, \bfZiQ)}{C_{\bfZ}} p_{\bfZ}(d \bfZiB \mid H^{\bfB}_{\bfZiB, \bfZiS,\bfZiQ}) N_{(\mathbf{0}_{n \times \tilde{\nu}}, \tilde{\bfPhi}^{-1}, \bfI_{\tilde{\nu}})}(d\bfZiS) N_{(\mathbf{0}_{n \times n}, \bfI_n, \bfI_n)}(d\bfZiQ).$}
\end{gathered}
\end{equation*}
Finally, Equations~\eqref{al: Q_repres}, \eqref{al: Sigma_repres} and \eqref{al: B_repres} together imply that we can write:
\begin{equation*}
\adjustbox{max width=\textwidth}{$\displaystyle \smashoperator[r]{\int_{\mathbb{R}^{n\times n}\times\mathbb{R}^{n\times \tilde{\nu}}\times\mathbb{R}^{m\times n}}} \mathbf{1}_{\mathcal{A}}(\bfZiB)\mathbf{1}_{\mathcal{B}}(\bfZiS)\mathbf{1}_{\mathcal{C}}(\bfZiQ)\, p_{\bfZ}(d \bfZ^{i} \mid H_{\bfZiB, \bfZiS,\bfZiQ}),$}
\end{equation*}
which is equal to:
\begin{equation*}
\Pr\left((\bfZiB, \bfZiS,\bfZiQ)
\in \mathcal{A}\times \mathcal{B} \times \mathcal{C}\right).
\end{equation*}
Thus, we have shown that:
\begin{equation*}
\begin{gathered}
\Pr\left((\bfZiB, \bfZiS,\bfZiQ)\in \mathcal{A}\times \mathcal{B} \times \mathcal{C}\right)=\\
\adjustbox{max width=\textwidth}{$\displaystyle \smashoperator[r]{\int_{\mathbb{R}^{n\times n}\times\mathbb{R}^{n\times \tilde{\nu}}\times\mathbb{R}^{m\times n}}} \Pr\left((\bfZiB, \bfZiS,\bfZiQ)\in \mathcal{A} \times \mathcal{B} \times \mathcal{C} \mid  \bfZimB, \bfZimS,\bfZimQ\right) p_{\bfZ}(\bfZ^{i-1}  \mid H_{\bfZimB, \bfZimS,\bfZimQ})\, d \bfZimB\, d \bfZimS\, d \bfZimQ.$}
\end{gathered}
\end{equation*}
From this we can deduce that the distribution after a single transition from $\bfZ^{i-1}$ to $\bfZ^{i}$ does not change; hence, the chain is stationary. This proves Lemma~\ref{lemma:stationarity}.

\section{Efficient Initialization}\label{app:initialization}
As discussed in Section~\ref{sec:newalgorithm}, our algorithm must start from parameter values that satisfy the sign restrictions. Depending on the sign restrictions imposed, finding such parameter values to initialize the algorithm can be computationally costly. For example, in our applications in Section~\ref{sec:applications}, it can take up to 2 hours to find valid initial values using the state-of-the-art accept-reject sampler developed by \cite{chan2025largestructuralvarsmultiple}. Hence, it is desirable to have a robust and more efficient method to initialize the algorithm.

To achieve this goal, inspired by adaptive Sequential Monte Carlo (SMC), we begin with draws (particles) from the unrestricted posterior and move them toward the posterior conditional on the sign restrictions. 

More specifically, consider a sequence of posterior distributions indexed by $n \in \left\lbrace 1,\dots,N \right\rbrace$:
\begin{equation*}
\pi_{n}(\bfZ) = p_{\bfZ}(\bfZ \mid \bfyT, \mathbf{S}_{R}(T(\bfZ)) > {\bf{c}}_{n}) \\
\propto p_{\bfZ}(\bfZ \mid \bfyT) \, \left[\mathbf{S}_{R}(T(\bfZ)) > {\bf{c}}_{n}\right],
\end{equation*}
where  $p_{\bfZ}(\bfZ \mid \bfyT)$ is the unrestricted posterior density of $\bfZ$ in the transformed space and $\{\bfc_{n}\}^N_{n=1}$ is a sequence of threshold vectors with ${\bf{-\infty}} = {\bf{c}}_{1} < {\bf{c}}_{2} < \ldots < {\bf{c}}_{N}  \leq {\bf{0}}$. For instance, under the conjugate normal-inverse-Wishart prior, $p_{\bfZ}(\bfZ \mid \bfyT) = N_{(\tilde\bfPsi, \tilde\bfOme, \zeta(\bfZS))}(\bfZB) \, N_{(\mathbf{0}_{n \times \tilde{\nu}}, \tilde{\bfPhi}^{-1}, \bfI_{\tilde{\nu}})}(\bfZS) \, N_{(\mathbf{0}_{n \times n}, \bfI_n, \bfI_n)}(\bfZQ)$ as defined in Section~\ref{sec:changeofvariable}, Equation~\eqref{eqn:transf_post_dens}. Under alternative priors, this density can be adjusted accordingly; see Appendix~\ref{app:alt_priors}. When ${\bf{c}}_{1} = {\bf{-\infty}}$, the indicator is always one, so that $\pi_{1}(\bfZ)=p_{\bfZ}(\bfZ \mid \bfyT)$. As ${\bf{c}}_{n}$ increases, the restrictions tighten and $\pi_{n}$ converges to the targeted posterior that imposes the exact sign restrictions at ${\bf{c}}_{N} = {\bf{0}}$. That is, we have $\pi_{N}(\bfZ)=p_{\bfZ}(\bfZ \mid \bfyT, \mathbf{S}_R(T(\bfZ)) > {\bf0})$. Rather than specifying the sequence of threshold vectors in advance, we devise an adaptive rule that $\bfc_{n}$ is determined based on particles from $\pi_{n-1}$, which we will discuss after we introduce the algorithm.

\begin{Algorithm} \label{alg:initialization} The following algorithm finds $\bfZ^0 \in \mathcal{P}$ that satisfies sign restrictions and can be used to initialize Algorithm~\ref{alg:algorithmnovel}.

\begin{enumerate}
    \item Draw $P$ particles $\{\bfZ^i_1\}^P_{i=1}$ independently from $\pi_{1}(\bfZ)$, and let $n=2$.

    \item Exit if there is at least one particle that satisfies the sign restriction. That is, there exist $i^{*}$ such that $\mathbf{S}_{R}(T(\bfZ^{i^{*}}_{n-1})) > {\bf{0}}$. Then, set $\bfZ^{0} = \bfZ^{i^{*}}_{n-1}$ and exit the algorithm.

    \item Select $\bfc_{n}$ such that $\bfc_{n-1} < \bfc_{n} \leq 0$.

    \item Set weights for particles from stage $n-1$ by:
\begin{equation*}
    \tilde{w}_n^i = \frac{\pi_{n}(\bfZ^i_{n-1})}{\pi_{n-1}(\bfZ^i_{n-1})} \propto \frac{\left[\mathbf{S}_{R}(T(\bfZ^i_{n-1})) > {\bf{c}}_{n}\right]}{\left[\mathbf{S}_{R}(T(\bfZ^i_{n-1})) > {\bf{c}}_{n-1}\right]}.
\end{equation*}

\item Resample the particles. Let $\{\widehat\bfZ^i_{n}\}_{i=1}^P$ denote $P$ independent draws from a multinomial distribution characterized by support points $\{\bfZ^i_{n-1}\}_{i=1}^P$ and weights proportional to $\{ \tilde{w}_n^i\}_{i=1}^P$ for $i=1,\ldots,P$.

\item Propagate the particles $\{\widehat\bfZ^i_{n}\}_{i=1}^P$ via $M$ steps of Algorithm~\ref{alg:algorithmnovel} with $\left[\mathbf{S}_{R}(T(\bfZ)) > {\bf{c}}_{n}\right]$ to obtain $\{\bfZ^i_{n}\}_{i=1}^P$.

\item Let $n=n+1$, and go to Step 2.
\end{enumerate}
\end{Algorithm}

\bigskip

Our algorithm for the initialization starts from drawing particles from the unrestricted posterior. Then, we move through the ladder of intermediate posteriors using an importance-resampling-mutation scheme.  We terminate as soon as we find a particle $\bfZ_{n}^{i^{*}}$ that satisfies the original sign restrictions. Conceptually, this scheme nests the traditional accept-reject method as a special case with $N=2$ and $\bfc_{2} = {\bf{0}}$: if a feasible draw appears at the first stage, the outcome is identical to accept-reject; otherwise, our procedure proceeds through intermediate posteriors with progressively tighter thresholds until feasibility is reached. Several remarks on Algorithm~\ref{alg:initialization} are described in Online Appendix \ref{app:remarksinitilization}.

\renewcommand{\thesection}{\Roman{section}}
\setcounter{section}{0}

\section*{Online Appendix}

\section{Remarks on Algorithm~\ref{alg:initialization}}\label{app:remarksinitilization}
\paragraph{Remarks on Step 1.} As $\pi_{1}(\bfZ)=p_{\bfZ}(\bfZ \mid \bfyT)$, drawing from $\pi_{1}$ can be done by directly drawing from the unrestricted posterior.

\paragraph{Remarks on Step 2.} This step checks whether any of the particles satisfy the sign restrictions. As our goal is to initialize our Gibbs sampler, it is sufficient to find \textit{one} draw that satisfies the restrictions. Note that combining Step 1 and Step 2 yields the traditional accept-reject algorithm.

\paragraph{Remarks on Step 3.} This step chooses the next threshold level, $\bfc_{n}$. We choose the threshold sequence adaptively to ensure that at least a $\tau$ fraction of the current draws satisfy the restrictions at ${\bfc}_{n}$. Suppose we have particles $\{\bfZ^{i}_{n-1}\}_{i=1}^{P} \sim \pi_{n-1}(\bfZ)$. We then choose $q_{n} \in (0,1)$ as the largest quantile such that
\begin{equation*}
\left( \frac{1}{P}\sum_{i=1}^{P} \prod^R_{j=1}\left[\mathbf{S}_{R,j}(T(\bfZ_{n-1}^{i})) > \mathcal{C}_{j}(q_{n}) \right] \right) \geq \tau,
\end{equation*}
where $\mathcal{C}_{j}(q)$ is the empirical $q$-quantile of $\{\mathbf{S}_{R,j}(T(\bfZ_{n-1}^{i}))\}_{i=1}^{P}$, truncated above by $0$, and $\mathbf{S}_{R,j}(T(\bfZ_{n-1}^{i}))$ is the $j$th sign restriction from $\mathbf{S}_{R}(T(\bfZ_{n-1}^{i}))$ with $j=1,\ldots,R$. Note that we reduce the $R$-dimensional selection problem (choosing $\bfc_{n}$) to a one-dimensional problem by selecting a single quantile value $q_{n}$. We then set $\bfc_{n} = [\mathcal{C}_{1}(q_{n}), \mathcal{C}_{2}(q_{n}), \ldots, \mathcal{C}_{R}(q_{n})]'$.

The computation required for this step should be minimal, since Step 2 already evaluates the sign restrictions for all particles. We create 100 grid points over $(0,1)$ to find $q_{n}$. Note that $\tau$ controls the effective sample size of the particle system. In fact, the effective sample size is $P \times \tau$. By choosing a large $\tau$, the resulting particle system contains more draws that satisfy the restrictions for stage $n$. In such a case, however, the incremental step size becomes very small, and therefore the probability of finding a valid initial value increases only slowly. On the other hand, if $\tau$ is too small, the particle system moves quickly toward the target distribution, but the importance approximation to the stage-$n$ posterior will be poor. In practice, we choose $\tau$ to be 0.20.

\paragraph{Remarks on Step 4.} To obtain the $n$th posterior from the $i-1$th, we reweight the current particles via,
\begin{equation*}
    \tilde{w}_n^i \propto \frac{\left[\mathbf{S}_{R}(T(\bfZ^i_{n-1})) > {\bf{c}}_{n}\right]}{\left[\mathbf{S}_{R}(T(\bfZ^i_{n-1})) > {\bf{c}}_{n-1}\right]} = \left[\mathbf{S}_{R}(T(\bfZ^i_{n-1})) > {\bf{c}}_{n}\right].
\end{equation*}
Because particles are drawn from $\pi_{n-1}(\bfZ)$, the denominator equals one for every draws. Thus, this step simply checks which particles satisfy the tighter stage-$n$ restrictions, similar to accept-reject sampling, but with proposals from $\pi_{n-1}$ rather than from the unrestricted posterior, bringing proposals closer to the target. Again, the computation required for this step should be minimal as $\{ \mathbf{S}_{R}(T(\bfZ_{n-1}^{i}))\}_{i=1}^{P}$ has already been calculated in Step 2.

\paragraph{Remarks on Step 5.} As noted in the previous remarks, the weights in our algorithm take values of either 0 or 1. Therefore, the resampling step consists of randomly sampling the surviving particles without replacement.

\paragraph{Remarks on Step 6.} This step propagates the particle system using a Markov transition kernel whose stationary distribution is $\pi_{n}$. The purpose of this step is to mitigate the weight degeneracy issues frequently encountered in Sequential Monte Carlo algorithms. Although Algorithm~\ref{alg:algorithmnovel} is written for $\pi_N(\bfZ)$, it can be easily adapted to $\pi_n(\bfZ)$ for any $n \in \{2, \ldots, N-1\}$. The adaptation simply involves moving $\bfc_n$ from one side of the inequality to the other.

\paragraph{Final Remarks.} The algorithm without Step 2 is known as a Sequential Monte Carlo algorithm for parameter estimation. To determine the endpoint of the threshold sequence for SMC, one can set $\bfc_{n} = {\bf{0}}$ when the proportion of particles that satisfy the sign restrictions exceeds a pre-specified level such as $\tau$ in Step 3. This modification would make the above algorithm a formal SMC algorithm, and the final particles would approximate the target posterior distribution. As illustrated with several examples in the main text, we find our proposed Gibbs sampler easier to implement and sufficiently efficient. The SMC sampling algorithm may be particularly attractive when many computational cores are available, as the most computationally demanding part (Step 6) is straightforward to parallelize. The last stage of the SMC algorithm can be viewed as running $P$ instances of our Gibbs sampler from different initial values to generate $M$ draws.

Lastly, our initialization method, as well as our Gibbs sampling algorithm, can handle very general forms of sign restrictions on $T(\bfZ)$. Importantly, our restrictions do not need to be linear combinations of the elements from one of the columns of $\gamma(\bfZQ)$. These include restrictions involving multiple shocks simultaneously (e.g., the effect of one shock on inflation being greater than that of another shock), non-linear restrictions (e.g., the sum of squared impulse responses of the first two variables exceeding that of the last two variables), and restrictions pertaining only to $\bfZB$ and $\bfZS$ (e.g., covariance stationarity).

\paragraph{Example.}
Table \ref{tab:time_initial_value} reports the computational time (in seconds) required to find a draw $\bfZ$ that satisfies the restrictions. All calculations are based on the sample period from 1977Q4 to 2020Q3. Our initialization uses 500 particles and sets $\tau = 0.20$. The acceptance–rejection algorithm is based on the computational times reported in Table \ref{tab:infeasibility}.

\begin{table}[t!]
\centering
\begin{tabular}{ccc}
\toprule
\textbf{Number of Shocks} & \textbf{Our Initialization} & \textbf{Accept-reject algorithm} \\
\midrule
8  & 17 & 6\\
9  & 21 & 53\\
10 & 29 & 665\\
\bottomrule
\end{tabular}
\caption{Computation Time (Seconds) to Find an Initial Value by Number of Identified Shocks}
\label{tab:time_initial_value}
\end{table}

\section{Alternative Prior Specifications}\label{app:alt_priors}
Our Gibbs sampler algorithm can be adapted to work with two popular priors in SVAR analysis: the independent normal-inverse-Wishart prior, and the asymmetric priors proposed by \citeOA*{Chanassymetric}. 

\subsection{Independent Normal-Inverse-Wishart Prior}
Let us begin showing how to adapt our Gibbs sampler to the case in which a researcher aims to use an independent normal-inverse-Wishart prior for $(\bfB,\bfS)$ of the form $IW_{(\bar{\nu},\bar\bfPhi)}(\bfS)N_{(\text{vec}(\bar{\bfmu}_{\bfB}), \bar{\bfV}_{\bfB})}(\text{vec}(\bfB))$. When using this prior, the final objective of the researcher is to sample from the following posterior, which is equivalent to Equation~\eqref{eqn:posteriorSigns} in the main text:
\begin{equation}\label{eqn:posteriorGibbsIndependentPriors}
\begin{gathered}
   p(\bfB,\bfS,\bfQ \mid \bfyT,\mathbf{S}_R(\bfB,\bfS,\bfQ) > {\bf0}) = \\
   \frac{\left[\mathbf{S}_R(\bfB,\bfS,\bfQ)> {\bf0}\right] NIW_{(\hat{\nu},\hat\bfPhi,\hat\bfPsi,\hat\bfOme)}(\bfB,\bfS) IW_{(\bar{\nu},\bar\bfPhi)}(\bfS) N_{(\text{vec}(\bar{\bfmu}_{\bfB}), \bar{\bfV}_{\bfB})}(\text{vec}(\bfB)) \kappa}{\Pr\left(\mathbf{S}_R(\bfB,\bfS,\bfQ) > {\bf0} \mid \bfyT\right)}.
\end{gathered}
\end{equation}
where
\begin{equation*}
\begin{gathered}
    \Pr\left(\mathbf{S}_R(\bfB,\bfS,\bfQ) > {\bf0} \mid \bfyT\right) = \\
    \adjustbox{max width=\textwidth}{$\displaystyle \smashoperator[r]{\int_{\mathbb{R}^{m\times n}\times\mathcal{S}(n)\times\mathcal{O}(n)}} \left[\mathbf{S}_R(\bfB,\bfS,\bfQ) > {\bf0}\right] NIW_{(\hat{\nu},\hat\bfPhi,\hat\bfPsi,\hat\bfOme)}(\bfB,\bfS) IW_{(\bar{\nu},\bar\bfPhi)}(\bfS)N_{(\text{vec}(\bar{\bfmu}_{\bfB}),\bar{\bfV}_{\bfB})}(\text{vec}(\bfB))\kappa\, d\bfB\,d\bfS\,d\bfQ,$}
\end{gathered}
\end{equation*}
and $NIW_{(\hat{\nu},\hat\bfPhi,\hat\bfPsi, \hat\bfOme)}(\bfB,\bfS)$ denotes the likelihood, with
$\hat{\nu}=T-m-(n+1)$, $\hat\bfOme=(\bfX^{\prime}\bfX)^{-1}$,
$\hat\bfPsi=\hat\bfOme\bfX^{\prime}\bfY$, and
$\hat\bfPhi=\bfY^{\prime}\bfY-\hat\bfPsi^{\prime}\hat\bfOme^{-1}\hat\bfPsi$,
where $\bfY^{\prime}=(\bfy_{1},\dots,\bfy_t)$ and $\bfX^{\prime}=(\bfx_{1},\dots,\bfx_t)$. Table \ref{tab:independentpriors} describes the conditional posterior distributions obtained from Equation~\eqref{eqn:posteriorGibbsIndependentPriors}---using similar arguments to those given in Section \ref{sec:newalgorithm}---where 
\begin{gather}
\tilde{\nu}=\bar{\nu}+T,  \nonumber \\ 
\tilde\bfPhi=\bar\bfPhi+(\bfY-\bfX \bfB)^{\prime}(\bfY-\bfX \bfB)               ,\nonumber \\
\tilde{\bfV}_{\bfB}^{-1}=\bar{\bfV}_{\bfB}^{-1}+ (\bfS^{-1}\otimes \bfX^{\prime}\bfX),  \text{and} \nonumber  \\ \tilde{\bfmu}_{\bfB}=\tilde{\bfV}_{\bfB}(\bar{\bfV}_{\bfB}^{-1}\bar{\bfmu}_{\bfB}+(\bfS^{-1}\otimes \bfX^{\prime}\bfX)\hat\bfPsi).\nonumber 
\end{gather}
To obtain draws from Equation \eqref{eqn:posteriorGibbsIndependentPriors}, we use Algorithm~\ref{alg:algorithmnovel} adapted to consider each conditional distribution in Table~\ref{tab:independentpriors}.

\begin{table}[!htbp]
\centering
\begin{tabular}{ll}
\toprule
$p_{\bfZ}(\bfZQ \mid H^{\bfQ}_{\bfZB,\bfZS,\bfZQ})$ &  $\propto \left[\mathbf{S}_R(\bfZB,\zeta(\bfZS),\gamma(\bfZQ))> {\bf0}\right] N_{(\mathbf{0}_{n \times n}, \bfI_n, \bfI_n)}(\bfZQ)$ \\
$p_{\bfZ}(\bfZS \mid H^{\bfS}_{\bfZB,\bfZS,\bfZQ})$ &  $\propto \left[\mathbf{S}_R(\bfZB,\zeta(\bfZS),\gamma(\bfZQ))> {\bf0}\right] N_{(\mathbf{0}_{n \times \tilde{\nu}}, \tilde{\bfPhi}^{-1}, \bfI_{\tilde{\nu}})}(\bfZS)$ \\ 
$p_{\bfZ}(\bfZB \mid H^{\bfB}_{\bfZB,\bfZS,\bfZQ})$  &
$\propto \left[\mathbf{S}_R(\bfZB,\zeta(\bfZS),\gamma(\bfZQ))> {\bf0}\right] N_{(\text{vec}(\tilde{\bfmu}_{\bfB}), \tilde{\bfV}_{\bfB})}(\text{vec}(\bfZB))$\\ \bottomrule
\end{tabular} 
\caption{Conditional Posterior Distributions}
\label{tab:independentpriors}
\end{table}

\subsection{Asymmetric Priors}
Instead of directly working with the typical conjugate normal-inverse-Wishart prior over the reduced-form parameters ($\bfB,\bfS$), \citeOA*{Chanassymetric} works with structural parameters subject to a recursive identification scheme and denotes the resulting parameterization as $(\bftheta,\bfsigma^{2})$. This parameterization implies a prior over the reduced-form parameters $(\bfB,\bfS)$ that allows the shrinkage strength for coefficients on own lags to be different than the shrinkage strength for coefficients on the lag of other variables.

For completeness, we will describe the mapping from $(\bftheta,\bfsigma^{2})$ to ($\bfB,\bfS$). To this end, it is useful to begin by considering the recursive SVAR proposed by \citeOA*{Chanassymetric}:
\begin{eqnarray}\label{eqn:chanrecursivestructural}
\bfA^{S} \bfy_t &=& \bfb^{S} + \bfB_1^{S} \bfy_{t-1} +  \cdots +  \bfB_1^{S} \bfy_{t-p} + \bfe_t^{y}, \quad  \bfe_t^{y}\sim N(\mathbf{0},\bfS^{S})
\end{eqnarray}
where $\bfA^{S}$ is a lower triangular matrix with ones along the diagonal, $\bfb^{S}$, $\bfB_1^{S}$,\ldots, $\bfB_p^{S}$, denote a constant term and the slope coefficients, and $\bfS^{S}=\text{diag}(\bfsigma^{2})$, where $\bfsigma^{2}=(\sigma_1^{2},\dots,\sigma_n^{2})$. Given this representation, we use the following notation: $b_i^{S}$ denotes the $i$-th element of $\bfb^{S}$, $\bfb_{j,i}^{S}$ represents the $i$-th row of $\bfB_j^{S}$, $\bfbeta_i=(b_i^{S},\bfb_{1,i}^{S},\dots,\bfb_{p,i}^{S})^{\prime}$ collects the constant and all the VAR slope coefficients associated with the $i$-th equation, and $\bfalpha_i=(A_{i,1},\dots,A_{i,i-1})^{\prime}$ is a vector containing the unrestricted elements of the $i$-th row of $\bfA^{S}$, where $A_{i,j}$ denotes the $i$-th row and the $j$-th column of $\bfA^{S}$.

For $i=1,\dots,n$, let $\bftheta_{i}  = \left[\bfbeta_{i}^{\prime},\bfalpha_{i}^{\prime}\right]^{\prime}$. Furthermore, set $\bftheta =(\bftheta_{1}^{\prime},\dots,\bftheta_n^{\prime})^{\prime}$.  If we define $\bfw_{i,t}^{S}=(-y_{1,t},\dots,-y_{i-1,t})$, $\bfx_t^{S}=(\bfy_{t-1}^{\prime},\dots,\bfy_{t-p}^{\prime},1)$, $\bfx_{i,t}=(\bfx_t^{S},\bfw_{i,t}^{S})$, $\bfy_i=(y_{i,1},\dots,y_{i,T})^{\prime}$, and $\bfX_i=(\bfx_{i,1}^{\prime},\dots,\bfx_{i,T}^{\prime})^{\prime}$, then $\bfy_{i} = \bfX_i \bftheta_i + \bfe_{i}^{y}, \quad  \bfe_{i}^{y}\sim N(\mathbf{0},\sigma_{i}^{2}\bfI_T)$. Importantly, Equation~\eqref{eqn:chanrecursivestructural} can be written as
\begin{eqnarray}\label{eqn:chanrf}
\bfy_t &=& \bfd^{\prime} +  \bfB_1^{\prime}  \bfy_{t-1} + \cdots +  \bfB_p^{\prime}  \bfy_{t-p} + \bfe_t^{y}, \quad  \bfe_t^{y}\sim N(\mathbf{0},\bfS)
\end{eqnarray}
where $\bfd^{\prime}=(\bfA^{S})^{-1}\bfb^{S}$, $\bfB_j^{\prime}=(\bfA^{S})^{-1} \bfB_j^{S}$ for $j=1,\dots,p$, $\bfB=\left[\bfB_1^{\prime},\dots,\bfB_p^{\prime},\bfd^{\prime}\right]^{\prime}$, and $\bfS=(\bfA^{S})^{-1}\bfS^{S}((\bfA^{S})^{-1})^{\prime}$. Hence, we have implicitly defined a mapping from $(\bftheta,\bfsigma^{2})$ to $(\bfB,\bfS)$.

The mapping is invertible. Given $(\bfB,\bfS)$ we can obtain $(\bftheta,\bfsigma^{2})$ by using the Cholesky composition to obtain $\bfA^{S}$ and $\bfS^{S}$, and then we can construct $\bfbeta_i$ by using $\bfb^{S}=\bfA^{S}\bfd^{\prime}$,  $\bfB_1^{S}=\bfA^{S}\bfB_1^{\prime}$, $\dots$ $\bfB_p^{S}=\bfA^{S}\bfB_p^{\prime}$,  for $i=1,\dots,n$. Putting together the mapping defined above with the mapping $\phi$ defined in Equation \eqref{eqn:RQtoL}, we obtain a one-to-one mapping between the impulse responses, that is $(\bfL_0,\bfL_+)$, and $(\bftheta,\bfsigma^{2},\bfQ)$. 

Exploiting Proposition 2 in \citetOA*{arias2025uniform}, we will combine the product of normal-inverse-gamma over $(\bftheta,\bfsigma^{2})$ proposed by \citeOA*{Chanassymetric}, i.e., 
\begin{equation}\label{eqn:asymmetricprior}
p(\bftheta,\bfsigma^{2})=\prod_{i=1}^{n}p(\bftheta_i,\sigma_i^{2})=\prod_{i=1}^{n}IG_{(\nu_i,S_i)}(\sigma^{2}_i)N_{(\bfm_i,\sigma_i^{2}\bfV_i)}(\bftheta_i),
\end{equation}
with a uniform prior over $\bfQ$ under the Haar measure in order to induce a prior over $(\bfL_0,\bfL_+)$. Then, as in the case in which we work with the orthogonal reduced-form parameters ($\bfB,\bfS,\bfQ$), we will conduct posterior inference over $(\bfL_0,\bfL_+)$ subject to sign restrictions by sampling from the following posterior:
\begin{equation}\label{eqn:posteriorGibbsAsymmetricPriors}
\begin{gathered}
    p(\bftheta,\bfsigma^2,\bfQ \mid \bfyT,\mathbf{S}(\bftheta,\bfsigma^2,\bfQ) > {\bf0}) = \\
    \adjustbox{max width=\textwidth}{$\displaystyle \frac{\left[\mathbf{S}(\bftheta,\bfsigma^2,\bfQ)> {\bf0}\right]\prod_{i=1}^{n} IG_{(\tilde{\nu}_i,\tilde{S}_i)}(\sigma_{i}^{2})N_{(\tilde{\bftheta}_i,\sigma_{i}^{2}\bfK_{\bftheta_i}^{-1})}(\bftheta_i)}{\Pr\left(\mathbf{S}(\bftheta,\bfsigma^2,\bfQ) > {\bf0} \mid \bfyT\right)},$}
\end{gathered}
\end{equation}
where
\begin{equation*}
\begin{gathered}
    \Pr\left(\mathbf{S}(\bftheta,\bfsigma^2,\bfQ) > {\bf0} \mid \bfyT\right) = \\
    \adjustbox{max width=\textwidth}{$\displaystyle \smashoperator[r]{\int_{\left(\prod_{i=1}^{n}\mathbb{R}^{k_i}\right)\times\mathbb{R}_{+}^{n}\times\mathcal{O}(n)}} \left[\mathbf{S}(\bftheta,\bfsigma^2,\bfQ) > \bf0\right]\prod_{i=1}^{n} IG_{(\tilde{\nu}_i,\tilde{S}_i)}(\sigma_{i}^{2})N_{(\tilde{\bftheta}_i,\sigma_{i}^{2}\bfK_{\bftheta_i}^{-1})}(\bftheta_i)\, d\bftheta\, d\bfsigma^{2}\, d\bfQ$}
\end{gathered}
\end{equation*}
and then converting the draws to the desired parameterization; where $\tilde{\nu}_i=\nu_i+\frac{T}{2}$, $\tilde{S}_i = S_i + (\bfy_i^{\prime}\bfy_i + \bfm_i^{\prime}\bfV_{i}^{-1}\bfm_i - \tilde{\bftheta}_{i}^{\prime}\bfK_{\bftheta_i}\tilde{\bftheta}_{i})/2$, $\bfy_i=(\bfy_{i1},\dots,\bfy_{iT})^{\prime}$, $\bfy_{it}$ denotes the i-th entry of $\bfy_t$, $\bfK_{\bftheta_i}=\bfV_i^{-1} + \bfX_i^{\prime}\bfX_i$,  and $\tilde{\bftheta}_i=\bfK_{\bftheta_i}^{-1}(\bfV_{i}^{-1}\bfm_{i} + \bfX_i^{\prime}\bfy_i)$. Next, we show how to adapt the ideas in Sections~\ref{sec:changeofvariable} and~\ref{sec:algorithmmainGibbs} to conduct inference based on the posterior distribution in Equation~\eqref{eqn:posteriorGibbsAsymmetricPriors}. 

\subsubsection*{Change of Variables}\label{app:asymmetric_changeofvariable}
Let $\bfZ_{\bftheta}=(\bfZ_{\bftheta_1},\dots,\bfZ_{\bftheta_n})$ with $\bfZ_{\bftheta_i}\in \mathbb{R}^{k_i}$, where $k_i=\dim(\bftheta_i)$, for $i=1,\dots,n$, and $\bfZ_{\bfsigma^{2}}=(\bfZ_{\sigma_1^{2}},\dots,\bfZ_{\sigma_n^{2}})$ with $\bfZ_{\sigma_i^{2}}\in \mathbb{R}^{1\times 2\tilde{\nu}_{i}}$ for $i=1,\dots,n$. Given these definitions, we can proceed as in Section~\ref{sec:changeofvariable} and let $\bfZ_{A}=(\bfZ_{\bftheta},\bfZ_{\bfsigma^{2}},\bfZQ)\in \wtZ_{A}$, with $\wtZ_{A} = \left(\prod_{i=1}^{n}\mathbb{R}^{k_i}\right)\times\left(\prod_{i=1}^{n}\mathbb{R}^{1\times 2\tilde{\nu}_{i}}\right)\times \mathbb{R}^{n \times n}$, and let $p_{\bfZ_{A}}(\bfZ_{A}\mid \bfyT,\mathbf{S}(T_{A}(\bfZ_{A})) > \bf0)$ denote the following conditional Lebesgue density, which is equivalent to Equation~\eqref{eqn:transf_post_dens} in the main text:
\begin{equation}\label{eqn:transf_post_dens_asymmetric}
 \adjustbox{max width=\textwidth}{$\displaystyle \frac{\left[\mathbf{S}(T_{A}(\bfZ_{A})) > \bf0\right]\left(\prod_{i=1}^{n} N_{(\tilde{\bftheta}_i,\zeta(\bfZ_{\sigma_i^{2}})\bfK_{\bftheta_i}^{-1})}(\bfZ_{\bftheta_i})\right)\left(\prod_{i=1}^{n} N_{(\mathbf{0}_{1 \times 2\tilde{\nu}_{i}}, (2\tilde{S}_{i})^{-1}, \bfI_{2\tilde{\nu}_{i}})}(\bfZ_{\sigma_i^{2}})\right) N_{(\mathbf{0}_{n \times n}, \bfI_{n}, \bfI_{n})}(\bfZQ)}{\Pr\left(\mathbf{S}(T_{A}(\bfZ_{A})) > {\bf0} \mid \bfyT\right)},$}
\end{equation}
where $\Pr\left(\mathbf{S}(T_{A}(\bfZ_{A})) > {\bf0} \mid \bfyT\right)$ equals:
\begin{equation*}
\adjustbox{max width=\textwidth}{$\displaystyle \smashoperator[r]{\int_{\mathcal{Z}_A}}\left[\mathbf{S}(T_{A}(\bfZ_{A})) > \bf0\right] \left(\prod_{i=1}^{n} N_{(\tilde{\bftheta}_i,\zeta(\bfZ_{\sigma_i^{2}})\bfK_{\bftheta_i}^{-1})}(d\bfZ_{\bftheta_i})\right)\left(\prod_{i=1}^{n} N_{(\mathbf{0}_{1 \times 2\tilde{\nu}_{i}}, (2\tilde{S}_{i})^{-1}, \bfI_{2\tilde{\nu}_{i}})}(d\bfZ_{\sigma_i^{2}})\right) N_{(\mathbf{0}_{n \times n}, \bfI_{n}, \bfI_{n})}(d\bfZQ)$}
\end{equation*}
and 
\begin{equation*}
T_{A}(\bfZ_{A})= \left(\bfZ_{\bftheta},\varphi(\bfZ_{\bfsigma^{2}}),\gamma(\bfZQ)\right),
\end{equation*}
with $\varphi(\bfZ_{\bfsigma^{2}})=(\varphi_{1}(\bfZ_{\sigma_1^{2}}),\dots,\varphi_{n}(\bfZ_{\sigma_n^{2}}))$ and $\varphi_{i}(\bfZ_{\sigma_i^{2}})=(\bfZ_{\sigma_i^{2}}\bfZ_{\sigma_i^{2}}^{\prime})^{-1}\in \mathbb{R}_{+}$.

Considering the transformed random vector $T_A(\bfZ_A)$, the measurable sets $\mathcal{A}_{\bftheta}\subseteq \prod_{i=1}^{n}\mathbb{R}^{k_i}$, $\mathcal{A}_{\bfsigma^2}\subseteq \mathbb{R}_{+}^{n}$, and $\wtC\subseteq \mathcal{O}(n)$, and defining $(\bftheta,\bfsigma^2,\bfQ)=T_{A}(\bfZ_{A})$, we have the following expression, which is equivalent to Equation~\eqref{eqn:distBSQ} in the main text:
\begin{eqnarray}
\label{eqn:distThetaSigmaQ}
&& \Pr\left(T_{A}(\bfZ_A)\in \mathcal{A}_{\bftheta}\times \mathcal{A}_{\bfsigma^2}\times \wtC \mid  \bfyT, \mathbf{S}(T_{A}(\bfZ_{A})) > \bf0\right)= \\
&& \adjustbox{max width=\textwidth}{$\displaystyle \smashoperator[r]{\int_{\wtC\times\mathcal{A}_{\bfsigma^2}\times\mathcal{A}_{\bftheta}}} \frac{\left[\mathbf{S}(\bftheta,\bfsigma^2,\bfQ) > \bf0\right]\left(\prod_{i=1}^{n} N_{(\tilde{\bftheta}_i,\sigma_{i}^{2}\bfK_{\bftheta_i}^{-1})}(d\bftheta_i)\right)\left(\prod_{i=1}^{n} IG_{(\tilde{\nu}_{i},\tilde{S}_{i})}(d\sigma_{i}^{2})\right)\kappa\, d\bfQ}{\Pr\left(\mathbf{S}(T_{A}(\bfZ_A)) > {\bf0} \mid \bfyT\right)},$}\nonumber
\end{eqnarray}
where
\begin{equation*}
IG_{(\tilde{\nu}_{i},\tilde{S}_{i})}(d\sigma_{i}^{2}) = (\varphi_{i\#} N(\mathbf{0}_{1 \times 2\tilde{\nu}_{i}}, (2\tilde{S}_{i})^{-1}, \bfI_{2\tilde{\nu}_{i}}))(d\sigma_{i}^{2})
\quad \text{and} \quad
\kappa\, d\bfQ = (\gamma_{\#} N(\mathbf{0}_{n \times n}, \bfI_n, \bfI_n))(d\bfQ),
\end{equation*}
denote the push-forward measures associated with the mappings $\varphi_{i}\colon \mathbb{R}^{1\times 2\tilde{\nu}_{i}} \to \mathbb{R}_{+}$ and $\gamma\colon \mathbb{R}^{n\times n} \to\mathcal{O}(n)$. As a consequence, the density of the random vector $T_A(\bfZ_A)$ associated with Equation~\eqref{eqn:distThetaSigmaQ} coincides with the density in Equation~\eqref{eqn:posteriorGibbsAsymmetricPriors}, that is, the posterior distribution of interest.

\subsubsection*{Elliptical Slice within Gibbs Sampler}\label{app:asymmetric_sampler}
We now describe an elliptical slice within Gibbs sampler that targets the conditional density $p_{\bfZ_A}$ in Equation~\eqref{eqn:transf_post_dens_asymmetric} and returns draws from the posterior of interest via the transformation $T_A$.

\paragraph{Conditional Posterior for $\bfZQ$.}
Using Bayes' rule and independence of $(\bfZ_{\bftheta},\bfZ_{\bfsigma^{2}})$ and $\bfZQ$, Equation~\eqref{eqn:transf_post_dens_asymmetric} implies:
\begin{equation*}
\adjustbox{max width=\textwidth}{$\displaystyle p_{\bfZ_A}(\bfZQ \mid \bfZ_{\bftheta}, \bfZ_{\bfsigma^{2}}, \bfyT,\mathbf{S}(T_{A}(\bfZ_A)) > {\bf0}) \propto \left[\mathbf{S}(T_{A}(\bfZ_A)) > {\bf0}\right]N_{(\mathbf{0}_{n \times n}, \bfI_n, \bfI_n)}(\bfZQ).$}
\end{equation*}

\paragraph{Conditional Posterior for $\bfZ_{\bfsigma^2}$.}
Similarly, conditioning on $(\bfZ_{\bftheta},\bfZQ)$ yields:
\begin{equation*}
\begin{gathered}
p_{\bfZ_A}(\bfZ_{\bfsigma^{2}} \mid \bfZ_{\bftheta}, \bfZQ, \bfyT,\mathbf{S}(T_{A}(\bfZ_A)) > {\bf0}) \propto \\
\adjustbox{max width=\textwidth}{$\displaystyle \left[\mathbf{S}(T_{A}(\bfZ_A)) > {\bf0}\right]\left(\prod_{i=1}^{n} N_{(\mathbf{0}_{1 \times 2\tilde{\nu}_{i}}, (2\tilde{S}_{i})^{-1}, \bfI_{2\tilde{\nu}_{i}})}(\bfZ_{\sigma_i^2})\right)\left(\prod_{i=1}^{n} N_{(\tilde{\bftheta}_i,\zeta(\bfZ_{\sigma_i^{2}})\bfK_{\bftheta_i}^{-1})}(\bfZ_{\bftheta_i})\right).$}
\end{gathered}
\end{equation*}

\paragraph{Conditional Posterior for $\bfZ_{\bftheta}$.}
Finally, conditioning on $(\bfZ_{\bfsigma^2},\bfZQ)$ gives:
\begin{equation*}
\adjustbox{max width=\textwidth}{$\displaystyle p_{\bfZ_A}(\bfZ_{\bftheta} \mid \bfZ_{\bfsigma^{2}}, \bfZQ, \bfyT,\mathbf{S}(T_{A}(\bfZ_A)) > {\bf0}) \propto \left[\mathbf{S}(T_{A}(\bfZ_A)) > {\bf0}\right]\prod_{i=1}^{n} N_{(\tilde{\bftheta}_i,\zeta(\bfZ_{\sigma_i^{2}})\bfK_{\bftheta_i}^{-1})}(\bfZ_{\bftheta_i}).$}
\end{equation*}

\paragraph{Algorithm.}
Crucially, each of the conditional densities defined above is proportional to the product of a multivariate Gaussian and an arbitrary function that we can evaluate. This is exploited in an elliptical slice within Gibbs sampler analogous to Algorithm~\ref{alg:algorithmnovel}. We derive a Markov chain $(\bfZ_{A}^{j})_{j\in\mathbb{N}}=(\bfZ_{\bftheta}^{j},\bfZ_{\bfsigma^{2}}^{j},\bfZ_{\bfQ}^{j})_{j\in\mathbb{N}}$ and the transformed chain $(\bftheta^{j},(\bfsigma^2)^{j},\bfQ^{j})_{j\in\mathbb{N}}=(T_A(\bfZ_{A}^{j}))_{j\in\mathbb{N}}$.

\begin{Algorithm}\label{alg:algorithmnovelAsymmetricPriors}
This algorithm draws samples $(\bfZ_{A}^{j})_{0\leq j \leq J}$ from an elliptical slice within Gibbs sampler and returns transformed samples $(\bftheta^{j},(\bfsigma^2)^{j},\bfQ^{j})_{1\leq j \leq J}$.
\begin{enumerate}
\item  Set $J > 1$, initialize $j = 1$, and choose initial values $\bfZ_{A}^{0}\in \wtZ_A$ such that $\mathbf{S}(T_A(\bfZ_{A}^{0})) > {\bf0}$.
\item \label{algo:asymZ-stepQ}
Use elliptical slice sampling to approximately draw $\bfZ_{\bfQ}^{j}$, targeting
\begin{equation*}
p_{\bfZ_{A}}(\bfZQ \mid \bfZ_{\bftheta}^{j-1}, \bfZ_{\bfsigma^{2}}^{j-1}, \bfyT,\mathbf{S}(\bfZ_{\bftheta}^{j-1},\varphi(\bfZ_{\bfsigma^{2}}^{j-1}),\gamma(\bfZQ)) > {\bf0}),
\end{equation*}
and set $\bfQ^{j}=\gamma(\bfZ_{\bfQ}^{j})$.
\item \label{algo:asymZ-stepS}
Use elliptical slice sampling to approximately draw $\bfZ_{\bfsigma^{2}}^{j}$, targeting
\begin{equation*}
p_{\bfZ_{A}}(\bfZ_{\bfsigma^{2}} \mid \bfZ_{\bftheta}^{j-1}, \bfZ_{\bfQ}^{j}, \bfyT,\mathbf{S}(\bfZ_{\bftheta}^{j-1},\varphi(\bfZ_{\bfsigma^{2}}),\gamma(\bfZ_{\bfQ}^{j})) > {\bf0}),
\end{equation*}
and set $(\bfsigma^{2})^{j}=\varphi(\bfZ_{\bfsigma^{2}}^{j})$.
\item \label{algo:asymZ-stepB}
Use elliptical slice sampling to approximately draw $\bfZ_{\bftheta}^{j}$, targeting
\begin{equation*}
p_{\bfZ_{A}}(\bfZ_{\bftheta} \mid \bfZ_{\bfsigma^{2}}^{j}, \bfZ_{\bfQ}^{j}, \bfyT,\mathbf{S}(\bfZ_{\bftheta},\varphi(\bfZ_{\bfsigma^{2}}^{j}),\gamma(\bfZ_{\bfQ}^{j})) > {\bf0}),
\end{equation*}
and set $\bftheta^{j}=\bfZ_{\bftheta}^{j}$.
\item If $j < J$, increment $j$ and return to Step~\ref{algo:asymZ-stepQ}.
\end{enumerate}
\end{Algorithm}

\section{Large-SVAR with Asymmetric Priors}\label{app:largeSVARasymmetric}
To conclude the appendix, we reproduce the analysis in Section \ref{sec:largeSVAR} using the asymmetric priors proposed by \citetOA*{Chanassymetric}. In this case, we restrict the sample to the Great Moderation period 1985Q1--2019Q4 in order to have a direct comparison with the impulse responses reported in their paper. Section \ref{sec:irfsasymmetric} shows that both the accept-reject and the Gibbs sampler deliver the same impulse responses. Section \ref{ref:asymmetric} shows that the timing of the {accept-reject algorithm} increases exponentially as the identified sets become tighter. If we extend the sample to the period 1977Q4--2020Q3, as in an earlier version of the replication files that the authors shared with us, the timing of the accept-reject algorithm increases to the point of becoming practically infeasible while our Gibbs sampler remains equally efficient, see Section \ref{sec:infeasibility}. 

\subsection{Impulse Responses} \label{sec:irfsasymmetric}
We begin by reporting the impulse responses {for each of} the eight structural shocks identified by \citetOA*{chan2025largestructuralvarsmultiple} obtained using the Gibbs sampler described in  Algorithm \ref{alg:algorithmnovelAsymmetricPriors}. The results are reported in Figures \ref{fig:asymmmetricdemand}-\ref{fig:asymmmetricwagebargaining}. The solid red lines and the orange-shaded areas depict the point-wise posterior median and 68 percent probability bands implied by the Gibbs sampler algorithm; the dashed gray lines and the gray shaded areas depict the point-wise 68 percent posterior probability bands implied by the accept-reject algorithm. Analogous to the results relying on the Minnesota prior, when using the Gibbs sampler, we obtain one million draws and retain one every 10, and when using the accept-reject method, we obtain 1,000 draws. {As can be seen,} the results are nearly identical to those reported in \citetOA*{chan2025largestructuralvarsmultiple}. Notice that in this case we can extend the horizon of the impulse responses beyond 20 quarters without a large increase in posterior uncertainty. This is because the asymmetric priors used in \citetOA*{chan2025largestructuralvarsmultiple} allow for more flexibility in terms of shrinkage of the slope coefficients of the reduced-form representation of the SVAR.

\subsection{Timing}\label{ref:asymmetric}
\begin{figure}[!htbp]
\vspace*{-1cm}
  \centering
 \begin{subfigure}[t]{0.48\linewidth}
    \centering
    \caption{Demand shock}
    \includegraphics[width=0.95\linewidth]{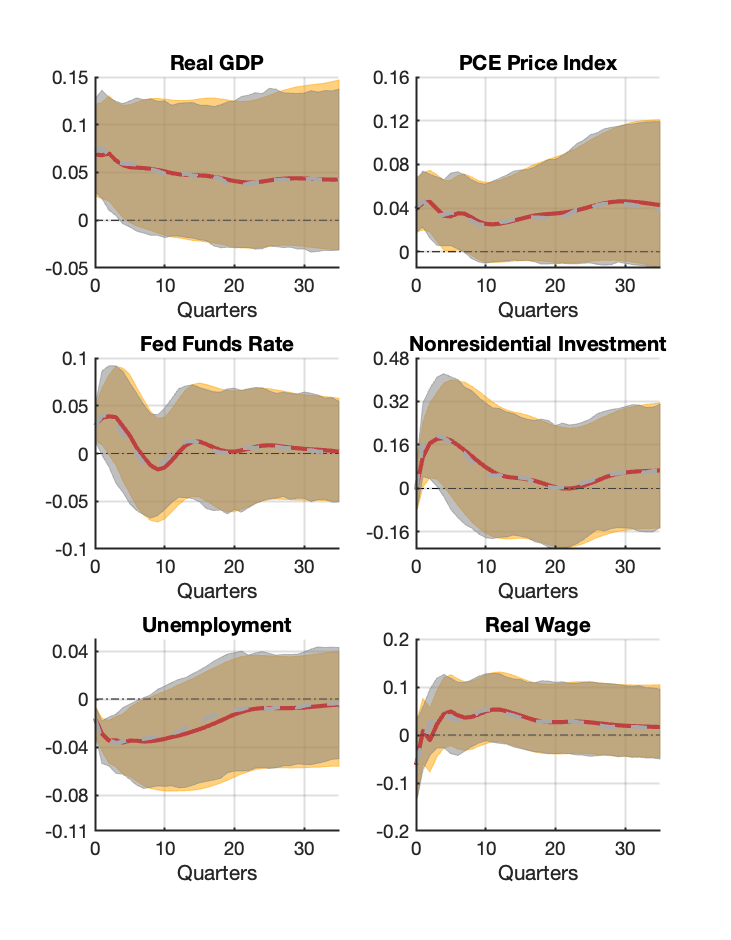}
    \label{fig:asymmmetricdemand}
  \end{subfigure}
  \hfill
  \begin{subfigure}[t]{0.48\linewidth}
    \centering
    \caption{Investment shock}
    \includegraphics[width=0.95\linewidth]{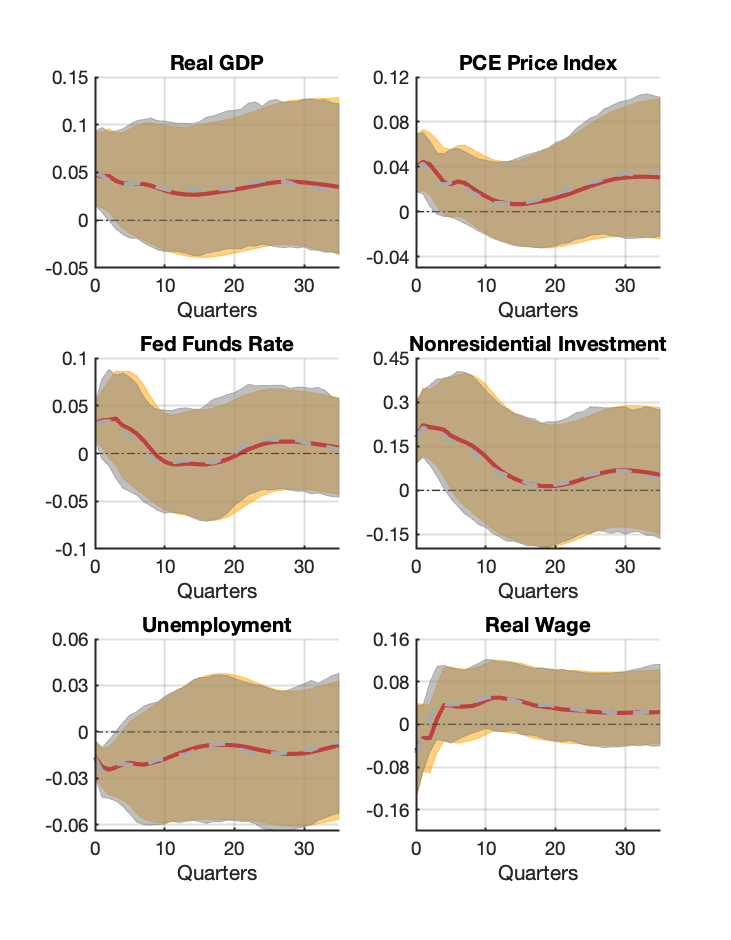}
    \label{fig:asymmmetricinvestment}
  \end{subfigure}
  
  \vspace{0pt} 
  
  \begin{subfigure}[t]{0.48\linewidth}
    \centering
    \caption{Financial shock}
    \includegraphics[width=0.95\linewidth]{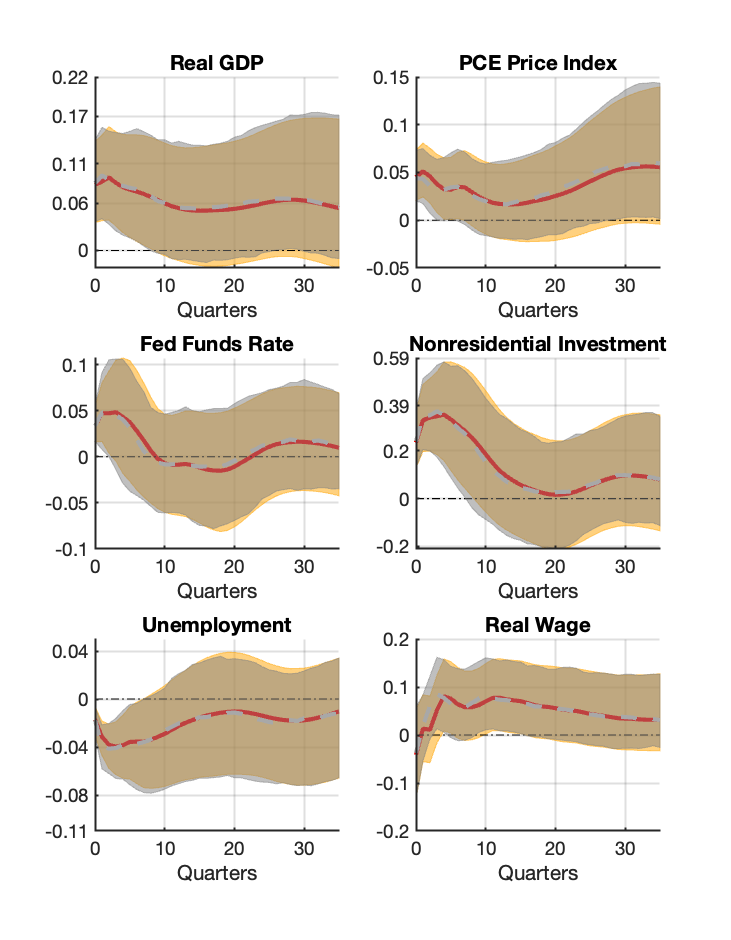}
    \label{fig:asymmmetricfinancial}
  \end{subfigure}
  \hfill
  \begin{subfigure}[t]{0.48\linewidth}
    \centering
    \caption{Monetary policy shock}
    \includegraphics[width=0.95\linewidth]{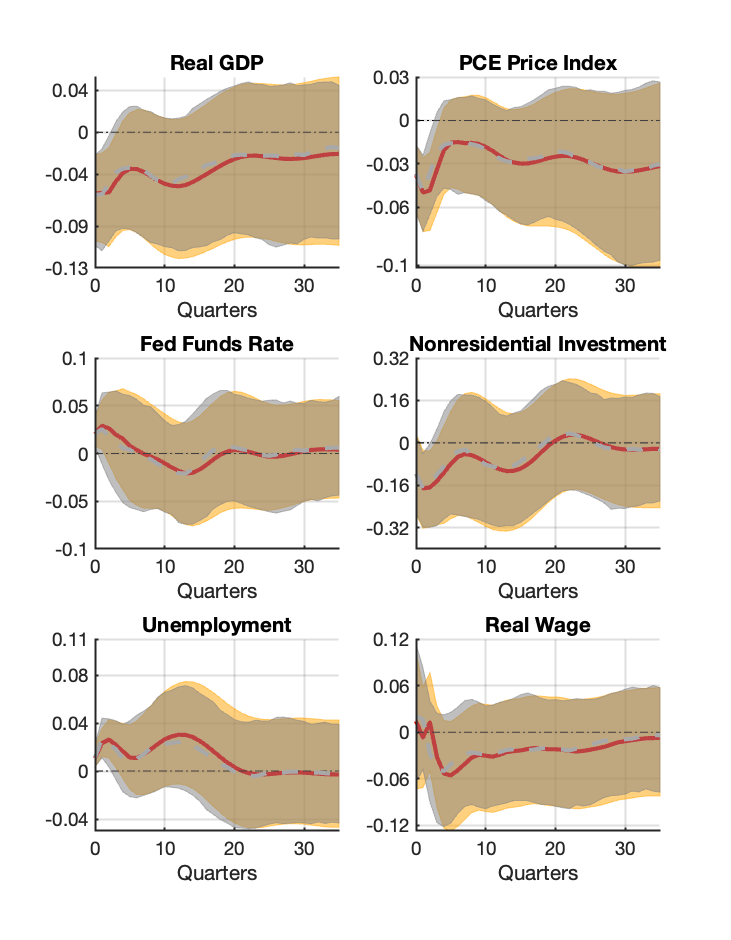}
    \label{fig:asymmmetricmonetary}
  \end{subfigure}
  
  \vspace{-20pt} 
  \caption{Impulse Responses}
  \label{fig:asymmmetricdemand_financial_monetary_shocks}
      \begin{minipage}{\textwidth}
\footnotesize\textit{Note:} The solid red lines and the orange-shaded areas depict the point-wise posterior median and 68 percent probability bands implied by the Gibbs sampler algorithm; the dashed gray lines and the gray shaded areas depict the point-wise 68 percent posterior probability bands implied by the accept-reject algorithm.
\end{minipage}
\end{figure}

\begin{figure}[!htbp]
\vspace*{-1cm}
  \centering
  \begin{subfigure}[t]{0.48\linewidth}
    \centering
    \caption{Government spending shock}
    \includegraphics[width=0.95\linewidth]{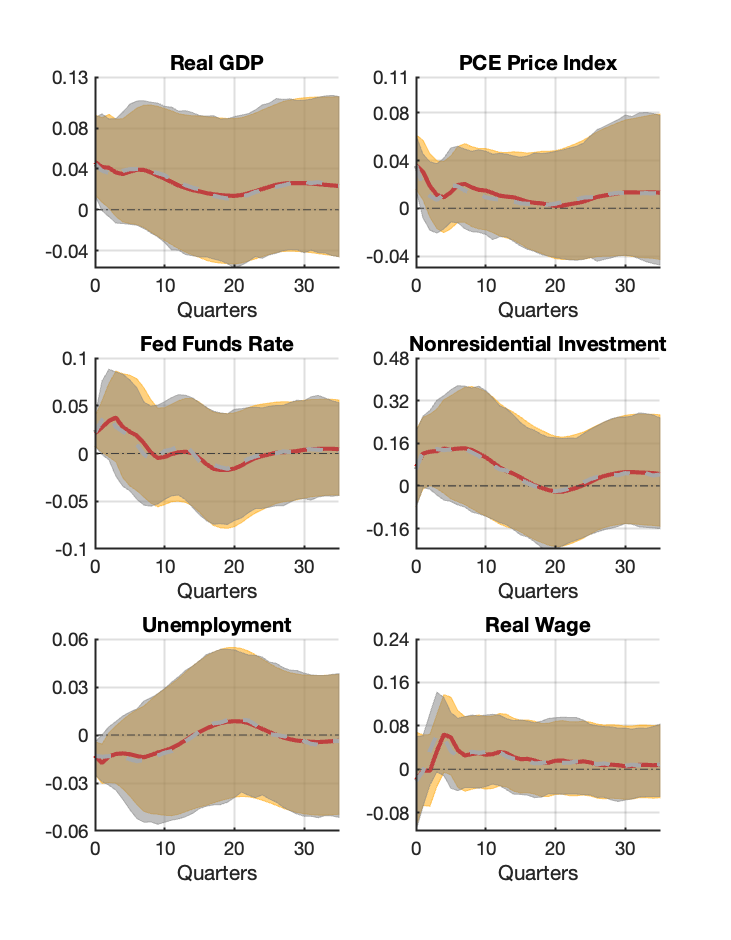}
    \label{fig:asymmmetricGspending}
  \end{subfigure}
  \hfill
  \begin{subfigure}[t]{0.48\linewidth}
    \centering
   \caption{Technology shock}
    \includegraphics[width=0.95\linewidth]{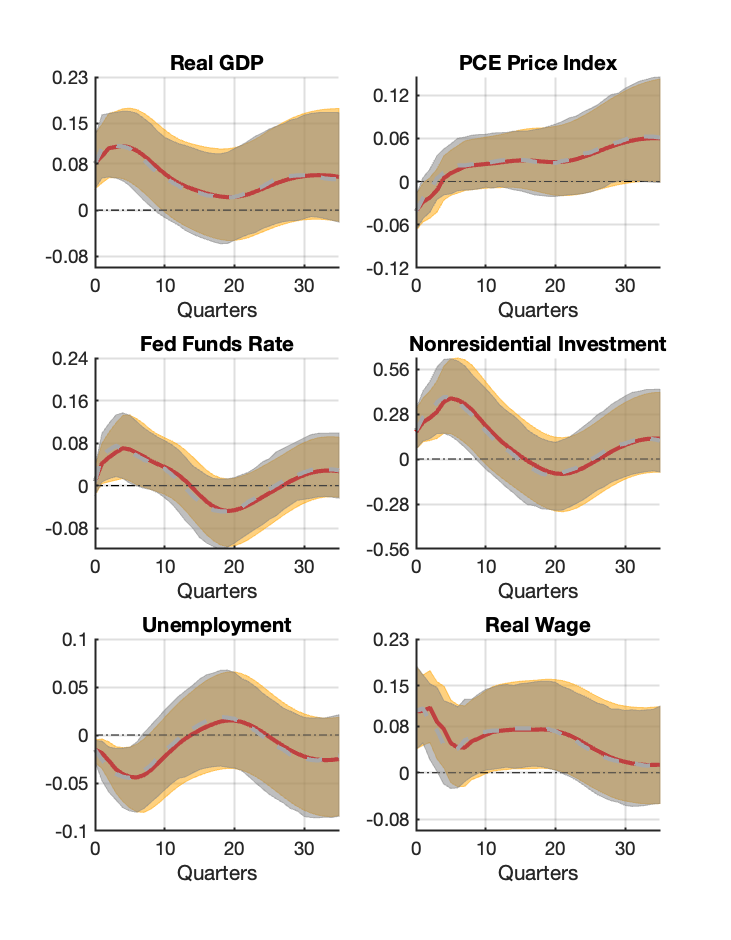}
    \label{fig:asymmmetrictechnology}
  \end{subfigure}
  
  \vspace{0pt} 
  
  \begin{subfigure}[t]{0.48\linewidth}
    \centering
    \caption{Labor supply shock}
    \includegraphics[width=0.95\linewidth]{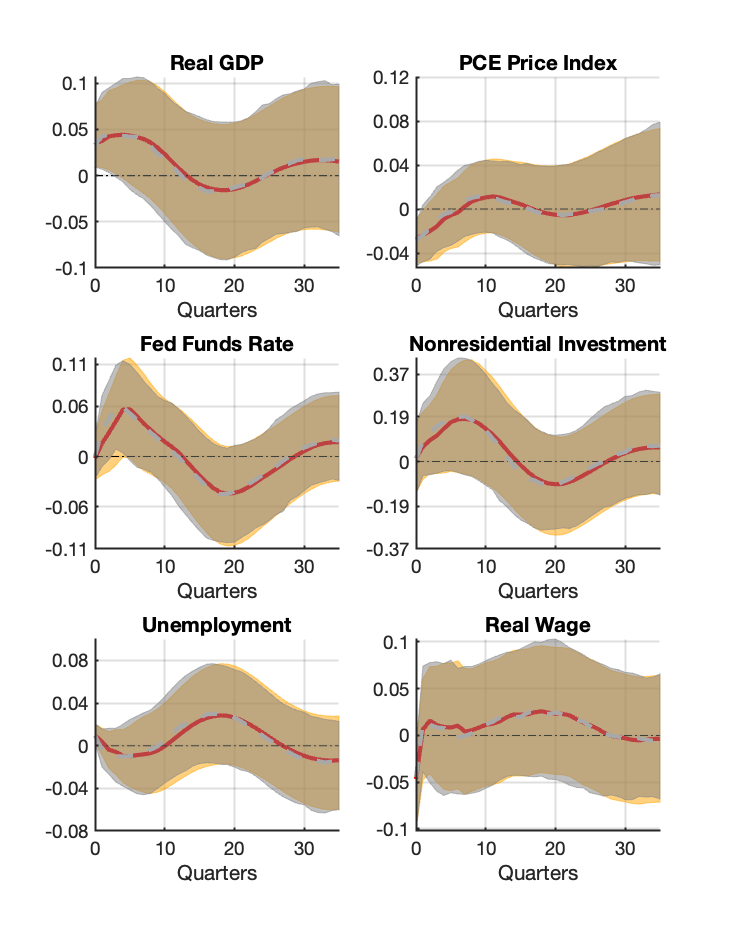}
    \label{fig:asymmmetriclaborsupply}
  \end{subfigure}
  \hfill
  \begin{subfigure}[t]{0.48\linewidth}
    \centering
    \caption{Wage bargaining shock}
    \includegraphics[width=0.95\linewidth]{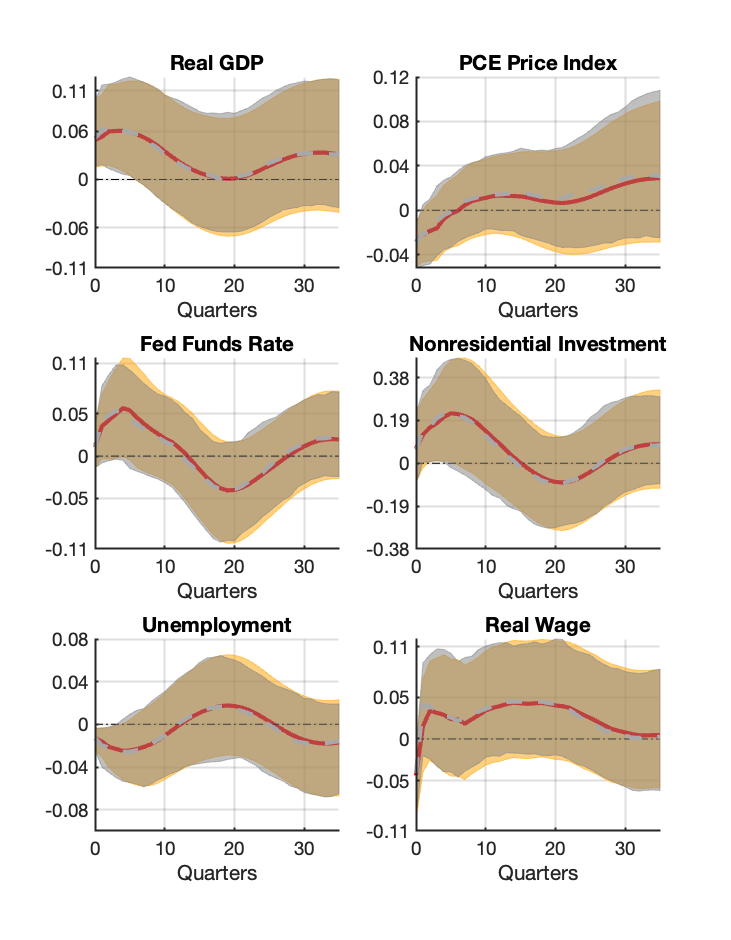}
    \label{fig:asymmmetricwagebargaining}
  \end{subfigure}
  
  \vspace{-20pt} 
  \caption{Impulse Responses}
  \label{fig:asymmmetricmore}    \begin{minipage}{\textwidth}
\footnotesize\textit{Note:} The solid red lines and the orange-shaded areas depict the point-wise posterior median and 68 percent probability bands implied by the Gibbs sampler algorithm; the dashed gray lines and the gray shaded areas depict the point-wise 68 percent posterior probability bands implied by the accept-reject algorithm.
\end{minipage}
\end{figure}

\begin{figure}[!htbp]
  \centering
  \begin{subfigure}[t]{0.325\linewidth}
    \centering
    \caption{Gibbs Sampler}
    \label{fig:timing_ess_large_SVAR_asymmetric}
    \includegraphics[width=\linewidth]{figures_replication/timing_ess_large_SVAR_1to8.eps}
  \end{subfigure}
  \hfill
  \begin{subfigure}[t]{0.325\linewidth}
    \centering
    \caption{Accept-Reject}
    \label{fig:timing_Talgo_large_SVAR_asymmetric}
    \includegraphics[width=\linewidth]{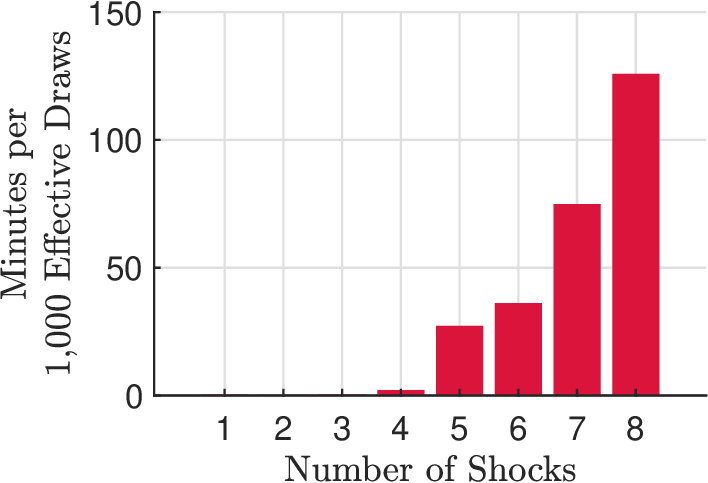}
  \end{subfigure}
  \hfill
  \begin{subfigure}[t]{0.325\linewidth}
    \centering
    \caption{Comparison}
    \label{fig:timing_BOTH_large_SVAR_asymmetric}
    \includegraphics[width=\linewidth]{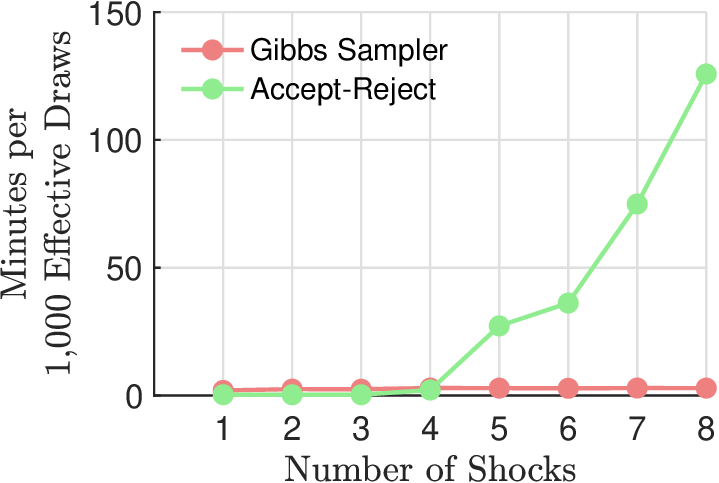}
  \end{subfigure}
  \caption{Time Per 1,000 Effective Draws}
  \label{fig:timing_three_panel_asymmetric}
\end{figure}

\begin{figure}[!htbp]
  \centering
  \begin{subfigure}[t]{0.435\linewidth}
    \centering
    \caption{Gibbs Sampler Shocks (1 to 9)}
    \label{fig:timing_ess_large_SVAR_1to9_asymmetric}
    \includegraphics[width=\linewidth]{figures_replication/timing_ess_large_SVAR_1to9.eps}
  \end{subfigure}
  \hfill
  \begin{subfigure}[t]{0.435\linewidth}
    \centering
    \caption{Gibbs Sampler Shocks (1 to 10)}
    \label{fig:timing_ess_large_SVAR_1to10_asymmetric}
    \includegraphics[width=\linewidth]{figures_replication/timing_ess_large_SVAR_1to10.eps}
  \end{subfigure}
  
  \vspace*{0.5cm} 
  
  \begin{subfigure}[t]{0.435\linewidth}
    \centering
    \caption{Accept-Reject Shocks (1 to 9)}
    \label{fig:timing_Talgo_large_SVAR_1to9_asymmetric}
    \includegraphics[width=\linewidth]{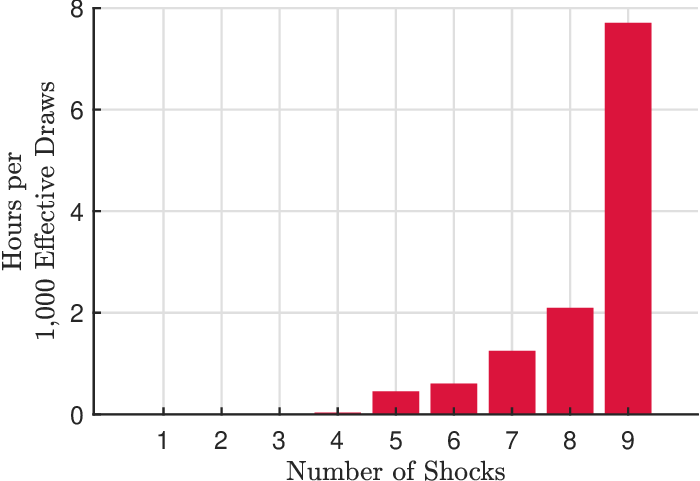}
  \end{subfigure}
  \hfill
  \begin{subfigure}[t]{0.435\linewidth}
    \centering
    \caption{Accept-Reject Shocks (1 to 10)}
    \label{fig:timing_Talgo_large_SVAR_1to10_asymmetric}
    \includegraphics[width=\linewidth]{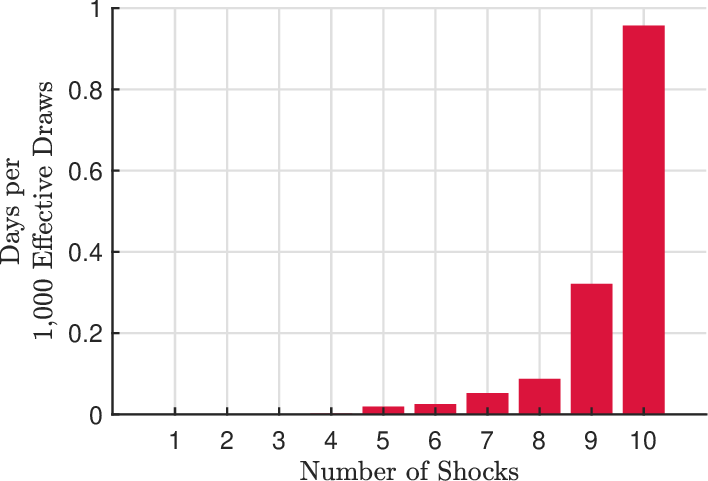}
  \end{subfigure}
  \caption{Gibbs Sampler vs. Accept-Reject}
  \label{fig:timing_comparison_all_asymmetric}
  \vspace{0.5em}
    \begin{minipage}{\textwidth}
\small\textit{Note:} The time of the accept-reject algorithm for shocks 9 and 10 {is extrapolated from 10 draws}.
\end{minipage}
\end{figure}

\begin{figure}[t!]
  \centering
  \begin{subfigure}[t]{0.435\linewidth}
    \centering
    \caption{Comparison Shocks (1 to 9)}
    \label{fig:timing_BOTH_large_SVAR_1to9_asymmetric}
    \includegraphics[width=\linewidth]{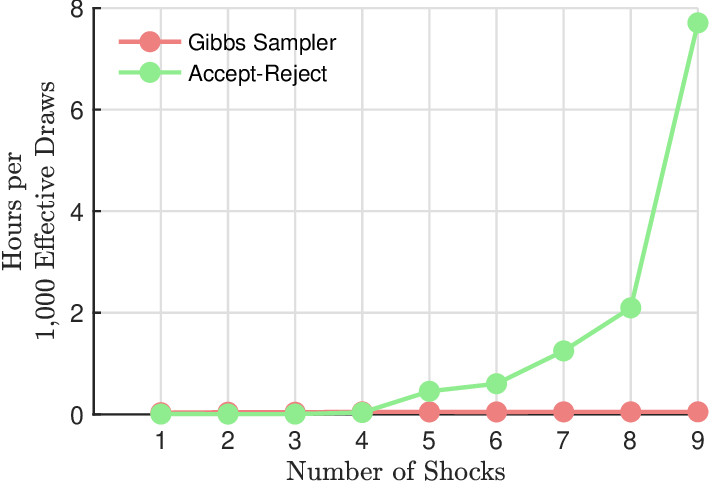}
  \end{subfigure}
  \hfill
  \begin{subfigure}[t]{0.435\linewidth}
    \centering
    \caption{Comparison Shocks (1 to 10)}
    \label{fig:timing_BOTH_large_SVAR_1to10_asymmetric}
    \includegraphics[width=\linewidth]{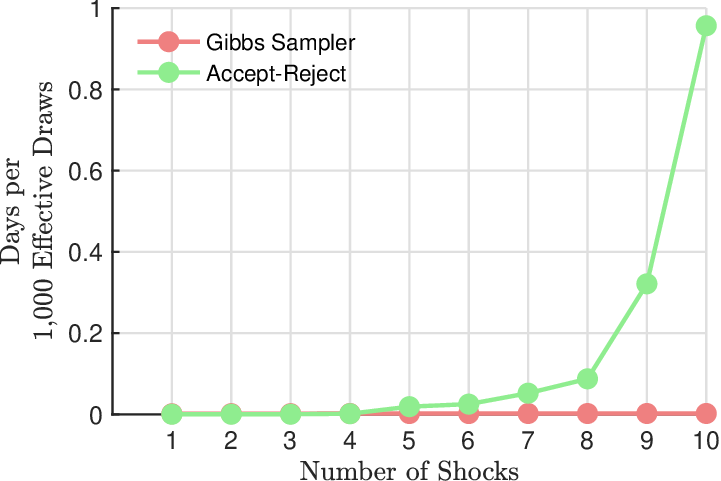}
  \end{subfigure}
  \caption{Gibbs Sampler vs. Accept-Reject}
  \label{fig:timing_both_1to9_1to10_asymmetric}
  \begin{minipage}{\textwidth}
\end{minipage}
\end{figure}

In this section, we compare the efficiency of our Gibbs sampler algorithm {to that of the accept-reject algorithm when using the asymmetric priors}. We highlight three results. First, there are cases in which when using asymmetric priors with a relatively small number of sign restrictions, the accept-reject algorithm is somewhat more efficient than our approach, see for example the timings for the cases where up to 3 shocks are identified. Second, as shown in Figure \ref{fig:timing_three_panel_asymmetric}, the advantage of the accept-reject vanishes as we increase the number of shocks. Notice that the Gibbs sampler algorithm is faster than the baseline specification in \citetOA*{chan2025largestructuralvarsmultiple}, which features 8 structural shocks. Third, as shown in Figure \ref{fig:timing_comparison_all_asymmetric}, as we continue to increase the number of sign restrictions by adding shocks 9 and 10, respectively, the accept-reject algorithm reproduces the explosive patterns in Figures \ref{fig:simple}-\ref{fig:timing_both_1to9_1to10_asymmetric}. In contrast, the efficiency  of the Gibbs sampler algorithm remains nearly unchanged. 

\subsection{Practical Infeasibility for Real-Time Applications}\label{sec:infeasibility}
Table \ref{tab:infeasibility} compares the computational time it takes to obtain 1,000 effective draws when identifying 8, 9, and 10 structural shocks over the period 1977Q4--2020Q3, respectively. {The timings reported in Table~\ref{tab:infeasibility} demonstrate that the accept-reject algorithm could become practically infeasible for real-time applications.} In contrast, the Gibbs Sampler remains computationally tractable.

\clearpage
\vspace*{0pt} 

\begin{table}[!htbp]
    \centering
    \begin{tabular}{ccc}\toprule
Number of Shocks     &  Gibbs Sampler   & Accept-Reject  \\ \hline
8    &   3.1 Minutes&    100.2 Minutes \\
9     &  3.2 Minutes &     14.6 Hours \\
10    & 3.1 Minutes  &  7.7  Days\\ \hline \hline
    \end{tabular}
    \caption{Time Per 1,000 Effective Draws}
    \label{tab:infeasibility}
      \vspace{0.5em}
    \begin{minipage}{\textwidth}
\small\textit{Note:} The times of the accept-reject algorithm  are extrapolated based on 10 draws.
\end{minipage}
\end{table}

\bibliographystyleOA{chicago}
\bibliographyOA{myrefsappendix}

\end{document}